\documentclass[12pt, fleqn]{article}

\usepackage[T1]{fontenc}

\usepackage[nottoc]{tocbibind}

\usepackage{pifont}
\usepackage{multirow}

\usepackage{tikz}
\usetikzlibrary{calc}
\usetikzlibrary{decorations.pathreplacing}
\usetikzlibrary{decorations.markings}
\usetikzlibrary{intersections} 
\usetikzlibrary{er,positioning}
\usetikzlibrary{arrows,shapes}
\usetikzlibrary{decorations.markings}
\usetikzlibrary{decorations.pathreplacing,calligraphy}
\usepackage{BOONDOX-cal} 

\usepackage[title]{appendix}
\usepackage[margin=2cm]{geometry}

\usepackage{amsmath, amssymb, amsthm, amsfonts}

\makeatletter
\newcommand{\doublewidetilde}[1]{{%
  \mathpalette\double@widetilde{#1}%
}}
\newcommand{\double@widetilde}[2]{%
  \sbox\z@{$\m@th#1\widetilde{#2}$}%
  \ht\z@=.9\ht\z@
  \widetilde{\box\z@}%
}
\makeatother

\usepackage[colorlinks=true, linktoc=all, linkcolor=black, citecolor=red, urlcolor=blue, backref=page]{hyperref}

\usepackage{etoolbox}
\makeatletter
\patchcmd{\BR@backref}{\newblock}{\newblock(}{}{}
\patchcmd{\BR@backref}{\par}{)\par}{}{}
\makeatother

\numberwithin{equation}{section}
\usepackage{empheq}

\usepackage{cite} 

\newcommand{\n}{n}

\newcommand{\TL}{\mathcal{T\!\!L}}
\newcommand{\JTL}{\mathcal{J\!T\!\!L}}
\newcommand{\uJTL}{u\mathcal{J\!T\!\!L}}
\newcommand{\PTL}{P\!\mathcal{T\!\!L}}
\newcommand{\B}{\mathcal{B}}
\newcommand{\PP}{\mathcal{P}}
\newcommand{\ATL}{\mathcal{A\!T\!\!L}}
\newcommand{\RB}{\mathcal{R\!B}}
\newcommand{\dTL}{d\mathcal{T\!\!L}}
\newcommand{\MM}{\mathcal{M}}

\DeclareRobustCommand{\stirling}{\genfrac\{\}{0pt}{}}

\begin{document}

\begin{flushleft}
{\bfseries\sffamily\Large 
Spaces of states
\\
of the two-dimensional $O(n)$ and Potts models
\vspace{1.5cm}
\\
\hrule height .6mm
}

\vspace{1.5cm}

{\bfseries\sffamily 
Jesper Lykke Jacobsen$^{1,2}$, Sylvain Ribault, Hubert Saleur$^3$
}
\vspace{4mm}

{\textit{\noindent
Institut de physique th\'eorique, CEA, CNRS, 
Universit\'e Paris-Saclay
\\ \vspace{2mm}
$^1$ Also at: Laboratoire de Physique de l'\'Ecole Normale Sup\'erieure, ENS, Universit\'e PSL,
CNRS, Sorbonne Universit\'e, Universit\'e de Paris, F-75005 Paris, France
\\ 
$^2$ Also at: Sorbonne Universit\'e, \'Ecole Normale Sup\'erieure, CNRS,
Laboratoire de Physique (LPENS), F-75005 Paris, France
\\ 
$^3$ Also at: 
Department of Physics and Astronomy, University of Southern California, Los Angeles
}}
\vspace{2mm}

{\textit{E-mail:} \texttt{jesper.jacobsen@ens.fr,
sylvain.ribault@ipht.fr, hubert.saleur@ipht.fr
}}
\end{flushleft}
\vspace{7mm}

{\noindent\textsc{Abstract:}
We determine the spaces of states of the two-dimensional $O(n)$ and $Q$-state Potts models with generic parameters $n,Q\in \mathbb{C}$ as representations of their known symmetry algebras. While the relevant representations of the conformal algebra were recently worked out, it remained to determine the action of the global symmetry groups: the orthogonal group for the $O(n)$ model, and the symmetric group $S_Q$ for the $Q$-state Potts model.

We do this by two independent methods. First we compute the twisted torus partition functions of the models at criticality. The twist in question is the insertion of a group element along one cycle of the torus: this breaks modular invariance, but allows the partition function to have a unique decomposition into characters of irreducible representations of the global symmetry group. 

Our second method reduces the problem to determining branching rules of certain diagram algebras. For the $O(n)$ model, we decompose representations of the Brauer algebra into representations of its unoriented Jones--Temperley--Lieb subalgebra. For the $Q$-state Potts model, we decompose representations of the partition algebra into representations of the appropriate subalgebra. We find explicit expressions for these decompositions as sums over certain sets of diagrams, and over standard Young tableaux. 

We check that both methods agree in many cases. Moreover, our spaces of states are consistent with recent bootstrap results on four-point functions of the corresponding CFTs. 
}

\pagebreak

\tableofcontents

\vspace{5mm}
\hrule
\vspace{5mm}

\hypersetup{linkcolor=blue}

\section{Introduction}

\subsubsection{Lattice models and conformal field theories}

The $O(n)$ model and the $Q$-state Potts model are one-parameter families of statistical models, which have quite a few compelling features:
\begin{itemize}
 \item For special values of $n$ or $Q$, they reduce to important models such as the Ising model or percolation. We consider generic values $n,Q\in\mathbb{C}$ of the parameters, which interpolate between these important models, and offer the hope of a unified approach. 
 \item Each model has a number of variants, depending on the lattice geometry, the interactions, or the choice of a dense or dilute phase. However, each model also has a critical limit, where all these variants are described by the same conformal field theory: the $O(n)$ and Potts CFTs. 
 \item In two dimensions, early exact results on critical exponents \cite{Baxter73, nie82} raise the hope of analytically solving the models.
\end{itemize}
In this article, we take a significant step towards exactly solving these models, by determining their spaces of states. We build on earlier work such as the calculation of torus partition functions \cite{fsz87} and the analysis of enlarged symmetries \cite{rs07}, and complete the picture by determining the action of the global symmetry groups: the orthogonal group $O(n)$ for the $O(n)$ model, and the symmetric group $S_Q$ for the Potts model. 

Our analysis will start with specific lattice realizations, namely the $O(n)$ model on a hexagonal lattice in the dense phase, and the Potts model on a square lattice. We will then take a critical limit $\underset{\text{critical}}{\lim}$, which means 
\begin{itemize}
 \item sending the lattice size $L$ to infinity,
 \item sending the coupling $K$ to a critical value $K_c$,
 \item restricting to scaling states via a double limit procedure. 
\end{itemize}
In the case of symmetry algebras or spaces of states, we will write the critical limit in a heuristic sense, and not attempt to define it precisely. In the case of the main objects that we will compute, namely partition functions and branching coefficients, the critical limit can be defined precisely. Actually, branching coefficients behave in a particularly simple way, as each coefficient becomes $L$-independent for $L$ large enough. 

It is only for technical convenience that we choose specific lattice realizations, and focus on the critical limit. Our results actually hold off-criticality, and for a wide class of lattice realizations. In particular, our algebraic results only assume rather generic symmetry requirements on the interactions.

\subsubsection{A preview of the main results}

Let us sketch our results for the $O(n)$ model, which are slightly simpler than for the Potts model. The action of $O(n)$ on the space of states is described by a family of representations $\Lambda_{(r,s)}$ with $r\in\frac12 \mathbb{N}^*$ and $s\in \frac{1}{r}\mathbb{Z}$. A representation $\Lambda_{(r,s)}$ is a multiplicity space, which describes how the group $O(n)$ acts on a space of fields with given properties under conformal transformations.
Our two independent methods lead to two different expressions:
\begin{align}
 \Lambda_{(r,s)} &= \delta_{r,1}\delta_{s\in 2\mathbb{Z}+1}[] + \frac{1}{2r} \sum_{r'=0}^{2r-1} e^{\pi ir's} U_{\mathrm{gcd}(2r, r')}\left(
 \delta_{\frac{2r}{\mathrm{gcd}(2r, r')}\in 2\mathbb{N}}[] + \sum_{k=0}^{\frac{2r}{\mathrm{gcd}(2r, r')}-1}(-1)^k \left[\tfrac{2r}{\mathrm{gcd}(2r, r')}-k,1^k\right]
 \right)\ ,
 \label{prer}
 \\
 &= \sum_{\substack{|\lambda|\leq 2r \\ |\lambda| \equiv 2r\bmod 2}}
 \sum_{T\in T_\lambda}
 \sum_{\omega\in \Omega^{(2r)}_{|\lambda|}} 
 \delta_{e^{\pi i\omega\left(s - \frac{\operatorname{ind}(T)}{r}\right)}, 1}
  \lambda\ .
  \label{prea}
\end{align}
The first expression, reproduced from Eq.~\eqref{lrs}, involves irreducible representations of $O(n)$ written as Young diagrams $[], \left[\frac{2r}{\mathrm{gcd}(2r, r')}-k,1^k\right]$, together with Chebyshev polynomials $U_{\mathrm{gcd}(2r, r')}$. The second expression, reproduced from Eqs.~\eqref{stou} and \eqref{clrs}, is a sum over Young diagrams $\lambda$, Young tableaux $T$, and over orbit lengths $\omega$ for the action of the cyclic group $\mathbb{Z}_{2r}$ on certain sets $\Omega^{(2r)}_{|\lambda|}$ of perfect matchings. The first few examples are given explicitly in Eqs. \eqref{lxxx-all}.

While both formulas are reasonably compact, each one has its advantages. The first formula is more explicit, as it does not involve sums over combinatorial sets. However, it involves not only direct sums of irreducible representations, but also formal combinations with negative and even complex coefficients. The second formula makes it manifest that the coefficient of each irreducible representation $\lambda$ is a positive integer. 

Similarly, the action of the symmetric group $S_Q$ on the space of states of the Potts model is described by a family of representations $\Xi_{(r,s)}$ with $r\in \mathbb{N}+2$ and $s\in\frac{1}{r}\mathbb{Z}$. The analog of the first formula is Eq. \eqref{xirs}, while the analog of the second formula is found in Eqs. \eqref{stos} and \eqref{dlrs}. The formulas for $\Xi_{(r,s)}$ and for $\Lambda_{(r,s)}$ are quite similar. In fact, their characters are related (up to simple contributions of $[],[1]$) as $\chi_{\Xi_{(r,s)}}^{S_Q} = PS_{p_k\to p_k-2}\left(\chi_{\Lambda_{(\frac{r}{2},2s)}}^{O(n)}\right)$, where $PS_{p_k\to p_k-2}$ is the plethystic substitution that subtracts $2$ from any power sum polynomial $p_k$. 

Both formulas \eqref{prer} and \eqref{prea} for $\Lambda_{(r,s)}$ are derived as limits of $L$-dependent quantities when the lattice size $L\in\mathbb{N}$ goes to infinity. At finite $L$, the formulas are not expected to agree, because they arise from two different lattice realizations of the $O(n)$ model. The formulas for $\Lambda_{(r,s)}$ are conjectural because they involve simple but unproven assumptions: on the large $L$ limit of sums over lattice configurations with a fixed topology in one case, and on the $L$-independence of branching coefficients in the other case. (The rest of the derivations may be sketchy, but they amount to proofs.) The equality between the two formulas for $\Lambda_{(r,s)}$ is therefore a conjecture, motivated physically by the fact that our two lattice realizations belong to the same universality class. 

The resulting spaces of states have already been used for interpreting conformal bootstrap results on the $O(n)$ model \cite{gnjrs21} and on the $Q$-state Potts model \cite{niv22}. In both cases, the bootstrap results are consistent with our spaces of states. 

\subsubsection{Conformal field theories and their symmetries}

The $O(n)$ and Potts CFTs are two-dimensional CFTs with local conformal symmetry. This symmetry is described by the conformal algebra $\mathfrak{C}_c$, i.e.\ the product of two copies of the Virasoro algebra, called the left-moving and right-moving copies. 
The central charge $c$ of the conformal algebra is related to the parameters $n,Q$ of the models via a variable $\beta^2$ such that 
\begin{align}
 c= 13-6\beta^2 - \frac{6}{\beta^2}\ , 
\end{align}
together with
\begin{align}
 n = -2\cos(\pi \beta^2) \quad , \quad Q=4\cos^2(\pi \beta^2) \ . 
 \label{nqb}
\end{align}
In particular, the dilute $O(n)$ model with $n\in [-2,2]$ leads to $\beta^2\in  [1,2]$, while the dense $O(n)$ model with $n\in (-2,2)$ leads to $\beta^2\in (0,1)$ \cite{nie82}.
Actually, the $O(n)$ model (dense or dilute) is expected to have a critical limit for any $n$ in a fairly large region of the complex plane, whose intersection with the real line is the interval $n\in (-2,2)$ \cite{gnjrs21}. Recent numerical results in a related model suggest that for the dense $O(n)$ model, this region might be defined by the condition that the energy operator is irrelevant \cite{bjjz22},
\begin{align}
 \Re \Delta_{(1,3)} > 1 \iff \Re \beta^{-2} > 1\ . 
\end{align}
Similarly, the critical $Q$-state Potts model is usually defined for $Q\in (0,4)$, which leads to $\beta^2\in (0,1)$ in the critical limit \cite{Baxter73}. A dilute version can also be defined by adding vacancies, leading to $\beta^2\in [1,2]$ in the critical limit --- this is called the Potts tri-critical universality class \cite{PhysRevLett.43.737}. We expect that the model and its critical limit still exist for complex values of $Q$.

Large as they may seem, the regions where the $O(n)$ and Potts models have critical limits are still smaller than the regions where the corresponding CFTs make sense, because a lattice model may flow away from the CFT under renormalization due to the presence of relevant operators. From the CFT point of view, the values of $\beta^2$ are only constrained by the convergence of the OPE, which requires \cite{prs16}
\begin{align}
 \Re \beta^2 > 0 \iff \Re c < 13 \ .
\end{align}
Conformal bootstrap studies are consistent with idea that the $O(n)$ and Potts CFTs can be analytically continued to this large region of parameter space \cite{gnjrs21, niv22}.

In addition to constraints from conformal symmetry, the spaces of states and correlation functions of the $O(n)$ and Potts CFTs are constrained by the existence of the degenerate fields $V_{\langle 1,3\rangle}$ and $V_{\langle 1,2\rangle}$ respectively. These further constraints can be interpreted in terms of extensions of the conformal algebra, called interchiral algebras \cite{hjs20}. Calling $T,\bar T$ the left- and right-moving energy-momentum tensors, we define the interchiral algebras $\widetilde{\mathfrak{C}}_{\beta^2}$ and $\doublewidetilde{\mathfrak{C}}_{\beta^2}$ by
\begin{align}
 \mathfrak{C}_c = \text{Span}\left(T, \bar T\right) \quad \subset  \quad \widetilde{\mathfrak{C}}_{\beta^2} = \text{Span}\left(T,\bar T, V_{\langle 1, 3\rangle}\right) 
 \quad \subset  \quad \doublewidetilde{\mathfrak{C}}_{\beta^2} = \text{Span}\left(T,\bar T, V_{\langle 1, 2\rangle}\right) \ .
\end{align}
The inclusion $\widetilde{\mathfrak{C}}_{\beta^2}\subset \doublewidetilde{\mathfrak{C}}_{\beta^2}$ is because the degenerate field $V_{\langle 1,3\rangle}$ can be obtained from $V_{\langle 1, 2\rangle}$ by fusing it with itself. The interchiral algebras no longer depend on $c$ but on $\beta^2$, i.e.\ they are not invariant under $\beta\to \beta^{-1}$ --- just like the $O(n)$ and Potts models themselves.

Due to interchiral symmetry, the conformal characters in the $O(n)$ and Potts CFTs' partition functions come in infinite families, which can be summed into interchiral characters. However, this is not crucial for the calculation of the partition functions, which we will write in terms of conformal characters. Interchiral symmetry is more crucial in the algebraic approach, because interchiral representations have natural counterparts in the corresponding lattice models, whereas conformal representations do not. 

\subsubsection{Partition functions and twisted partition functions}

In two-dimensional CFT, the torus partition function can constrain or even determine the space of states. A first approach to determining the torus partition function, which may be called the modular bootstrap, is to solve the constraint of modular invariance. This approach has been used for classifying minimal models, and determining their spaces of states. Another approach is to compute the partition function using a dynamical definition of the theory. In this article, we will compute the $O(n)$ and Potts CFT's partition functions using their definitions as lattice models.

Once we know the partition function, it remains to decompose it into characters of the conformal algebra, and deduce the structure of the space of states. In simple cases such as minimal models, the space of states is completely determined by the partition function. This is no longer true in the $O(n)$ and Potts CFTs, because the partition function only encodes the conformal dimensions of states. These conformal dimensions are not enough for understanding the action of the conformal algebra in logarithmic representations, where Virasoro operator $L_0$ is not diagonalizable. Moreover, these conformal dimensions tell us nothing about the action of the global symmetry group $O(n)$ or $S_Q$, which come with their own quantum numbers. 

While the structure of logarithmic representations has been recently determined \cite{nr20, glhjs20}, it remained to determine the action of the global symmetry groups. In the case of $O(n)$, this was recently conjectured \cite{gnjrs21}. We will prove the conjecture by computing the twisted partition function, and perform the analogous calculation in the case of the Potts CFT. 

Our twisted partition functions take into account not only the action of the Virasoro algebra, but also the action of the global symmetry group. Introducing such twists is indeed a standard method for determining the action of symmetry groups or algebras beyond conformal symmetry. Technically, the twist is a topological defect that depends on an element of the global symmetry group. This allows the twisted partition function to be a combination not just of conformal characters, but also of characters of the global symmetry group. The price to pay is that the twist breaks modular invariance. However, modular invariance plays no role in our approach.

\subsubsection{Algebraic approach}

We start with a lattice model whose space of states is of the type 
\begin{align}
 \mathcal{S}_L=V^{\otimes L}\ ,
\end{align}
where $L$ is the lattice size, and $V$ is a representation of the global symmetry group $G\in\{O(n),S_Q\}$: the defining representation of $O(n)$ in the case of the $O(n)$ model, the natural permutation representation of $S_Q$ in the case of the Potts model. 

By construction, we know how the global symmetry acts on $\mathcal{S}_L$. A first challenge is to define the counterpart of conformal symmetry. By the Koo--Saleur construction, this is described by a diagram algebra, which generalizes the symmetric group $S_L$ that acts by permuting the $L$ factors of $\mathcal{S}_L=V^{\otimes L}$ \cite{KOO1994459}. This construction relies on the emergence of geometrical objects (loops or clusters) in the high or low temperature expansions of the partition and other correlation functions. Technically, it uses the local transfer matrices in the Euclidian version of the model. The resulting algebra is generated by the local energy and momentum densities. 

Our space of states describes a one-dimensional chain with $L$ sites. Arbitrary permutations of the sites would break the chain: only the cyclic subgroup $\mathbb{Z}_L\subset S_L$ is a symmetry of the model. More generally, our diagram algebras have to be planar, i.e. lines cannot cross. 

In the case of the $O(n)$ model, the relevant diagram algebra is called the unoriented Jones--Temperley--Lieb algebra $\uJTL_L(n)$ \cite{rs07}. Its structure is dictated by the rules that loops do not intersect and have weight $n$. Let us momentarily write $\mathfrak{A}_L$ for this algebra or its Potts model counterpart. One might  expect that the critical limit of $\mathfrak{A}_L$ is the conformal algebra $\mathfrak{C}_c$.  While this is difficult to put on solid mathematical footing in general \cite{zw18}, there is also evidence that the limit is  in fact larger than  $\mathfrak{C}_c$, and coincides with the interchiral algebra \cite{grs12}, 
\begin{align}
 \lim_{\text{critical}} \mathfrak{A}_L = \widetilde{\mathfrak{C}}_{\beta^2}\ .
\end{align}
The algebraic problem is now to decompose the space of states $\mathcal{S}_L$ into representations of $\mathfrak{A}_L\times G$, before taking the critical limit and recovering the CFT results. To do this, let us consider the algebra $G^*_L$ of linear maps on $\mathcal{S}_L$ that commute with the action of $G$, i.e.\ the commutant of $G$ in the space $\mathcal{S}_L$. By Schur--Weyl duality, we know how to decompose $\mathcal{S}_L$ into irreducible representations of $G^*_L$ and $G$,
\begin{align}
 \mathcal{S}_L \underset{G^*_L \times G}{=} \bigoplus_{|\lambda|\leq L} V_{\lambda}^{(L)}\otimes \lambda \ ,
\end{align}
where the sum is over partitions, which we identify with representations of $G$. Since $\mathfrak{A}_L\subset G^*_L$, our problem boils down to decomposing the representations $V_{\lambda}^{(L)}$ of $G^*_L$ into representations of $\mathfrak{A}_L$, schematically 
\begin{align}
 V_{\lambda}^{(L)} \underset{\mathfrak{A}_L}{=} \bigoplus_q c^\lambda_q \mathcal{A}_q^{(L)}\quad \implies \quad \mathcal{S}_L \underset{\mathfrak{A}_L \times G}{=} \bigoplus_q\bigoplus_{|\lambda|\leq L} c^\lambda_q \mathcal{A}_q^{(L)}\otimes \lambda \ ,
\end{align}
where $c^\lambda_q\in\mathbb{N}$ is called a branching coefficient from the branching rule $G^*_L\downarrow\mathfrak{A}_L$, and turns out to be $L$-independent for $L$ large enough. In the critical limit, $\mathcal{A}_q^{(L)}$ becomes an indecomposable representation of the interchiral algebra, and $\bigoplus_\lambda c^\lambda_q \lambda$ a representation of the global symmetry group. In the case of the $O(n)$ model, $\bigoplus_\lambda c^\lambda_q \lambda$ was displayed as $\Lambda_{(r,s)}$ in Eq. \eqref{prea}.

The representation $\bigoplus_\lambda c^\lambda_q \lambda$ of $G$ may be viewed as an irreducible representation of the larger algebra $\mathfrak{A}_L^*$, the commutant of $\mathfrak{A}_L$ in the algebra of linear maps on $\mathcal{S}_L$. It would be tempting to interpret $\mathfrak{A}_L^*$ as a larger global symmetry algebra of the model. However, we find that the set $\{\bigoplus_\lambda c^\lambda_q \lambda\}_q$ of representations of $G$ does not close under tensor products, which makes such an interpretation difficult if not impossible. The situation is more favourable if we impose open (rather than periodic) boundary conditions on our chain of $L$ sites. In this case, which corresponds in the critical limit to a CFT with boundaries, we do find closure under tensor products, and therefore the possibility of a larger global symmetry \cite{rs07}.

\subsubsection{Related material: SageMath code and Wikipedia articles}

SageMath 9.2+ code that computes the spaces of states of the $O(n)$ and Potts models in terms of irreducible representations of the global symmetry groups is available at GitLab \cite{rz22}. The code implements:
\begin{itemize}
 \item Formula \eqref{lrsp} for the space of states of the $O(n)$ model. 
 \item Formula \eqref{xirsp} for the space of states of the $Q$-Potts model.
 \item Formula \eqref{clrs} for the branching rules that also lead to the space of states of the $O(n)$ model. 
\end{itemize}
Formula \eqref{dlrs} for the branching rules that lead to the space of states of the $Q$-Potts model is not implemented because there is little doubt that it would agree with Formula \eqref{xirsp}. 

In this article, we do not attempt to review the relevant mathematical background in detail. 
For basic information on representation theory and diagram algebras, we refer the reader to the following Wikipedia articles:
\begin{itemize}
 \item For representations of $O(n)$: \href{https://en.wikipedia.org/wiki/Representations_of_classical_Lie_groups}{Representations of classical Lie groups} (recently updated).
 \item For representations of $S_Q$: \href{https://en.wikipedia.org/wiki/Representation_theory_of_the_symmetric_group}{Representation theory of the symmetric group} (recently updated).
 \item For the partition algebra: \href{https://en.wikipedia.org/wiki/Partition_algebra}{Partition algebra} (recently created).
 \item For the Brauer algebra: \href{https://en.wikipedia.org/wiki/Brauer_algebra}{Brauer algebra} (recently updated).
 \item For the Temperley--Lieb algebra: \href{https://en.wikipedia.org/wiki/Temperley-Lieb_algebra}{Temperley--Lieb algebra} (recently updated). 
\end{itemize}
In Appendix \ref{app:diag}, we further review diagram algebras and their representations, including some not-so-basic results. See in particular Appendix \ref{app:sum} for synthetic tables of algebras and representations.

\section{Space of states of the \texorpdfstring{$O(n)$}{O(n)} CFT}\label{sec:aog}

The torus partition function of the $O(n)$ CFT \cite{fsz87} strongly constrains the space of states, but does not determine it. Nevertheless, we have recently formulated a conjecture for the space of states \cite{gnjrs21}. We will review the conjecture, before proving it by computing a twisted partition function.

\subsection{Representations and characters of the orthogonal group}

\subsubsection{Representations and their tensor products}

For generic complex values of the parameter $n$, the irreducible finite-dimensional representations $\lambda$ of the orthogonal group $O(n)$ are labelled by Young diagrams i.e.\ integer partitions. We write a Young diagram in terms of its row lengths, $\lambda = [\lambda_1\lambda_2\lambda_3\cdots]$, that is, a decreasing sequence $\lambda_1 \ge \lambda_2 \ge \lambda_3 \ge \cdots$ of natural integers, with $|\lambda| = \sum_i \lambda_i$ denoting the total number of boxes in $\lambda$. We condense the notation for repeated integers, such as $[53321111]=[53^221^4]$. In particular, $[k]$ is a fully symmetric representation, $[1^k]$ is a fully antisymmetric representation, and $[]$ is the identity representation. More generally, a partition of the type $[k1^\ell]$ is called a hook partition of size $k+\ell$, and the corresponding representation may be called a hook representation. 

The tensor product of two representations can be written as 
\begin{align}
 \lambda \otimes \mu = \sum_\nu N_{\lambda,\mu,\nu} \nu \ , 
\end{align}
where the Newell--Littlewood numbers $N_{\lambda,\mu,\nu}\in\mathbb{N}$ are $n$-independent, and invariant under permutations of the three indices $\lambda,\mu,\nu$. For example, the simplest non-trivial tensor product is 
\begin{align}
 [1] \otimes [1] = [2] + [1^2] + []\ .
\end{align}
For $n\in \mathbb{N}$, the tensor products of $O(n)$ representations involve $n$-dependent coefficients. These coefficients tend toward Newell--Littlewood numbers as $n$ becomes large. Therefore, the limit $n\to \infty$ gives us access to generic values of $n$ from integer values.

\subsubsection{Characters of the general linear group}

We will now define characters of $O(n)$ for generic $n\in\mathbb{C}$ as $n$-independent objects. We start with the case of the general linear group $GL(n)$.
The character of a representation is originally defined as the trace of an element $g\in GL(n)$ in that representation. This is a symmetric polynomial of the eigenvalues $x_1,x_2,\dots,x_n$ of $g$. The character of the irreducible representation of $GL(n)$ associated to a Young diagram $\lambda$ is the corresponding Schur polynomial $s_\lambda$, which is a homogeneous polynomial of degree $|\lambda|$:
\begin{align}
 \chi_\lambda^{GL(n)}(x_1,x_2,\dots, x_n) = \operatorname{Tr}_\lambda g = s_\lambda(x_1,x_2,\dots,x_n)\ . 
\end{align}
Schur polynomials provide a basis of the space of symmetric polynomials. In particular, the power sum polynomial $p_r(x_1,x_2,\dots,x_n) =\sum_{i=1}^n x_i^r$ is a combination of Schur polynomials indexed by hook partitions:
\begin{align}
 p_r = \sum_{k=0}^{r-1}  (-1)^k   s_{[r-k,1^k]}\ .
 \label{prs}
\end{align}
This well-known relation may be viewed as a special case of the Murnaghan--Nakayama rule, see \cite{vo96}(Section 8) or \cite{jam78}(Theorem 21.4). This relation has integer coefficients that do not depend on the rank $n$ of the linear group. 
Actually, we may view Schur polynomials and other symmetric polynomials as abstract, $n$-independent symmetric functions that may be evaluated on arbitrary numbers of variables.

\subsubsection{Characters of the orthogonal group, and alternating hook representations}

$GL(n)$ representations can be decomposed into $O(n)$ representations. For generic values of $n$, the resulting branching rule is \cite{htw03}
\begin{align}
 \nu^{GL(n)} \underset{O(n)}{=} \sum_{\lambda,\mu} c_{\lambda,2\mu}^\nu \lambda^{O(n)}\ , 
\end{align}
where $2\mu=[2\mu_1,2\mu_2,\dots , 2\mu_k]$ is an even integer partition, and $c_{\lambda,\mu}^\nu\in\mathbb{N}$ is a Littlewood--Richardson coefficient, i.e. $\lambda^{GL(n)}\otimes \mu^{GL(n)} = \sum_\nu c_{\lambda,\mu}^\nu \nu^{GL(n)}$. 
Writing the branching rule in terms of characters, and inserting it in Eq. \eqref{prs}, we obtain a decomposition of the power sum polynomial $p_r$ into characters of $O(n)$,
\begin{align}
 p_r = \sum_{\lambda,\mu} \sum_{k=0}^{r-1}(-1)^k c_{\lambda, 2\mu}^{[r-k,1^k]} \chi^{O(n)}_\lambda\ .
 \label{pre}
\end{align}
Now it turns out that $\sum_{k=0}^{r-1}(-1)^k c_{\lambda, \mu}^{[r-k,1^k]} \neq 0 \implies |\lambda||\mu|=0$. This is easily seen if $\mu=[\ell]$, in which case the Littlewood--Richardson coefficients are given by Pieri's rule, and $c_{\lambda, [\ell]}^{[r-k,1^k]}$ can be nonzero only if $\lambda$ is a hook partition. Then this generalizes to arbitrary $\mu$ by associativity of tensor products of $GL(n)$ representations. As a result, the decomposition \eqref{pre} simplifies,
\begin{align}
 p_r = \delta_{r\in 2\mathbb{N}} \chi^{O(n)}_{[]} + \sum_{k=0}^{r-1}(-1)^k \chi^{O(n)}_{[r-k,1^k]}\ ,
\end{align}
where the first term results from $|\lambda|=0$ in Eq. \eqref{pre}, while the sum results from $|\mu|=0$. This motivates us to introduce the alternating hook representations of $O(n)$,
\begin{align}
 \boxed{\lambda_r = \delta_{r\in 2\mathbb{N}}[] + \sum_{k=0}^{r-1}(-1)^k [r-k,1^k]} \ .
 \label{lr}
\end{align}
This is a formal combination of irreducible representations, with integer coefficients that can however be negative. For example,
\begin{align}
 \lambda_4 = [] + [4] - [31] + [21^2] - [1^4]\ .
\end{align}
The defining feature of $\lambda_r$ is that its character is the power sum polynomial, $\chi^{O(n)}_{\lambda_r} = p_r$. For $n\in\mathbb{N}$, this means
\begin{align}
 \boxed{\operatorname{Tr}_{\lambda_r}(g) = \operatorname{Tr}_{[1]}(g^r)}\ . 
 \label{tlr}
\end{align}
In particular, the dimension of an alternating hook representation is 
\begin{align}
 \dim^{O(n)}(\lambda_r) = \operatorname{Tr}_{\lambda_r}(\text{id}) = n\ .
 \label{diaho}
\end{align}

\subsection{Action of the orthogonal group on the space of states}

The symmetries of the $O(n)$ CFT include the orthogonal group $O(n)$ itself, and the conformal algebra $\mathfrak{C}_c$, which is a product of the left- and right-moving Virasoro algebras. The CFT's space of states is therefore a representation of $O(n) \times \mathfrak{C}_c$, which we want to decompose into indecomposable representations. 

\subsubsection{General structure of the space of states}

Together with the existence of degenerate fields, the torus partition function is enough for determining the representations of the conformal algebra that appear in the space of states \cite{nr20}. The partition function determines the conformal dimensions of the primary states, which are of the type 
\begin{align}
 \Delta_{(r,s)} = P_{(r,s)}^2-P_{(1,1)}^2 \quad \text{with} \quad P_{(r,s)} = \frac12\left(\beta r - \beta^{-1}s\right)\ , 
 \label{drs}
\end{align}
where $r,s\in\mathbb{Q}$ are called Kac table indices. We now introduce the relevant representations of the conformal algebra, including some that are zero by convention:
\begin{align}
\renewcommand{\arraystretch}{1.5}
\renewcommand{\arraycolsep}{8pt}
 \begin{array}{|l|l|l|l|}
 \hline
 \text{Notation} & \text{Indices} & \text{Name} & \text{Primary states}
 \\
 \hline\hline
  \mathcal{R}_{\langle 1,s\rangle} & s\in \mathbb{N}^* &  \text{Degenerate} & (\Delta_{(1,s)},\Delta_{(1,s)})
  \\
  \hline
  \mathcal{W}_{(r,s)} & r\notin\mathbb{Z}^*\text{ or } s\notin\mathbb{Z}^* & \text{Verma module} & (\Delta_{(r,s)},\Delta_{(r,-s)})
  \\
  \hline 
  \mathcal{W}_{(r,s)} & r,s\in\mathbb{N}^* & \text{Logarithmic} & 
  \renewcommand{\arraystretch}{1}
  \begin{array}{@{}l@{}}  (\Delta_{(r,s)},\Delta_{(r,-s)}) \\ (\Delta_{(r,-s)},\Delta_{(r,s)}) \\ (\Delta_{(r,-s)},\Delta_{(r,-s)}) \end{array} 
  \renewcommand{\arraystretch}{1.5}
  \\
  \hline
  \mathcal{W}_{(r,s)} & r,-s\in\mathbb{N}^* & 0 & 
  \\
  \hline
 \end{array}
 \label{vwww}
\end{align}
The logarithmic representation $\mathcal{W}_{(r,s)}$ is generated by the field $\partial_P V_{P_{(r,-s)}} - \mathcal{L}_{(r,s)}\bar{\mathcal{L}}_{(r,s)}\partial_P V_{P_{(r,s)}}$, where $V_P$ is a diagonal primary field of momentum $P$ (i.e. its left and right conformal dimensions are $\Delta=\bar\Delta = P^2-P_{(1,1)}^2$), and $\mathcal{L}_{(r,s)}$ is the creation operator that appears when writing the singular vector at level $rs$, i.e. $V_{P_{(r,-s)}} = \mathcal{L}_{(r,s)}\bar{\mathcal{L}}_{(r,s)} V_{P_{(r,s)}}$ \cite{nr20}. (Our conventions differ from those of \cite{nr20}: the degenerate representations of the $O(n)$ CFT are of the type $\mathcal{R}_{\langle 1, s\rangle}$ here and $\mathcal{R}_{\langle r, 1\rangle}$ there.) 

The space of states of the $O(n)$ CFT can be written in terms of these representations, together with representations of $O(n)$: 
\begin{align}
 \boxed{\mathcal{S}^{O(n)} \underset{\mathfrak{C}_c\times O(n)}{=} \bigoplus_{s\in 2\mathbb{N}+1} \mathcal{R}_{\langle 1,s\rangle}\otimes [] \oplus \bigoplus_{r\in\frac12\mathbb{N}^*}\bigoplus_{s\in\frac{1}{r}\mathbb{Z}} \mathcal{W}_{(r,s)}\otimes \Lambda_{(r,s)}}\ .
 \label{son}
\end{align}
Here the representations of $O(n)$ are the trivial representation $[]$, and a family of representations $\Lambda_{(r,s)}$: linear combinations of irreducible finite-dimensional representations, with positive integer coefficients that do not depend on $n$ \cite{grz18, br19}.

The symmetry of the model under exchanging the left- and right-moving Virasoro algebras leads to the relation 
\begin{align}
 \Lambda_{(r,s)} = \Lambda_{(r,-s)}\ .
 \label{lms}
\end{align}
Further constraints come from the existence of a degenerate field $V_{\langle 1,3\rangle}$ that transforms in the degenerate representation $\mathcal{R}_{\langle 1,3\rangle}$ of the conformal algebra. The fusion of such a field with a non-diagonal primary field $V^N_{(r,s)}\in \mathcal{W}_{(r,s)}$ yields \cite{mr17}
\begin{align}
 V_{\langle 1,3\rangle} \times V^N_{(r,s)} \simeq V^N_{(r,s-2)} + V^N_{(r,s)} + V^N_{(r,s+2)} \ . 
\end{align}
In the $O(n)$ CFT, $V_{\langle 1,3\rangle}$ is an $O(n)$ singlet. This implies that $V^N_{(r,s\pm 2)}$ transform in the same way as $V^N_{(r,s)}$ under $O(n)$, therefore
\begin{align}
\Lambda_{(r,s)} = \Lambda_{(r,s+2)}\ .
\label{lspt}
\end{align}
This suggests that we combine the representations of the conformal algebra into the larger representations
\begin{align}
 \widetilde{\mathcal{R}}_{\langle 1,1\rangle} = \bigoplus_{s\in 2\mathbb{N}+1} \mathcal{R}_{\langle 1,s\rangle}  \quad , \quad \widetilde{\mathcal{W}}_{(r,s)} = \bigoplus_{s'\in 2\mathbb{Z}+s} \mathcal{W}_{(r,s)}\ .
\end{align}
These can be interpreted as indecomposable representations of the interchiral algebra $\widetilde{\mathfrak{C}}_{\beta^2}$, which is obtained from the conformal algebra by adding the degenerate field $V_{\langle 1,3\rangle}$  \cite{hjs20}. The space of states can then be rewritten as  
\begin{align}
 \mathcal{S}^{O(n)} \underset{\widetilde{\mathfrak{C}}_{\beta^2}\times O(n)}{=}\widetilde{\mathcal{R}}_{\langle 1,1\rangle}\otimes [] \oplus \bigoplus_{r\in\frac12\mathbb{N}^*}\bigoplus_{\substack{s\in\frac{1}{r}\mathbb{Z}\\ -1<s\leq 1}} \widetilde{\mathcal{W}}_{(r,s)}\otimes  \Lambda_{(r,s)}\ .
 \label{soni}
\end{align}

\subsubsection{Action of the orthogonal group}

The action of the orthogonal group on the space of states is encoded in the family of representations $\Lambda_{(r,s)}$. In order to write these representations, we need to introduce the Chebyshev  polynomials $U_d$ such that 
\begin{align}
 U_d(q+q^{-1}) = q^d + q^{-d} \ , \qquad (d\in\mathbb{N})\ .
 \label{xd}
\end{align}
(To be precise, the polynomial $\frac12 U_d(2z)$ is called a $d$-th order Chebyshev polynomial of the first kind.)
These polynomials can be characterized by the recurrence relation 
\begin{align}
 U_0(z) = 2 \quad , \quad U_1(z) = z \quad , \quad zU_d(z) = U_{d-1}(z)+U_{d+1}(z)\ ,
\end{align}
and they obey
\begin{align}
 U_{2d}(z) = U_d(z^2-2)\quad , \quad U_d(0) = 2\cos\left(\tfrac{\pi}{2}d\right)\ . 
 \label{xtd}
\end{align}
The conjectured form of $\Lambda_{(r,s)}$ is \cite{gnjrs21}
\begin{align}
\renewcommand{\arraystretch}{1.3}
 \boxed{\Lambda_{(r,s)} = \delta_{r,1}\delta_{s\in 2\mathbb{Z}+1}[] + \frac{1}{2r} \sum_{r'=0}^{2r-1} e^{\pi ir's} U_{\mathrm{gcd}(2r, r')}\left(\lambda_{\frac{2r}{\mathrm{gcd}(2r, r')}}\right)}\ , \quad \left\{\begin{array}{l}r\in\frac12\mathbb{N}^*\ ,\\ s\in\frac{1}{r}\mathbb{Z}\ , \end{array}\right.
 \label{lrs}
\end{align}
where the arguments of the Chebyshev polynomials are alternating hook representations $\lambda_r$ \eqref{lr}. According to Eq. \eqref{tlr}, the character of this representation is 
\begin{align}
 \chi_{\Lambda_{(r,s)}}^{O(n)}(g) = \delta_{r,1}\delta_{s\in 2\mathbb{Z}+1} + \frac{1}{2r} \sum_{r'=0}^{2r-1} e^{\pi ir's}U_{\mathrm{gcd}(2r, r')}\left(\operatorname{Tr}_{[1]}g^\frac{2r}{\mathrm{gcd}(2r, r')}\right)\ . 
 \label{chirs}
\end{align}
From its definition, it is possible to write $\Lambda_{(r,s)}$ as a linear combination of finite-dimensional irreducible representations, by replacing each monomial $z^k$ appearing in the polynomial $U_d(z)$ with the $k$-fold
$O(\n)$ tensor product of the corresponding representation $\lambda_r$.
It is however not manifest that the resulting coefficients are positive integers. Let us at least show that they are rational numbers. To do this, we notice that the combinations of roots of unity that appear in our formula are Ramanujan's sums, i.e. they are of the type  
\begin{align}
 \varphi_k(r) = \sum_{\ell=1}^{r} \delta_{\mathrm{gcd}(r, \ell), 1} e^{2\pi i \frac{k\ell}{r}}\ .
 \label{vfkr}
\end{align}
Ramanujan's sums are always integer, although not necessarily positive, as follows from the properties
\begin{align}
\mathrm{gcd}(r_1, r_2) = 1 &\implies 
 \varphi_k(r_1r_2) = \varphi_k(r_1)\varphi_k(r_2) \ ,
 \\
 p \text{ prime} &\implies \varphi_k(p^r) = p^r \delta_{k\equiv 0\bmod p^r} - p^{r-1}\delta_{k\equiv 0\bmod p^{r-1}} \ .
 \end{align}
Moreover, $\varphi_k(r)$ only depends on $k$ through $\mathrm{gcd}(k, r)$.
Special cases include Euler's totient function $\varphi_0(r)$. In terms of Ramanujan's sums, our expression for $\Lambda_{(r,s)}$ becomes 
\begin{align}
 \Lambda_{(r,s)} = \delta_{r,1}\delta_{s\in 2\mathbb{Z}+1}[] +\frac{1}{2r}\sum_{\substack{g,g'|2r\\ gg'=2r}}\varphi_{rs}(g)U_{g'}(\lambda_g)\ ,
 \label{lrsp}
\end{align}
where the sum is now over the positive integer divisors of $2r$, including $2r$ itself. This shows that the coefficient of any irreducible representation in $\Lambda_{(r,s)}$ is a rational number whose denominator divides $2r$. For computer calculations, this is more convenient than dealing with the complex numbers of the original formula \eqref{lrs}.

\subsubsection{Explicit examples}

Given the relations \eqref{lms} and \eqref{lspt}, the cases $0\leq s\leq 1$ already sample all different representations $\Lambda_{(r,s)}$. 
The number of different cases is further restricted by the fact that $\Lambda_{(r,s)}$ only depends on $s$ through $\mathrm{gcd}(2r, rs)$.
We now write these representations according to Eq.~\eqref{lrs}. We check that the decompositions into irreducible representations always involve positive integer coefficients. Here we display the results for $r\leq \frac72$. The computer code \cite{rz22} can easily reach at least $r\simeq 10$. 
\begingroup
\allowdisplaybreaks
\begin{subequations}
\label{lxxx-all}
\begin{align}
\Lambda_{(\frac12,0)} &= [1]\ , \label{lhz}
 \\
 \Lambda_{(1,0)} &= [2]\ ,
 \\
 \Lambda_{(1,1)} &= [1^2]\ , \label{loo}
 \\
\Lambda_{(\frac32,0)}&= [3]+[1^3]\ ,
\label{l320}
\\
\Lambda_{(\frac32,\frac23)} &= [21]\ ,
\label{l3223}
\\
\Lambda_{(2,0)}&= [4]+[2^2]+[21^2]+[2]+[]\ ,
\label{l20}
\\
\Lambda_{(2,\frac12)}& = [31]+[21^2]+[1^2]\ ,
\\
\Lambda_{(2,1)} &= [31]+[2^2]+[1^4]+[2]\ ,
\label{l21}
\\
\Lambda_{(\frac52,0)} &= [5]+[32]+2[31^2]+[2^21]+[1^5]+[3]+2[21]+[1^3]+[1]\ ,
\label{l520}
\\
\Lambda_{(\frac52,\frac25)} &= [41]+[32]+[31^2]+[2^21]+[21^3]+[3]+2[21]+[1^3]+[1]\ , 
\\
\Lambda_{(3, 0)} &= 
[6]+2[42]+2[41^2]+[3^2]+2[321]+2[31^3]+2[2^3]+[2^21^2]+[21^4]
\nonumber \\ 
& \quad +2[4]+4[31]+4[2^2]+4[21^2]+2[1^4]+4[2]+2[1^2]+2[]\ ,
\label{l30}
\\
\Lambda_{(3,\frac13)} &= [51]+[42]+2[41^2]+[3^2]+3[321]+[31^3]+2[2^21^2]+[21^4]
\nonumber \\ 
& \quad
+[4]+5[31]+2[2^2]+5[21^2]+[1^4]+2[2]+4[1^2]\ ,
\\
\Lambda_{(3,\frac23)} &= [51]+2[42]+[41^2]+3[321]+2[31^3]+[2^3]+[2^21^2]+[21^4]
\nonumber \\ 
& \quad
+2[4]+4[31]+4[2^2]+4[21^2]+2[1^4]+4[2]+2[1^2]+[]\ ,
\\
\Lambda_{(3, 1)} &= [51]+[42]+2[41^2]+2[3^2]+2[321]+2[31^3]+[2^3]+2[2^21^2]+[1^6]
\nonumber \\ 
& \quad
+[4]+5[31]+2[2^2]+5[21^2]+[1^4]+2[2]+4[1^2]\ ,
\label{l310}
\\
\Lambda_{(\frac72, 0)} &= [7]+2[52]+3[51^2]+2[43]+5[421]+2[41^3]+3[3^21]+3[32^2]
\nonumber \\ 
& \quad
+5[321^2]+3[31^4]+2[2^31]+2[2^21^3]+[1^7]+2[5]+8[41]+10[32] \ ,
\nonumber \\ 
& \quad
+12[31^2]+10[2^21]+8[21^3]+2[1^5]+7[3]+14[21]+7[1^3]+5[1] \ ,
\\
\Lambda_{(\frac72, \frac27)} &=
[61]+2[52]+2[51^2]+2[43]+5[421]+3[41^3]+3[3^21]+3[32^2]+5[321^2]
\nonumber \\ 
& \quad
+2[31^4]+2[2^31]+2[2^21^3]+[21^5]+2[5]+8[41]+10[32]
\nonumber \\ 
& \quad
+12[31^2]+10[2^21]+8[21^3]+2[1^5]+7[3]+14[21]+7[1^3]+5[1]\ .
\end{align}
\end{subequations}
\endgroup

\subsection{Twisted partition function of the \texorpdfstring{$O(n)$}{O(n)} CFT}

\subsubsection{Definition and character decomposition}

Let us define the twisted partition function of the $O(n)$ CFT on a torus of modulus $\tau$ as 
\begin{align}
 Z^{O(n)}(\tau,g) = \operatorname{Tr}_{\mathcal{S}^{O(n)}}\left(e^{2\pi i \tau(L_0-\frac{c}{24})} e^{-2\pi i \bar\tau(\bar L_0-\frac{c}{24})}\cdot g\right) \ , 
 \label{ztg}
\end{align}
where the twist factor is an element $g\in O(n)$, the space of states $\mathcal{S}^{O(n)}$ is given by \eqref{soni},  and $L_0,\bar L_0$ are elements of the conformal algebra $\mathfrak{C}_c$. The decomposition \eqref{son} of the space of states $\mathcal{S}^{O(n)}$ into representations of $O(n)$ and $\mathfrak{C}_c$ leads to a decomposition of the partition function into characters,
\begin{align}
 Z^{O(n)}(\tau,g) =  \sum_{s\in 2\mathbb{N}+1} \chi_{\langle 1,s\rangle}(\tau) + \sum_{r\in\frac12\mathbb{N}^*}\sum_{s\in\frac{1}{r}\mathbb{Z}} \chi^N_{(r,s)}(\tau) \chi^{O(n)}_{\Lambda_{(r,s)}}(g)\ .
 \label{ztgchi}
\end{align}
Here we introduced the characters of representations of $\mathfrak{C}_c$,
\begin{align}
 \chi_{\langle 1,s\rangle}(\tau) = \left| \frac{e^{2\pi i\tau P^2_{(r,s)}} - e^{2\pi i\tau P^2_{(r,-s)}}}{\eta(\tau)}\right|^2\qquad , \qquad 
 \chi^N_{(r,s)}(\tau) = \frac{e^{2\pi i\tau P^2_{(r,s)}} e^{-2\pi i\bar\tau P^2_{(r,-s)}}}{|\eta(\tau)|^2}\ ,
 \label{chis}
\end{align}
where $\eta(\tau)$ is the Dedekind eta function. These are the characters of the representations that appear in Table \eqref{vwww},
\begin{align}
 \operatorname{Tr}_{\mathcal{R}_{\langle 1,s\rangle}}\left( e^{2\pi i \tau(L_0-\frac{c}{24})} e^{2\pi i \bar\tau(\bar L_0-\frac{c}{24})}\right) &= \chi_{\langle 1,s\rangle}(\tau)\ ,
 \\
 \renewcommand{\arraystretch}{2.5}
  \operatorname{Tr}_{\mathcal{W}_{(r,s)}}\left( e^{2\pi i \tau(L_0-\frac{c}{24})} e^{2\pi i \bar\tau(\bar L_0-\frac{c}{24})}\right) &= \left\{\begin{array}{ll} \chi^N_{(r,s)}(\tau) & \text{if} \quad r\notin \mathbb{Z}^*\text{ or } s\notin\mathbb{Z}^*\ ,                                                                                                                                                             \\ 
         \chi^N_{(r,s)}(\tau) + \chi^N_{(r,-s)}(\tau) & \text{if} \quad r,s\in\mathbb{N}^*\ ,
         \\ 0 & \text{if} \quad r,-s\in\mathbb{N}^*\ .
         \end{array}\right.
\end{align}
(See \cite{nr20} for the determination of the character of the logarithmic representation $\mathcal{W}_{(r,s)}$ with $r,s\in\mathbb{N}^*$.)

The modular invariant partition function of \cite{fsz87} is the special case $Z^{O(n)}(\tau,\text{id})$. In its decomposition into characters, the $O(n)$ characters reduce to $\chi^{O(n)}_{\Lambda_{(r,s)}}(\text{id}) = \dim^{O(n)}\Lambda_{(r,s)}$. These dimensions are polynomial functions of $n$, for example $\dim^{O(n)}\Lambda_{(1,0)} = \frac12(n+2)(n-1)$. Knowing the dimensions helped us guess the conjecture \eqref{lrs} for $\Lambda_{(r,s)}$ in \cite{gnjrs21}, although this is not enough for proving the conjecture---a representation is not determined by its dimension.

On the other hand, the character $\chi^{O(n)}_{\Lambda_{(r,s)}}(g)$ fully determines the representation $\Lambda_{(r,s)}$. In order to prove the conjecture, it is therefore enough to compute the twisted partition function $Z^{O(n)}(\tau,g)$ and check that it agrees with \eqref{ztgchi}. Due to the inclusion of a group element $g$, the proof will only be valid for $n\in\mathbb{N}$. However, just like Newell--Littlewood numbers, the structure of $\Lambda_{(r,s)}$ is $n$-independent for $n$ generic, and can be accessed from integer values of $n$ by taking the limit $n\to \infty$. 

\subsubsection{Loop representation of the $O(n)$ model}

For $n$ in a rather large region of the complex plane, including for $n\in(-2,2)$, the $O(n)$ CFT is the infinite-size limit of the $O(n)$ lattice model at the critical coupling constant, $K = K_c$ \cite{gnjrs21, bjjz22}. The simplest version of the lattice model is defined on a hexagonal lattice, in which case \cite{nie82}
\begin{align}
 K_c = \frac{1}{\sqrt{2+\sqrt{2-n}}} \ .
\end{align}
In the loop representation, the $O(n)$ model's partition function can be written as a combination of partition functions of free fields \cite{fsz87}. We will  
now perform a similar calculation in the case of the twisted partition function $Z^{O(n)}(\tau,g)$ \eqref{ztg}. 

In the torus $\mathbb{T}=\frac{\mathbb{C}}{\mathbb{Z}+\tau \mathbb{Z}}$, a loop is topologically characterized by a pair of integers $(m,m')\in \mathbb{Z}^2=\pi_1(\mathbb{T})$, its winding numbers with respect to the two periods of $\mathbb{T}$. A loop with $(m,m') = (0,0)$ is called a topologically trivial, or contractible loop. The loops of the loop representation are non-self-intersecting, which implies that for any topologically non-trivial loop $m$ and $m'$ are coprime, $\mathrm{gcd}(m, m') = 1$. As an example, let us draw the loop $(m, m')=(3, 2)$, together with the lattice $\mathbb{Z}+\tau\mathbb{Z}$, in the complex plane $\mathbb{C}$, where the torus may be identified with one of the basic parallelograms:
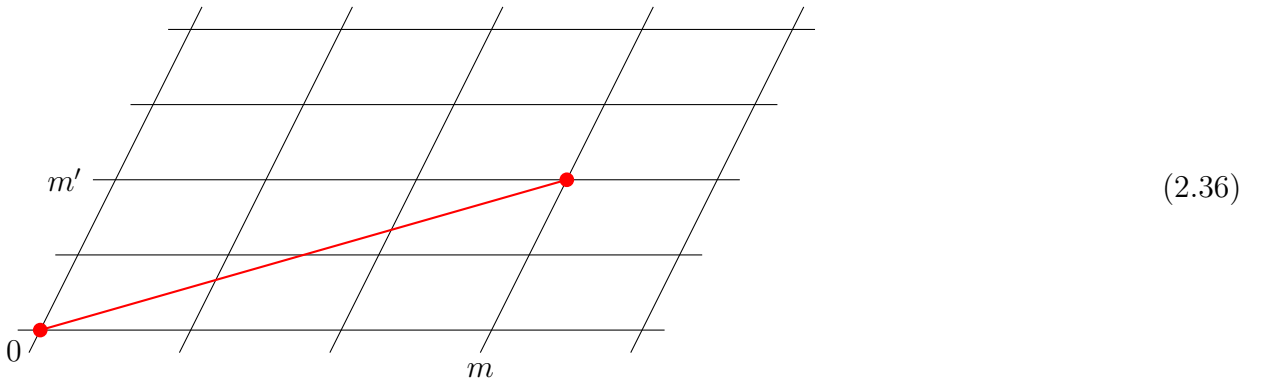
\begin{align}
\newcommand{\tilt}[2]{(#1+.5*#2, #2)}
 \begin{tikzpicture}[baseline=(current  bounding  box.center)]
  \foreach\j in {0,...,4}{
  \draw \tilt{2*\j}{-.3} -- \tilt{2*\j}{4.3};
  \draw \tilt{-.3}{\j} -- \tilt{8.3}{\j};
  }
  \node[red] at (0, 0) [fill, circle, scale = .5] {};
  \node[red] at \tilt{6}{2} [fill, circle, scale = .5] {};
  \draw[thick, red] (0, 0) -- \tilt{6}{2};
  \node[below left] at (-.1, 0) {$0$};
  \node[below] at \tilt{6}{-.3} {$m$};
  \node[left] at \tilt{-.3}{2} {$m'$};
 \end{tikzpicture}
 \label{pll}
\end{align}
In the loop model, a configuration is a finite set of non-intersecting loops on a lattice. Forbidding intersections implies that only one type of topologically non-trivial loops can exist in a given configuration. For a configuration $\mathcal{C}$, we write
\begin{itemize}
\item $N_0$ the number of contractible loops,
\item $N$ the number of topologically non-trivial loops,
 \item $(m,m')$ the class of the topologically non-trivial loops in $\pi_1(\mathbb{T})$ if $N\neq 0$,
 \item $|A|$ the number of lattice edges that belong to the union of all loops in $\mathcal{C}$. 
\end{itemize}
The partition function of the lattice $O(n)$ model is 
\begin{align}
 Z^{O(n)}_\text{lattice}(\tau,g|K) = \sum_{\mathcal{C}} K^{|A|(\mathcal{C})} n^{N_0(\mathcal{C})} \left(\operatorname{Tr}_{[1]} g^{m(\mathcal{C})}\right)^{N(\mathcal{C})} \ ,
 \label{zol}
\end{align}
where $K$ is the model's coupling constant. The idea is that each contractible loop has weight $n$, and each topologically non-trivial loop has weight $\operatorname{Tr}_{[1]} g^{m}$. Twisting the partition function by $g$ indeed amounts to picking a factor $g$ whenever a loop crosses the cycle $(0, 1)$ in one direction, and a factor $g^{-1}$ in the other direction. A topologically non-trivial loop therefore comes with the weight $ \operatorname{Tr}_{[1]} g^{m}=\operatorname{Tr}_{[1]} g^{-m}$. This weight reduces to the weight $n$ of contractible loops in the case $g=\text{id}$. 
Following \cite{fsz87}, let us relate this partition function to sums over oriented loop configurations. For an oriented loop configuration, we define 
\begin{itemize}
 \item $N_+,N_-$ the numbers of topologically non-trivial loops of each orientation, so that $N=N_++N_-$, 
 \item $M = m(N_+-N_-)$ and $M'=m'(N_+-N_-)$ the algebraic numbers of loops that cross each one of the two basic cycles $(0, 1)$ and $(1, 0)$. 
\end{itemize}
As we will shortly review, it is easy to compute the following sums over oriented loop configurations with fixed numbers $M,M'$:
\begin{align}
 Z^{M,M'}_\text{lattice}(\tau,n|K) = \sum_{\text{oriented }\mathcal{C}|(M(\mathcal{C}),M'(\mathcal{C}))=(M,M')} K^{|A|(\mathcal{C})} n^{N_0(\mathcal{C})} \ . 
 \label{zmmp}
\end{align}
Let us therefore rewrite $Z^{O(n)}_\text{lattice}(\tau,g|K)$ in terms of these sums. We want to interpret the weight $z=\operatorname{Tr}_{[1]} g^{m}$ of an unoriented topologically non-trivial loop as a sum over the two possible orientations. A priori, the two orientations can have arbitrary weights $q_+,q_-$ such that $q_++q_-=z$. However, we also impose 
$q_+q_-=1$, so that the weight $q_+^{N_+} q_-^{N_-}$ of a configuration only depends on the difference $N_+-N_-$. Using the Chebyshev polynomials \eqref{xd}, and the property $U_{-d}(z) = U_d(z)$ which follows from the definition, we have 
\begin{align}
 2z^{N} 
 = \sum_{\substack{N_+,N_-\in\mathbb{N} \\ N_++N_-=N}}
 \binom{N}{N_+} U_{N_+-N_-}(z) 
 \underset{N_+\equiv N_-\bmod 2}{=} \sum_{\substack{N_+,N_-\in\mathbb{N} \\ N_++N_-=N}} \binom{N}{N_+}
 U_{\frac12(N_+-N_-)}(z^2-2)\ , 
 \label{tzn}
\end{align}
where the second equality is written for later use, with the help of Eq. \eqref{xtd}. 
Since the values of $M,M'$ determine the difference $N_+-N_-=\mathrm{gcd}(M, M')$ as well as $m=\frac{M}{\mathrm{gcd}(M, M')}$, this allows us to express the $O(n)$ model's partition function in terms of the fixed $M,M'$ sums $Z^{M,M'}_\text{lattice}(\tau,n|K)$: 
\begin{align}
 Z^{O(n)}_\text{lattice}(\tau,g|K) = \frac12 \sum_{M,M'\in\mathbb{Z}} Z^{M,M'}_\text{lattice}(\tau,n|K) U_{\mathrm{gcd}(M, M')}\left(\operatorname{Tr}_{[1]} g^{\frac{M}{\mathrm{gcd}(M, M')}}\right) \ .
 \label{zonl}
\end{align}

\subsubsection{Critical limit}

In the critical limit, the sum $Z^{M,M'}_\text{lattice}(\tau,n|K)$ tends to the partition function of a compactified free boson, restricted to field configurations of a given topology. The compactification radius of the boson is $\frac14\beta^2$, where $\beta^2$ is defined as a function of $n$ in Eq. \eqref{nqb}. A field configuration is a map from the torus $\mathbb{T}$ to the circle $S^1$ of radius $\frac14\beta^2$, and its topology is specified by the images of the basic cycles $(1, 0)$ and $(0, 1)$, which we denote as $M, M'\in\pi_1(S^1)=\mathbb{Z}$. The Gaussian functional integral over such field configurations yields \cite{fsz87}
\begin{align}
 Z^{M,M'}(\tau, n) = \lim_\text{critical} Z^{M,M'}_\text{lattice}(\tau,n|K) = \frac{\beta}{2\sqrt{\Im\tau}|\eta(\tau)|^2} e^{ -\frac{\pi\beta^2}{4\Im\tau}\left(M^2|\tau|^2-2MM'\Re\tau +M'^2\right)}\ .
 \label{zmmc}
\end{align}
Let us insert this into Eq. \eqref{zonl} in order to compute $Z^{O(n)}(\tau,g) = \lim_\text{critical} Z^{O(n)}_\text{lattice}(\tau,g|K)$. We start with the term $M=0$. In this case, using Eq. \eqref{nqb}, we find $U_{\mathrm{gcd}(M, M')}\left(\operatorname{Tr}_{[1]} g^{\frac{M}{\mathrm{gcd}(M, M')}}\right)=U_{M'}(n)=(-1)^{M'}\sum_{\pm} e^{\pm \pi i \beta^2M'}$. Performing a Poisson resummation of the sum over $M'$, we obtain 
\begin{align}
 \left. Z^{O(n)}(\tau,g)\right|_{M=0} = \sum_{s\in 2\mathbb{Z}+1} \left|\frac{e^{2\pi i\tau P_{(1,s)}^2}}{\eta(\tau)}\right|^2 = \sum_{s\in 2\mathbb{N}+1} \chi_{\langle 1,s\rangle}(\tau) + \sum_{s\in 2\mathbb{Z}+1} \chi^N_{(1,s)}(\tau) \ ,
\end{align}
where the characters of the conformal algebra were defined in Eq. \eqref{chis}. Next we focus on the terms $M\neq 0$. In this case we write $M'=kM+r'$ with $r'\in\{0,1,\dots, M-1\}$, so that $U_{\mathrm{gcd}(M, M')}\left(\operatorname{Tr}_{[1]} g^{\frac{M}{\mathrm{gcd}(M, M')}}\right)$ does not depend on $k$. We perform a Poisson resummation of the Gaussian sum over $k$, which yields 
\begin{multline}
  \left. Z^{O(n)}(\tau,g)\right|_{M\neq 0} = \frac{1}{|\eta(\tau)|^2}\sum_{M=1}^\infty \sum_{r'=0}^{M-1}
  U_{\mathrm{gcd}(M, r')}\left(\operatorname{Tr}_{[1]} g^{\frac{M}{\mathrm{gcd}(M, r')}}\right)
  \\
  \times \frac{1}{M}
  \sum_{\ell\in\mathbb{Z}} e^{2\pi i \frac{\ell r'}{M}} e^{-\pi\Im\tau\left(\frac{4}{\beta^2}\frac{\ell^2}{M^2} +\frac{\beta^2}{4} M^2\right)} e^{2\pi i\ell \Re\tau}\ .  
\end{multline}
Let us rewrite this in terms of non-diagonal characters of the conformal algebra, while replacing the variables $M,\ell$ with $r=\frac{M}{2}$ and 
$s=\frac{2\ell}{M}$. We obtain 
\begin{align}
 \left. Z^{O(n)}(\tau,g)\right|_{M\neq 0} = \sum_{r\in\frac12\mathbb{N}^*}\sum_{s\in\frac{1}{r}\mathbb{Z}} \chi^N_{(r,s)}(\tau) \frac{1}{2r}\sum_{r'=0}^{2r-1} e^{\pi i r's} U_{\mathrm{gcd}(2r, r')}\left(\operatorname{Tr}_{[1]} g^{\frac{2r}{\mathrm{gcd}(2r, r')}}\right) \ . 
\end{align}
Therefore, we have proved that the expression for the twisted partition function $Z^{O(n)}(\tau,g) = \left. Z^{O(n)}(\tau,g)\right|_{M= 0} + \left. Z^{O(n)}(\tau,g)\right|_{M\neq 0}$ that we derived from the loop representation of the $O(n)$ model, is indeed of the form \eqref{ztgchi} predicted by the model's symmetries, with the $O(n)$ characters \eqref{chirs}. 

\section{Space of states of the Potts CFT}\label{sec:asg}

We will now determine the space of states of the Potts CFT, in the same way as we have done for the $O(n)$ CFT in Section \ref{sec:aog}. The comparison between the two models is often illuminating, and we encourage the reader to keep the $O(n)$ CFT in mind. However, it is also possible to read this section independently, and only refer to Section \ref{sec:aog} for a few definitions that we will not repeat. 

\subsection{Representations and characters of the symmetric group}\label{sec:rcsg}

\subsubsection{Representations and their tensor products}

Irreducible, finite-dimensional representations of the symmetric group $S_Q$ are parametrized by Young diagrams with $Q$ boxes. 
To define the representation theory of $S_Q$ with generic complex values of $Q$, we first have to parametrize these representations in a $Q$-independent way. This is done by distinguishing the first row in the Young diagram from the remaining rows. To $\lambda = [\lambda_1\lambda_2\lambda_3\cdots]$ an integer partition, and for integer $Q\geq |\lambda|+\lambda_1$, we associate the representation of $S_Q$ that is parametrized by the Young diagram $[Q-|\lambda|,\lambda]$:
\begin{align}
 \begin{tikzpicture}[baseline=(current  bounding  box.center), scale = .5]
  \draw (0, 0) -- (0, 7) -- (13, 7) -- (13, 6) -- (0, 6);
  \draw (0, 0) -- (2, 0) -- (2, 1) -- (3, 1) -- (3, 3) -- (6, 3) -- (6, 4) -- (7, 4) -- (7, 5) -- (9, 5) -- (9, 6);
  \draw [decorate, decoration = {calligraphic brace, amplitude=6pt}, ultra thick] (0, 7.3) -- (13, 7.3) node [pos = .5, above = .2] {$Q-|\lambda|$};
 \end{tikzpicture}
 \label{ydiagr}
\end{align}
With this notation, the trivial, one-dimensional representation is written $\lambda = []$. A $Q$-dimensional vector whose components are permuted by $S_Q$ belongs to the representation $[1]+[]$. More generally, a representation $\lambda$ can be interpreted in terms of tensors with $|\lambda|$ indices, each of which takes $Q$ values \cite{cjv17}.

The tensor product of two representations of $S_Q$ for generic $Q\in \mathbb{C}$ can be written as 
\begin{align}
 \lambda \otimes \mu = \sum_\nu M_{\lambda,\mu,\nu} \nu \ , 
 \label{losq}
\end{align}
where the reduced Kronecker coefficients $M_{\lambda,\mu,\nu}\in\mathbb{N}$ are $Q$-independent, and invariant under permutations of the three indices $\lambda,\mu,\nu$. For example, the simplest non-trivial tensor product is 
\begin{align}
 [1] \otimes [1] = [2] + [1^2] + [1] + []\ .
 \label{ooto}
\end{align}
Reduced Kronecker coefficients can be computed by at least two (not very easy) methods \cite{oz17, as18}. Fortunately, they are implemented in the computer algebra system SageMath \cite{oz15}. 

For $Q\in \mathbb{N}$, the tensor products of $S_Q$ representations involve $Q$-dependent coefficients called Kronecker coefficients, which tend toward reduced Kronecker coefficients as $Q$ becomes large. Therefore, the limit $Q\to \infty$ gives us access to generic values of $Q$ from integer values.

\subsubsection{Characters as symmetric polynomials}

By definition, characters of the symmetric group are functions of the finitely many elements of that group. However, it is possible to define symmetric polynomials of arbitrarily many variables, which encode the characters of $S_Q$ for large enough $Q\in\mathbb{N}$ \cite{oz16}. Calling these polynomials $\chi_\lambda^{S_Q}$ (rather than $\tilde s_\lambda$ in \cite{oz16}), this means that for any $Q\geq |\lambda|+\lambda_1$ and for any $g\in S_Q$ with eigenvalues $x_1,x_2,\dots, x_Q$, we have 
\begin{align}
 \chi_\lambda^{S_Q}(x_1,x_2,\dots, x_Q) = \operatorname{Tr}_\lambda(g) \ . 
\end{align}
We view the symmetric polynomials $\chi_\lambda^{S_Q}$ as the characters of the symmetric group for generic $Q\in\mathbb{C}$. They are $Q$-independent, and they form a ring whose structure constants are the reduced Kronecker coefficients \eqref{losq}, i.e. $\chi^{S_Q}_\lambda \chi^{S_Q}_\mu = \sum_\nu M_{\lambda,\mu,\nu} \chi^{S_Q}_\nu$. 

The characters $\chi_\lambda^{S_Q}$ are a basis of the space of symmetric polynomials, therefore they are linearly related to the Schur polynomials. In this basis, the decomposition of a symmetric polynomial $p_r(x_1,x_2,\dots, x_n)=\sum_{i=1}^n x_i^r$ is \cite{oz16}(Theorem 21)
\begin{align}
 p_r = \sum_{d|r}\left(1+ \sum_{k=0}^{d-1} (-1)^k \chi^{S_Q}_{[d-k,1^k]}\right) \ , 
 \label{prsq}
\end{align}
where the first sum is over the integer divisors of $r$, including $d=1$ and $d=r$. This motivates us to define the following formal representations of $S_Q$,
\begin{align}
\boxed{\xi_d = []+ \sum_{k=0}^{d-1} (-1)^k [d-k,1^k]}\ ,
 \label{xir}
\end{align}
such that $\sum_{d|r}\chi_{\xi_d}^{S_Q} = p_r$. (Conversely, it is possible to express $\chi_{\xi_d}^{S_Q}$ as a linear combination of $p_r$ using the M\"obius inversion formula.) 
For large enough values of $Q\in\mathbb{N}$, this implies 
\begin{align}
 \boxed{\sum_{d|r}\operatorname{Tr}_{\xi_d}(g) = \operatorname{Tr}_{[1]+[]}(g^r)}\ . 
 \label{stx}
\end{align}
In the particular case $g=\text{id}$, this reduces to 
\begin{align}
 \dim^{S_Q}(\xi_d) = Q\delta_{d,1} \ . 
\end{align}
The first few examples of the representations $\xi_d$ are 
\begin{align}
 \xi_1 = [] + [1] \quad , \quad \xi_2 = [] + [2] - [1^2] \quad , \quad\xi_3 = [] + [3] -[21] + [1^3] \ . 
\end{align}
These combinations of irreducible representations are formal, since the coefficients  may be negative, although they are integer.

\subsection{Action of the symmetric group on the space of states}

The symmetries of the Potts CFT include the symmetric group $S_Q$, and the conformal algebra $\mathfrak{C}_c$, which is a product of the left- and right-moving Virasoro algebras. The CFT's space of states is therefore a representation of $S_Q\times \mathfrak{C}_c$, which we want to decompose into indecomposable representations. 

\subsubsection{General structure of the space of states}

Together with the existence of degenerate fields, the torus partition function is enough for determining the representations of the conformal algebra that appear in the space of states \cite{nr20}. These representations belong to the same families $\mathcal{R}_{\langle 1,s\rangle}, \mathcal{W}_{(r,s)}$ \eqref{vwww} that we introduced for the $O(n)$ CFT. 
The space of states of the Potts CFT can be written in terms of these representations, together with representations of $S_Q$: 
\begin{align}
 \boxed{\mathcal{S}^{\text{$Q$-Potts}} \underset{\mathfrak{C}_c\times S_Q}{=} \bigoplus_{s\in \mathbb{N}^*} \mathcal{R}_{\langle 1,s\rangle}\otimes []\oplus\bigoplus_{s\in \mathbb{N}+\frac12} \mathcal{W}_{(0,s)}\otimes [1] \oplus \bigoplus_{r\in\mathbb{N}+2}\bigoplus_{s\in\frac{1}{r}\mathbb{Z}}  \mathcal{W}_{(r,s)}\otimes \Xi_{(r,s)}}\ .
 \label{spotts}
\end{align}
Here the representations of $S_Q$ are the trivial representation $[]$, the standard representation $[1]$, and a family of representations $\Xi_{(r,s)}$: linear combinations of irreducible finite-dimensional representations, with positive integer coefficients that do not depend on $Q$ \cite{grz18, br19}. 

The symmetry of the model under exchanging the left- and right-moving Virasoro algebras leads to the relation 
\begin{align}
 \Xi_{(r,s)} = \Xi_{(r,-s)}\ .
 \label{xms}
\end{align}
Further constraints come from the existence of a degenerate field $V_{\langle 1,2\rangle}$ that transforms in the degenerate representation $\mathcal{R}_{\langle 1,2\rangle}$ of the conformal algebra. The fusion of such a field with a non-diagonal primary field $V^N_{(r,s)}\in \mathcal{W}_{(r,s)}$ yields \cite{mr17}
\begin{align}
 V_{\langle 1,2\rangle} \times V^N_{(r,s)} \simeq V^N_{(r,s-1)} +  V^N_{(r,s+1)} \ . 
\end{align}
In the Potts CFT, $V_{\langle 1,2\rangle}$ is an $S_Q$ singlet. This implies that $V^N_{(r,s\pm 1)}$ transform in the same way as $V^N_{(r,s)}$ under $S_Q$, therefore
\begin{align}
\Xi_{(r,s)} = \Xi_{(r,s+1)}\ .
\label{xspt}
\end{align}
This suggests that we combine the representations of the conformal algebra into the larger representations
\begin{align}
 \doublewidetilde{\mathcal{R}}_{\langle 1,1\rangle} = \bigoplus_{s\in \mathbb{N}^*} \mathcal{R}_{\langle 1,s\rangle}  \quad , \quad 
 \doublewidetilde{\mathcal{W}}_{(0,\frac12)}  = \bigoplus_{s\in \mathbb{N}+\frac12} \mathcal{W}_{(0,s)}
 \quad , \quad 
 \doublewidetilde{\mathcal{W}}_{(r,s)} \underset{r\neq 0}{=} \bigoplus_{s'\in \mathbb{Z}+s} \mathcal{W}_{(r,s)}\ .
\end{align}
These can be interpreted as indecomposable representations of the interchiral algebra $\doublewidetilde{\mathfrak{C}}_{\beta^2}$, which is obtained from the conformal algebra by adding the degenerate field $V_{\langle 1,2\rangle}$  \cite{hjs20}. The space of states can then be rewritten as  
\begin{align}
 \mathcal{S}^{O(n)} \underset{\doublewidetilde{\mathfrak{C}}_{\beta^2}\times S_Q}{=} \doublewidetilde{\mathcal{R}}_{\langle 1,1\rangle}\otimes []
 \oplus \doublewidetilde{\mathcal{W}}_{(0,\frac12)}\otimes [1]
 \oplus \bigoplus_{r\in\mathbb{N}+2}\bigoplus_{\substack{s\in\frac{1}{r}\mathbb{Z}\\ -\frac12<s\leq \frac12}} \doublewidetilde{\mathcal{W}}_{(r,s)}\otimes \Xi_{(r,s)}\ .
 \label{spottsi}
\end{align}

\subsubsection{Action of the symmetric group}

The action of the orthogonal group on the space of states is encoded in the family of representations $\Xi_{(r,s)}$. Just as in the case of the $O(n)$ CFT, it is possible to guess these representations from the known torus partition function of \cite{fsz87}: the idea is to replace each occurrence of the parameter $Q$ with a well-chosen $Q$-dimensional representation of the form $\sum_{d|r} \xi_d$. We will not do this in detail, because we are about to derive the result from the twisted partition function. For the moment, we only display the result:
\begin{align}
 \boxed{\Xi_{(r,s)} = (-1)^r\delta_{s\in \mathbb{Z}+\frac{r+1}{2}}[1] + \frac{1}{r} \sum_{r'=0}^{r-1} e^{2\pi i r's} U_{\mathrm{gcd}(r, r')} \left(\textstyle{\sum}_{d|\frac{r}{\mathrm{gcd}(r, r')}} \xi_d-2\right)}
 \renewcommand{\arraystretch}{1.3}
 \ , \quad \left\{\begin{array}{l}r\in\mathbb{N}+2\ ,\\ s\in\frac{1}{r}\mathbb{Z}\ , \end{array}\right.
 \label{xirs}
\end{align}
where $\xi_d$ \eqref{xir} is a formal representation of $S_Q$, and $U_d$ \eqref{xd} is a Chebyshev polynomial. Alternatively, using Ramanujan's sums \eqref{vfkr}, this can be rewritten as 
\begin{align}
 \Xi_{(r,s)} = (-1)^r\delta_{s\in \mathbb{Z}+\frac{r+1}{2}}[1] + \frac{1}{r}
 \sum_{\substack{g,g'|r\\ gg'=r}} \varphi_{rs}(g) U_{g'}\left(\textstyle{\sum}_{d|g} \xi_d-2\right)\ .
 \label{xirsp}
\end{align}
Since $\varphi_{rs}(g)\in \mathbb{Z}$, and the Chebyshev polynomials have integer coefficients, $\Xi_{(r,s)}$ is a combination of irreducible, finite-dimensional representations with coefficients in $\frac{1}{r}\mathbb{Z}$. The coefficients should actually be positive integers: in examples, we can see that this is actually the case. 

\subsubsection{Explicit examples} 

Given the relations \eqref{xms} and \eqref{xspt}, the cases $0\leq s\leq \frac12$ already sample all different representations $\Xi_{(r,s)}$. The number of different cases is further restricted by the fact that $\Xi_{(r,s)}$ only depends on $s$ through $\mathrm{gcd}(r, rs)$.
We now write these reprentations according to Eq. \eqref{xirs}. Here we display the results for $r\leq 6$. The computer code \cite{rz22} can easily reach at least $r\simeq 10$. 
\begingroup
\allowdisplaybreaks
\begin{subequations}
\label{x-all}
\begin{align}
 \Xi_{(2, 0)} &= [2] \ , 
 \\
 \Xi_{(2,\frac12)} &= [1^2] \ ,
 \\
 \Xi_{(3, 0)} &= [3]+[1^3]  \ , 
 \\
 \Xi_{(3,\frac13)} &= [21]  \ ,
 \\
 \Xi_{(4, 0)} &= [4]+[2^2]+[21^2]+[3]+[21]+2[2]+[1]+[] \  ,
 \\
 \Xi_{(4,\frac14)} &= [31]+[21^2]+[21]+[1^3]+[1^2] \ ,
 \\
 \Xi_{(4,\frac12)} &= [31]+[2^2]+[1^4]+[3]+[21]+[2]+[1^2]+[1]\ , 
 \\
 \Xi_{(5, 0)} &= [5]+[32]+2[31^2]+[2^21]+[1^5]+[4]+3[31]+2[2^2]+3[21^2]+[1^4]
 \nonumber \\ 
& \quad
 +2[3]+4[21]+2[1^3]+2[2]+2[1^2]+[1] \ , 
 \\
 \Xi_{(5,\frac15)} &= [41]+[32]+[31^2]+[2^21]+[21^3]+[4]+3[31]+2[2^2]+3[21^2]+[1^4]
 \nonumber \\ 
& \quad
 +2[3]+4[21]+2[1^3]+2[2]+2[1^2]+[1] \ , 
 \\
 \Xi_{(6,0)} &= [6]+2[42]+2[41^2]+[3^2]+2[321]+2[31^3]+2[2^3]+[2^21^2]+[21^4]
 \nonumber \\ 
& \quad
 +2[5]+6[41]+8[32]+8[31^2]+8[2^21]+6[21^3]+2[1^5]+7[4]+14[31]+11[2^2]
 \nonumber \\ 
& \quad
 +13[21^2]+6[1^4]+11[3]+15[21]+8[1^3]+10[2]+6[1^2]+5[1]+3[] \  ,
 \\
 \Xi_{(6,\frac16)} &=
 [51]+[42]+2[41^2]+[3^2]+3[321]+[31^3]+2[2^21^2]+[21^4]
 \nonumber \\ 
& \quad
 +[5]+6[41]+7[32]+10[31^2]+7[2^21]+6[21^3]+[1^5]+3[4]+15[31]+8[2^2]
 \nonumber \\ 
& \quad
 +16[21^2]+4[1^4]+6[3]+16[21]+9[1^3]+5[2]+8[1^2]+2[1] \ ,
 \\
 \Xi_{(6,\frac13)} &= 
 [51]+2[42]+[41^2]+3[321]+2[31^3]+[2^3]+[2^21^2]+[21^4]
 \nonumber \\ 
& \quad
 +2[5]+6[41]+8[32]+8[31^2]+8[2^21]+6[21^3]+2[1^5]+6[4]+14[31]+12[2^2]
 \nonumber \\ 
& \quad
 +13[21^2]+5[1^4]+9[3]+17[21]+6[1^3]+9[2]+6[1^2]+4[1]+[] \ ,
 \\
 \Xi_{(6,\frac12)} &=
 [51]+[42]+2[41^2]+2[3^2]+2[321]+2[31^3]+[2^3]+2[2^21^2]+[1^6]
 \nonumber \\ 
& \quad
 +[5]+6[41]+7[32]+10[31^2]+7[2^21]+6[21^3]+[1^5]+4[4]+15[31]+7[2^2]
 \nonumber \\ 
& \quad
 +16[21^2]+5[1^4]+7[3]+15[21]+10[1^3]+5[2]+9[1^2]+3[1]\ .
\end{align}
\end{subequations}

\subsection{Twisted partition function of the Potts CFT}

\subsubsection{Definition and character decomposition}

Let us define the twisted partition function of the Potts CFT on a torus of modulus $\tau$ as 
\begin{align}
 Z^{\text{$Q$-Potts}}(\tau,g) = \operatorname{Tr}_{\mathcal{S}^{\text{$Q$-Potts}}}\left(e^{2\pi i \tau(L_0-\frac{c}{24})} e^{2\pi i \bar\tau(\bar L_0-\frac{c}{24})}\cdot g\right) \ , 
 \label{zptg}
\end{align}
where $g\in S_Q$, and $L_0,\bar L_0$ are elements of the conformal algebra $\mathfrak{C}_c$. The decomposition \eqref{spotts} of the space of states $\mathcal{S}^{\text{$Q$-Potts}}$ into representations of $S_Q$ and $\mathfrak{C}_c$ leads to a decomposition of the partition function into characters,
\begin{align}
 Z^{\text{$Q$-Potts}}(\tau,g) =  
 \sum_{s\in \mathbb{N}^*} \chi_{\langle 1,s\rangle}(\tau) + \sum_{s\in \mathbb{N}+\frac12} \chi^N_{(0,s)}(\tau) \chi^{S_Q}_{[1]}(g)+ \sum_{r\in\mathbb{N}+2}\sum_{s\in\frac{1}{r}\mathbb{Z}} \chi^N_{(r,s)}(\tau)\chi^{S_Q}_{\Xi_{(r,s)}}(g)
 \ ,
 \label{zptgchi}
\end{align}
where the characters $\chi_{\langle 1,s\rangle}(\tau),\chi^N_{(r,s)}(\tau)$ of the conformal algebra were defined in Eq. \eqref{chis}. 

The modular invariant partition function in \cite{fsz87} is the special case $Z^{\text{$Q$-Potts}}(\tau,\text{id})$. In its decomposition into characters, the $S_Q$ characters reduce to 
\begin{align}
 \chi^{S_Q}_{\Xi_{(r,s)}}(\text{id}) = \dim^{S_Q}\Xi_{(r,s)} = (-1)^r\delta_{s\in \mathbb{Z}+\frac{r+1}{2}}(Q-1) + \frac{1}{r} \sum_{r'=0}^{r-1} e^{2\pi i r's} U_{\mathrm{gcd}(r, r')} \left(Q-2\right)\ .
\end{align}
These dimensions are polynomial functions of $Q$. 
For example
\begingroup
\allowdisplaybreaks
\begin{subequations}
\label{dimXi-exl}
\begin{align}
   \dim^{S_Q}\Xi_{(2,0)} &= \tfrac12 Q(Q-3) \ , \\
   \dim^{S_Q}\Xi_{(2,\frac12)} &= \tfrac12 (Q-1)(Q-4) \ , \\
   \dim^{S_Q}\Xi_{(3,0)} &= \tfrac13 (Q-1)(Q-2)(Q-3) \ , \\
   \dim^{S_Q}\Xi_{(3,\frac13)} &= \tfrac13 Q(Q-2)(Q-4) \ , \\
   \dim^{S_Q}\Xi_{(4,0)} &= \tfrac14 Q(Q-2)(Q-3)^2 \ , \\
   \dim^{S_Q}\Xi_{(4,\frac14)} &= \tfrac14 Q(Q-1)(Q-3)(Q-4) \, , \\
   \dim^{S_Q}\Xi_{(4,\frac12)} &= \tfrac14 (Q-1)^2 (Q-2)(Q-4) \ ,
\end{align}
\end{subequations}
\endgroup
in agreement with \eqref{x-all}. However, knowing these dimensions is not enough for determining the corresponding representations.

On the other hand, the character $\chi^{S_Q}_{\Xi_{(r,s)}}(g)$ fully determines the representation $\Xi_{(r,s)}$. In order to prove the conjecture, it is therefore sufficient to compute the twisted partition function $Z^{\text{$Q$-Potts}}(\tau,g)$ and check that it agrees with \eqref{zptgchi}. Due to the inclusion of a group element $g$, the proof will only be valid for $Q\in\mathbb{N}$. However, just like reduced Kronecker coefficients, the structure of $\Xi_{(r,s)}$ is $Q$-independent for $Q$ generic, and can be accessed from integer values of $Q$ by taking the limit $Q\to \infty$. 

\subsubsection{Loop representation of the Potts model}

For $Q$ in a rather large region of the complex plane, including for $Q\in (0, 4)$, the Potts CFT is the infinite-size limit of the $Q$-state Potts model at the critical coupling constant, $J = J_c$ \cite{bjjz22}. The simplest version of the lattice model is defined on the square lattice, in which case \cite{Baxter73}
\begin{align}
 e^{J_c} = 1 \pm \sqrt{Q}\ .
 \label{pottsKc}
\end{align}
The Potts model's loop representation is very similar to that of the $O(n)$ model. This allows us to compute the Potts CFT's partition function as a modification of the $O(n)$ CFT's partition function \cite{fsz87}. We will now review this computation, while generalizing it to the twisted partition function. 

In its original formulation, the Potts model describes variables $s_i\in \{1,2,\dots, Q\}$ (called spins) on a square lattice, with an interaction energy $-J \delta_{s_i,s_j}$ between neighbouring sites. In order to compute the twisted partition function on the torus $\mathbb{T} =\frac{\mathbb{C}}{\mathbb{Z}+\tau\mathbb{Z}}$, we single out a topologically non-trivial oriented cycle on the torus, for example a cycle whose class in $\pi_1(\mathbb{T})=\mathbb{Z}^2$ is $(0,1)$. For a given $g\in S_Q$,
we then modify the interaction energy to $-J \delta_{s_i,g(s_{j})}$ whenever the edge $(ij)$ crosses the chosen cycle. We will now follow this modification when reformulating the Potts model as a random cluster model, and then as a loop model. 

The random cluster formulation of the Potts model is obtained by writing the edge Boltzmann weight as
\begin{align}
 e^{J \delta_{s_i,s_j}} = 1 + (e^J - 1) \delta_{s_i,s_j}
 \label{fkbol}
\end{align}
and replacing the sum over spins $\{s_i\}$ with a sum over $A$, where $A \subseteq E$ denotes the subset of edges for which the term $(e^J-1)\delta_{s_i,s_j}$ is taken in the expansion of $\prod_{(ij)\in E} e^{J\delta_{s_i,s_j}}$. For any lattice, this leads to \cite{jac12}
\begin{align}
 Z^{\text{$Q$-Potts}} = \sum_{A \subseteq E} (e^J - 1)^{|A|} Q^{k(A)} \ ,
 \label{FKmod}
\end{align}
where $k(A)$ denotes the total number of clusters in $A$.

Now on the torus, let us consider a topologically non-trivial cluster whose class in $\pi_1(\mathbb{T})$ is $(m,m')$. (See Figure \eqref{pll}.) Let us start at a site $i_0$ of spin $s_{i_0}$, and go once around the cluster. Along the way, 
the cycle where spins jump by the permutation $g$ is crossed $m$ times, and we come back to our starting point with the spin $g^m(s_{i_0})=s_{i_0}$. The number of possible configurations for this cluster is therefore
\begin{align}
 \sum_{s=1}^Q \delta_{s,g^m(s)} = \operatorname{Tr}_{[1]+[]}g^m \ ,
\end{align}
and this is the cluster weight in the twisted partition function.

The loop formulation of the Potts model is obtained by drawing the non-intersecting loops that separate clusters from dual clusters. In the following example the original square lattice is in black, the edge subset $A \subseteq E$ is in red, and the dual clusters in green live on the dual lattice. The loops in blue live on the corresponding medial graph, with each loop separating a cluster from a dual cluster:
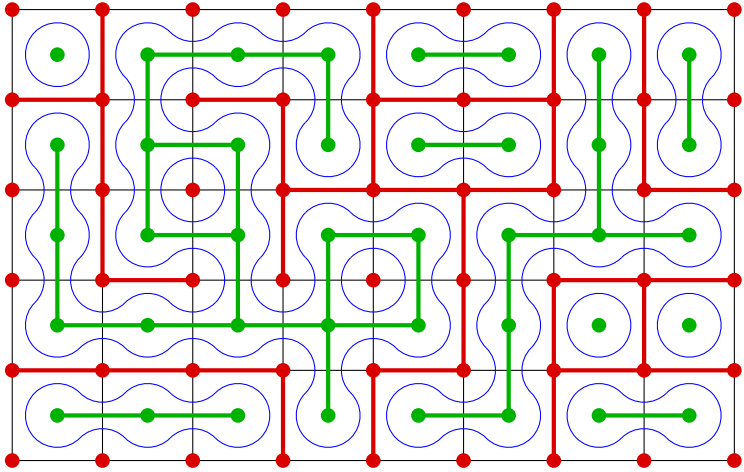
\begin{align}
\newcommand{\verti}[2]{
\draw[blue] (#1-.5, #2-.5) to [out=45, in=-45] (#1-.5, #2+.5);
\draw[blue] (#1+.5, #2-.5) to [out=135, in=-135] (#1+.5, #2+.5);
\ifodd#1\relax\draw[ultra thick, green!70!black] (#1, #2-1) -- (#1, #2+1);
\else\draw[ultra thick, red!85!black] (#1, #2-1) -- (#1, #2+1);\fi
}
\newcommand{\horiz}[2]{
\draw[blue] (#1-.5, #2-.5) to [out=45, in=135] (#1+.5, #2-.5);
\draw[blue] (#1-.5, #2+.5) to [out=-45, in=-135] (#1+.5, #2+.5);
\ifodd#1\relax\draw[ultra thick, red!85!black] (#1-1, #2) -- (#1+1, #2);
\else\draw[ultra thick, green!70!black] (#1-1, #2) -- (#1+1, #2);\fi
}
 \begin{tikzpicture}[baseline=(current  bounding  box.center), scale = .6]
 \clip (-.2, -.2) rectangle (16.2, 10.2);
 \foreach\i in {0,...,8}{
 \draw (2*\i, 0) -- (2*\i, 10);
 \draw[blue] (2*\i+.5, 9.5) to [out=45, in=135] (2*\i+1.5, 9.5);
\draw[blue] (2*\i+.5, .5) to [out=-45, in=-135] (2*\i+1.5, .5);
 }
 \foreach\j in {0,...,5}{
 \draw (0, 2*\j) -- (16, 2*\j);
 \draw[blue] (15.5, 2*\j+.5) to [out=45, in=-45] (15.5, 2*\j+1.5);
\draw[blue] (.5, 2*\j+.5) to [out=135, in=-135] (.5, 2*\j+1.5);
 }
 \foreach\i in {0,...,8}{\foreach\j in {0,...,5}{
 \node[red!85!black] at (2*\i, 2*\j) [fill, circle, scale = .5] {};}}
 \foreach\i in {0,...,7}{\foreach\j in {0,...,4}{
 \node[green!70!black] at (2*\i+1, 2*\j+1) [fill, circle, scale = .5] {};}}
 \horiz{1}{2};\horiz{3}{2};\horiz{5}{2};\verti{7}{2};\horiz{9}{2};\verti{11}{2};\horiz{13}{2};\horiz{15}{2};
 \verti{1}{4};\horiz{3}{4};\verti{5}{4};\verti{7}{4};\verti{9}{4};\verti{11}{4};\horiz{13}{4};\horiz{15}{4};
 \verti{1}{6};,\verti{3}{6};\verti{5}{6};\horiz{7}{6};\horiz{9}{6};\horiz{11}{6};
 \verti{13}{6};\horiz{15}{6};
 \horiz{1}{8};\verti{3}{8};\horiz{5}{8};\verti{7}{8};\horiz{9}{8};\horiz{11}{8};\verti{13}{8};\verti{15}{8};
 \horiz{2}{1};\horiz{2}{3};\verti{2}{5};\verti{2}{7};\verti{2}{9};
 \horiz{4}{1};\horiz{4}{3};\horiz{4}{5};\horiz{4}{7};\horiz{4}{9};
 \verti{6}{1};\horiz{6}{3};\verti{6}{5};\verti{6}{7};\horiz{6}{9};
 \verti{8}{1};\horiz{8}{3};\horiz{8}{5};\verti{8}{7};\verti{8}{9};
 \horiz{10}{1};\verti{10}{3};\verti{10}{5};\horiz{10}{7};\horiz{10}{9};
 \verti{12}{1};\verti{12}{3};\horiz{12}{5};\verti{12}{7};\verti{12}{9};
 \horiz{14}{1};\verti{14}{3};\horiz{14}{5};\verti{14}{7};\verti{14}{9};
 \end{tikzpicture}
 \label{plat}
\end{align}
On the torus, let us assume that there exists a cluster that includes a cycle of class $(m,m')\in \pi_1(\mathbb{T})$, and no other non-trivial cycle. Then all topologically non-trivial loops and clusters are of that class, with twice as many such loops as clusters. This is no longer the case for a configuration where a cluster contains at least two topologically independent cycles: this is called a cluster with cross topology. It is not hard to show that the first fundamental group of such a cluster includes the full $\pi_1(\mathbb{T})=\mathbb{Z}^2$, and that the weight of the cluster is $\operatorname{Tr}_{[1]+[]}g$. All loops are topologically trivial if and only if there exists a cluster or a dual cluster with cross topology. Representing the torus as a parallelogram, let us draw a cluster with cross topology:
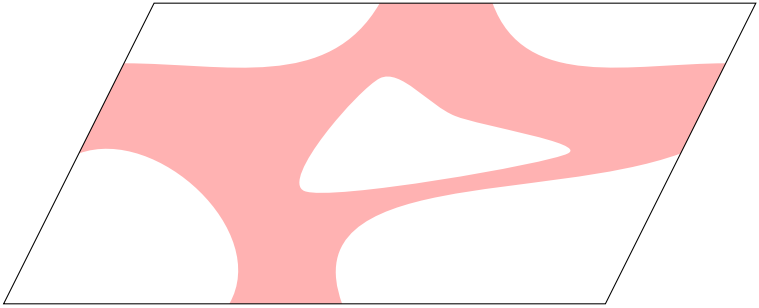
\begin{align}
 \begin{tikzpicture}[baseline=(current  bounding  box.center)]
  \fill [red!30] (3, 0) -- (4.5, 0) to [out = 110, in = -160] (9, 2) -- (9.6, 3.2) to [out = 180, in = -70] (6.5, 4) -- (5, 4) to [out = -120, in = 0] (1.6, 3.2) -- (1, 2) to [out = 20, in = 60] cycle;
  \fill [white] plot [smooth cycle] coordinates {(5, 3) (6, 2.5) (7.5, 2) (4, 1.5)};
  \draw (0, 0) -- (8, 0) -- (10, 4) -- (2, 4) -- cycle;
 \end{tikzpicture}
 \label{cwct}
\end{align}

\subsubsection{Twisted partition function}

For a configuration $\mathcal{C}$, we write
\begin{itemize}
\item $S_0$ the number of clusters whose images in $\pi_1(\mathbb{T})$ are trivial,
\item $S$ the number of clusters whose images are non-trivial, excluding clusters with cross topology,
\item $(m,m')$ the corresponding element of $\pi_1(\mathbb{T})$,
 \item $|A|$ the number of edges in the subset $A \subseteq E$.
\end{itemize}
Taking the critical value for the coupling, the twisted partition function of the $Q$-state Potts model on the torus $\mathbb{T} =\frac{\mathbb{C}}{\mathbb{Z}+\tau\mathbb{Z}}$ with $\Sigma$ lattice sites is 
\begin{align}
 Z_\text{lattice}^{\text{$Q$-Potts}}(\tau,g) = Q^{-\frac12\Sigma} \sum_{\mathcal{C}} 
 Q^{\frac12 |A|(\mathcal{C})} Q^{S_0(\mathcal{C})}  \left(\operatorname{Tr}_{[1]+[]}g^{m(\mathcal{C})} \right)^{S(\mathcal{C})} \left\{1 + \frac12\delta_{S(\mathcal{C}),0}\operatorname{Tr}_{[1]}g \right\} \ .
\end{align}
In the last factor, the term $\frac12\delta_{S(\mathcal{C}),0}\operatorname{Tr}_{[1]}g$ accounts for the cluster with cross topology. If such a cluster exists, then no other topologically non-trivial cluster can exist, therefore $S=0$. Conversely, if $S=0$, we have two mutually exclusive cases: there exists either a dual cluster with cross topology, or a cluster with cross topology. If $S=0$, the last factor can be rewritten as $\frac12\left( 1 + \operatorname{Tr}_{[1]+[]}g\right)$, with each term corresponding to one of the two cases. 

Let us rewrite the twisted partition function in terms of the numbers $N_0,N$ of loops, topologically trivial and non-trivial respectively. We obviously have $N=2S$, and using the Euler characteristic we can show $N_0= 2S_0+|A|-\Sigma$, which leads to
\begin{align}
\label{zlatt2z}
 Z_\text{lattice}^{\text{$Q$-Potts}}(\tau,g) = \sum_{\mathcal{C}|N(\mathcal{C})\in 2\mathbb{N}} 
 \sqrt{Q}^{N_0(\mathcal{C})} \left\{ \sqrt{\operatorname{Tr}_{[1]+[]}g^{m(\mathcal{C})} }^{N(\mathcal{C})} + \frac12 \delta_{N(\mathcal{C}),0}\operatorname{Tr}_{[1]}g \right\}
 \ .
\end{align}
Compared to the twisted partition function of the $O(n)$ model \eqref{zol}, the main differences are the restriction to even numbers of topologically non-trivial loops, and the term with $\delta_{N(\mathcal{C}),0}$. 

Let us relate this lattice partition function to the loop sums $Z^{M,M'}_\text{lattice}(\tau, n)$ \eqref{zmmp}, where we assume that the coupling $K$ takes its critical value. We split our loops into two possible orientations $N=N_++N_-$ with $N_+\equiv N_-\bmod 2$, and use Eq. \eqref{tzn} for the two weights that appear in the form \eqref{zlatt2z} of the partition function: $z =\sqrt{\operatorname{Tr}_{[1]+[]}g^{m(\mathcal{C})} }$ in the first term, and $z=0$ in the second term, which is proportional to $\delta_{N(\mathcal{C}),0}=0^{N(\mathcal{C})}$. Using Eq. \eqref{xtd} for $U_d(0)$ and 
writing $(M,M')=(m(N_+-N_-),m'(N_+-N_-))$, we obtain
\begin{multline}
 Z_\text{lattice}^{\text{$Q$-Potts}}(\tau,g) = \frac12 \sum_{M,M'\in 2\mathbb{Z}} Z^{M,M'}_\text{lattice}\left(\tau,\sqrt{Q}\right) \Bigg\{U_{\frac12 \mathrm{gcd}(M, M')}\left(\operatorname{Tr}_{[1]+[]} g^{\frac{M}{\mathrm{gcd}(M, M')}} - 2\right) 
 \\
 +\cos \tfrac{\pi}{2} (\mathrm{gcd}(M, M')) \operatorname{Tr}_{[1]}g  \Bigg\}\ . 
 \label{zqp}
\end{multline}

\subsubsection{Critical limit}

For the sake of simplicity, we have already taken the coupling to be critical, and we further send the lattice size to infinity. 
This allows us to use the expression \eqref{zmmc} for the critical limit of $Z^{M,M'}_\text{lattice}\left(\tau,\sqrt{Q}\right)$. We then perform a Poisson resummation over $M'$ if $M=0$, or over $k$ such that $M'=kM+2r'$ with $r'\in \{0,1,\dots,\frac{M}{2}-1\}$ if $M\neq 0$. Introducing the integer $r$ such that $M=2r$, and separately considering the two terms in Eq. \eqref{zqp}, we obtain four terms:
\begin{align}
 \left. Z^{\text{$Q$-Potts}}(\tau,g)\right|^\text{Term $1$}_{M=0} &= \sum_{s\in\mathbb{Z}} \left|\frac{e^{2\pi i\tau P_{(1,s)}^2}}{\eta(\tau)}\right|^2 = \sum_{s\in\mathbb{N}^*} \chi_{\langle 1,s\rangle}(\tau) + \sum_{s\in \mathbb{Z}} \chi_{(1,s)}^N(\tau)\ , 
 \\
 \left. Z^{\text{$Q$-Potts}}(\tau,g)\right|^\text{Term $1$}_{M\neq 0} &= \sum_{r\in\mathbb{N}^*}\sum_{s\in\frac{1}{r}\mathbb{Z}} \chi^N_{(r,s)}(\tau) \sum_{r'=0}^{r-1} e^{2\pi i r's} U_{\mathrm{gcd}(r, r')}\left(\operatorname{Tr}_{[1]+[]} g^{\frac{r}{\mathrm{gcd}(r, r')}} -2 \right) \ , 
 \\
 \left. Z^{\text{$Q$-Potts}}(\tau,g)\right|^\text{Term $2$}_{M=0} &= 
 \sum_{s\in\mathbb{Z}+\frac12} \chi_{(0,s)}^N(\tau)\operatorname{Tr}_{[1]}g \ , 
 \\
 \left. Z^{\text{$Q$-Potts}}(\tau,g)\right|^\text{Term $2$}_{M\neq 0} &= \left(\sum_{r\in 2\mathbb{N}^*} \sum_{s\in\mathbb{Z}+\frac12}\chi^N_{(r, s)}(\tau) - \sum_{r\in 2\mathbb{N}+1} \sum_{s\in\mathbb{Z}}\chi^N_{(r,s)}(\tau) \right)\operatorname{Tr}_{[1]}g \ .   
\end{align}
Summing the four terms, we recover the twisted partition function \eqref{zptgchi}. In particular, the terms that involve $\chi^N_{(1,s)}(\tau)$ all cancel.

\section{Algebraic approach to the \texorpdfstring{$O(n)$}{} model}

There are a number of different constructions of the $O(n)$ model on the lattice, which lead to the same critical limit \cite{nie82,Nienhuis89,rs07}. For simplicity, we consider the dense $O(n)$ universality class, which is contained in all these lattice models.
The space of states is 
\begin{align}
 \mathcal{S}^{O(n)}_L = [1]^{\otimes L}  \ .
 \label{SLOn}
\end{align}
This space describes the states of a chain of $L$ spins, where each spin lives in the vector representation $[1]$. (For the dilute model with spins in $[1]\oplus []$, see the conclusion.)
The corresponding model may be called the dense unoriented loop model \cite{rs07}. This model admits either periodic or non-periodic boundary conditions, and we refer to these cases as the closed and open spin chain, respectively. While our main interest is in the closed chain, we will also deal with the open chain, which has simpler algebraic properties.

\subsection{Action of diagram algebras on the space of states}\label{oada}

We now introduce the diagram algebras that act on the space of states of the $O(n)$ model. We will focus on the algebraic properties of these algebras and their representations. For readers not already familiar with diagram algebras, we now recommend reading Appendix \ref{app:diag}, which reviews diagram algebras and the eponymous diagrams.

\subsubsection{The relevant diagram algebras}

The algebra of linear maps on $\mathcal{S}^{O(n)}_L$ that commute with $O(n)$ is the Brauer algebra
\begin{align}
 O(n)^* = \mathcal{B}_L(n)\ . 
\end{align}
The Brauer algebra is generated by $2L-2$ elements $p_i, e_i$ with $i=1,\dots, L-1$. The generator $p_i$ permutes the factors $i$ and $i+1$ in $[1]^{\otimes L}$. For $n$ integer, the generator $e_i$ acts as follows on the factors $i$ and $i+1$ of $[1]^{\otimes L}$, while leaving the other factors invariant:
\begin{align}
 e_i :  \quad \begin{array}{lll}
               [1]_i\otimes [1]_{i+1} & \to & [1]_i\otimes [1]_{i+1}
               \\ w_k \otimes w_\ell & \mapsto & \delta_{k\ell}\sum_{m=1}^n w_m\otimes w_m 
              \end{array}
 \ , 
\end{align}
where $(w_k)_k$ is an orthonormal basis of $[1]$. As a diagram algebra, the Brauer algebra exists for any $n\in\mathbb{C}$, and the generators $p_i,e_i$ are described by the diagrams \eqref{s-gen}, \eqref{extra-gen}. 

The Brauer algebra includes the permutation group $S_L$, and ignores the geometry of the problem. If we assume that the $L$ spins are arrayed in a chain, we can no longer permute them. The symmetries of the system must then be described by a smaller algebra, which also takes into account the boundary conditions. If we have an open chain with free boundary conditions, the relevant algebra is the Temperley--Lieb algebra $\TL_L(n)$, which has $L-1$ generators $e_i$. For a closed chain with periodic boundary conditions, the relevant algebra is the unoriented Jones--Temperley--Lieb algebra $\uJTL_L(n)$, which is obtained from $\TL_L(n)$ by adding two generators: a generator $e_L$ \eqref{el} which contracts the sites $L,1$ in the same manner that $e_i$ contracts $i,i+1$, and a generator $u$ \eqref{u} which cyclically translates all the sites by one unit, and obeys $u^L=1$.  

The diagrammatic representations of the algebras $\TL_L(n)$ and $\uJTL_L(n)$ directly correspond to the representation of the $O(n)$ model as a loop model in the transfer matrix approach: the absence of the permutation generators means that loops do not cross, and the relations $e_i^2 = n e_i$ means that loops have weight $n$. However, in contrast to our construction of loops in Figure \eqref{plat}, the representation $[1]$ now lives on the loops themselves, rather than on sites between the loops. This difference does not affect the critical limit. 

Therefore, we expect that the algebras $\TL_L(n)$ and $\uJTL_L(n)$ describe symmetries of the $O(n)$ model. In particular, in the critical limit, we expect 
\begin{align}
 \lim_{\text{critical}} \uJTL_L(n) = \widetilde{\mathfrak{C}}_{\beta^2} \ , 
\end{align}
where the parameter $\beta^2$ of the interchiral algebra is related to $n$ by Eq. \eqref{nqb}. 

\subsubsection{Representations of the Brauer algebra}

By Schur--Weyl duality, the space of states \eqref{SLOn} decomposes into irreducible representations of $\mathcal{B}_L(n)\times O(n)$ as \cite{Br37}
\begin{align}
 \mathcal{S}^{O(n)}_L \underset{\mathcal{B}_L(n)\times O(n)}{=} \bigoplus_{\substack{|\lambda|\leq L\\ |\lambda|\equiv L\bmod 2}}  B^{(L)}_{\lambda}\otimes \lambda\ ,
 \label{sonl}
\end{align}
where the sum is over partitions $\lambda$, which we identify with irreducible representations of $O(n)$ with generic $n\in\mathbb{C}$. The irreducible finite-dimensional representations $B^{(L)}_{\lambda}$ of the Brauer algebra are also labelled by partitions: we will say more on their structure when it comes to computing branching rules in Section \ref{sec:obr}. For the moment, let us point out that their dimensions appear as coefficients in the decomposition of $\mathcal{S}^{O(n)}_L = [1]^{\otimes L} $ into irreducible representations of $O(n)$,
\begin{align}
 [1]^{\otimes L} \underset{O(n)}{=} \bigoplus_{\substack{|\lambda|\leq L\\ |\lambda|\equiv L\bmod 2}} \left(\dim B^{(L)}_{\lambda}\right)\lambda\ .
\end{align}
For example, $[1]^{\otimes 3} = [3]+2[21]+[111]+3[1]$ tells us that $\dim B^{(3)}_{[21]}=2$. 

\subsubsection{Representations of the Temperley--Lieb algebra}

We now want to decompose the representations of the Brauer algebra into irreducible representations of the smaller   $\TL_L(n)$ and $\uJTL_L(n)$ algebras. In the case of the Temperley--Lieb algebra $\TL_L(n)$, the branching rules are explicitly known \cite{bm05}. The irreducible representations of $\TL_L(n)$ are called standard modules, and parametrized by positive integers in $L-2\mathbb{N}$: for ease of comparison with the CFT, we will use half-integers $r\in \frac{L}{2}-\mathbb{N}$ instead, and write $W^{(L)}_{r}$ for the standard module of dimension \eqref{dimTL-loop}. 
The decomposition takes the form 
\begin{align}
 \boxed{B^{(L)}_{\lambda} \underset{\TL_L(n)}{=} f_\lambda \bigoplus^{\frac{L}{2}}_{r=\frac{|\lambda|}{2}} c^{|\lambda|}_r W^{(L)}_{r}}\ ,
 \label{bbtl}
\end{align}
where the sum runs by increments of $1$. Here, $f_\lambda$ is the number of standard Young tableaux of shape $\lambda$, given by the hook length formula
\begin{align}
 f_\lambda = \frac{|\lambda|!}{\prod_{b \in \lambda} \text{hook length}(b)} \ ,
 \label{hooklf}
\end{align}
where the product is over the boxes in the Young diagram. Moreover, the coefficents $c^\ell_r\in \mathbb{N}$ are
\begin{align}
 c^\ell_r =\sum_{r'=0}^{r-\frac{\ell}{2}}(-1)^{r'} \binom{2r-r'}{r'} \binom{2r-2r'}{\ell} (2r-2r'-\ell-1)!!\ .
 \label{clr}
\end{align}
It goes without saying that these integer coefficients do not depend on the complex number $n$. More remarkably, they also do not depend on the lattice size $L$: this makes it particularly easy to take the critical limit.

Inserting our decomposition of $B^{(L)}_{\lambda}$ into the decomposition \eqref{sonl} of the space of states, we obtain 
\begin{align}
 \boxed{\mathcal{S}^{O(n)}_L \underset{\TL_L(n)\times O(n)}{=} \bigoplus_{r=\left\{\frac{L}{2}\right\}}^{\frac{L}{2}} W^{(L)}_{r} \otimes \Lambda_r}  \qquad \text{with} \qquad \boxed{\Lambda_r = \bigoplus_{\substack{|\lambda|\leq 2r \\ |\lambda| \equiv 2r\bmod 2}} f_\lambda c^{|\lambda|}_r \lambda}\ ,
 \label{sto}
\end{align}
where the sum over $r$ runs by increments of $1$, starting from the fractional part of $\frac{L}{2}$.
This implies $\mathcal{S}^{O(n)}_L \underset{O(n)}{=} \bigoplus_{r=\left\{\frac{L}{2}\right\}}^{\frac{L}{2}} \left(\dim W^{(L)}_{r}\right) \Lambda_r$, from which $\Lambda_r$ may be recursively deduced, assuming we know the dimensions \eqref{dimTL-loop} of the Temperley--Lieb representations.  
The first few representations $\Lambda_r$ are 
\begin{subequations}
\begin{align}
 \Lambda_0 &= [] \ ,
 \\
 \Lambda_\frac12 &= [1] \ , 
 \\
 \Lambda_1 &= [2] + [1^2]\ , 
 \\
 \Lambda_\frac32 &= [3] + 2[21] + [1^3] + [1]\ ,
 \\
 \Lambda_2 &= [4] + 3 [31] + 2 [2^2] + 3 [21^2] + [1^4] + 3 [2] + 3 [1^2] + [] \ .
\end{align}
\end{subequations}
They obey the simple property \cite{bm05}(Proposition 3.2)
\begin{align}
 [1]\otimes \Lambda_r = \Lambda_{r-\frac12} \oplus \Lambda_{r+\frac12}\ . 
 \label{otl}
\end{align}
Using the associativity of the tensor product, this implies 
\begin{align}
 \Lambda_{r_1}\otimes \Lambda_{r_2} = \bigoplus_{r=|r_1-r_2|}^{r_1+r_2} \Lambda_r\ .
\end{align}
As the decomposition \eqref{sto} of the space of states suggests, $\Lambda_r$ can be interpreted as an irreducible representation of the commutant $\TL_L^*(n)$ of the Temperley--Lieb algebra \cite{rs07}. The closure of the representations $\Lambda_r$ under tensor products is a consequence of the existence of a Hopf algebra structure on $\TL_L^*(n)$, i.e. of a coassociative comultiplication. The comultiplication is constructed by joining two open chains by their end points: this will no longer be possible with closed chains \cite{gjs18}. 

\subsubsection{Representations of the unoriented Jones--Temperley--Lieb algebra}

The irreducible representations $W^{(L)}_{(r,s)}$ of $\uJTL_L(n)$ are parametrized by positive half-integers $r\in \frac{L}{2}-\mathbb{N}$, and by a rational number $s$ called the pseudo-momentum. The possible values of the pseudo-momentum depends on $r$: we have $s\in \frac{1}{r}\mathbb{Z} \cap (-1, 1]$ if $r\neq 0$, and by convention we set $s=0$ if $r=0$. The decomposition of an irreducible representation of the Brauer algebra reads 
\begin{align}
 \boxed{B^{(L)}_{\lambda} \underset{\uJTL_L(n)}{=} \bigoplus_{r=\frac{|\lambda|}{2}}^{\frac{L}{2}} \bigoplus_{\substack{s\in\frac{1}{r}\mathbb{Z} \\ -1< s\leq 1}} c^\lambda_{(r,s)} W^{(L)}_{(r,s)}}\ .
 \label{bcw}
\end{align}
The coefficients $c^\lambda_{(r,s)}$ are positive integer numbers, which do not depend on $n$. We moreover expect that they do not depend on $L$, as was the case in the decomposition of $B^{(L)}_{\lambda}$ into representations of the Temperley--Lieb algebra.  
We will give two methods for computing these coefficients: via a closed combinatorial formula in Section~\ref{sec:obr}, and via transfer matrix techniques in Appendix~\ref{app:branching}. For the moment, let us proceed with decomposing the space of states, assuming $c^\lambda_{(r,s)}$ to be given.
Inserting the decomposition of Brauer representations into Eq. \eqref{sonl}, we obtain 
\begin{align}
 \boxed{\mathcal{S}^{O(n)}_L \underset{\uJTL_L(n)\times O(n)}{=} \bigoplus_{r=\left\{\frac{L}{2}\right\}}^{\frac{L}{2}} \bigoplus_{\substack{s\in\frac{1}{r}\mathbb{Z} \\ -1< s\leq 1}} W^{(L)}_{(r,s)} \otimes \Lambda_{(r,s)}}  \qquad \text{with} \qquad \boxed{\Lambda_{(r,s)} = \bigoplus_{\substack{|\lambda|\leq 2r \\ |\lambda| \equiv 2r\bmod 2}} c^\lambda_{(r,s)} \lambda}\ .
 \label{stou}
\end{align}
The representations $\Lambda_{(r,s)}$ of $O(n)$ that we just defined should coincide with the representations $\Lambda_{(r,s)}$ \eqref{lrs} that appeared in the spectrum \eqref{son} of the $O(n)$ CFT. Moreover, the large $L$ limit of the representation $W^{(L)}_{(r,s)}$ should coincide with a representation of the interchiral algebra,
\begin{align}
 \lim_{\text{critical}} W^{(L)}_{(r,s)} = \widetilde{\mathcal{W}}_{(r,s)} \ . 
\end{align}
One may doubt the plausibility of this relation, on the grounds that $W^{(L)}_{ (r,s)}$ is always irreducible, while $\widetilde{\mathcal{W}}_{(r,s)}$ is logarithmic and therefore reducible for $(r,s)\in \mathbb{N}^*$. However, logarithmic features can emerge at $L=\infty$ while being absent for any finite $L$. A necessary requirement for this is that two eigenvalues of the Hamiltonian differ for finite $L$, while having the same limit \cite{glhjs20}.

We can obtain some nontrivial information on $\Lambda_{(r,s)}$ by comparing our decomposition \eqref{stou} of the space of states with the decomposition \eqref{sto} into representations of the Temperley--Lieb algebra. As reviewed in Eq. \eqref{wrswr}, the representation $W^{(L)}_{(r,s)}$ of $\uJTL_L(n)$ is constructed by gluing the representations $W^{(L)}_{r},W^{(L)}_{r+1},\dots, W^{(L)}_{\frac{L}{2}}$ of $\TL_L(n)$. The pseudo-momentum $s$ labels different ways of gluing these representations. (For the special case of $W_{(0,0)}^L$, see Eq. \eqref{w00w0h}.)
We will review the general definition of the pseudo-momentum in Appendix \ref{app:diag}. Here we only need to know that 
in the case $r=\frac{L}{2}$, the pseudo-momentum corresponds to an eigenvalue of the translation generator $u$ \eqref{u} of $\uJTL_L(n)$, 
\begin{align}
 \left(u-e^{\pi is}\right) W^{(L)}_{(\frac{L}{2}, s)} = 0\ .
 \label{us}
\end{align}
Inserting Eq. \ref{wrswr} into the decomposition \eqref{stou}, and comparing with the decomposition \eqref{sto}, we obtain
\begin{align}
 \boxed{\bigoplus_{\substack{s\in\frac{1}{r}\mathbb{Z} \\ -1< s\leq 1}}  \Lambda_{(r,s)} = \Lambda_r - \delta_{r\geq \frac32}\Lambda_{r-1}} \ . 
 \label{slrs}
\end{align}
Using the recursion relation \eqref{otl} for $\Lambda_r$, we deduce the identity
\begin{align}
 \bigoplus_{\substack{s\in\frac{1}{r}\mathbb{Z} \\ -1< s\leq 1}}  \Lambda_{(r,s)} = \left(\delta_{r,1}-\delta_{r,0}\right) [] \oplus U_{2r}([1])\ , 
\end{align}
where the Chebyshev polynomial $U_d$ was defined in Eq. \eqref{xd}. This identity is statisfied by our expression \eqref{lrs} for $\Lambda_{(r,s)}$: in fact, this was an important hint for guessing that expression in the first place. 

The algebraic definition of the representations $\Lambda_{(r,s)}$ of the group $O(n)$ that appear in the space of states also allows us to rederive their invariances under shifts \eqref{lspt} and reflection \eqref{lms} of the second index. The invariance under shifts is now due to the fact that the representation $W^{(L)}_{(r,s)}$ of the algebra $\uJTL_L(n)$ only depends on $s$ via $e^{\pi is}$. The invariance under reflection is due to the automorphism of the diagram algebras that acts on the Brauer generators as $e_i,p_i \to e_{L-i},p_{L-i}$. Under this automorphism, the representation $B^{(L)}_{\lambda}$ of the Brauer algebra is invariant, but representations of the unoriented Jones--Temperley--Lieb algebra change as $W^{(L)}_{(r,s)} \to W^{(L)}_{(r,-s)}$. It follows that the coefficients of the decomposition \eqref{bcw} obey 
\begin{align}
 c^\lambda_{(r,s)}=c^\lambda_{(r,-s)} \ .
\end{align}

Finally, let us discuss the tensor products of the representations $\Lambda_{(r,s)}$ \eqref{lxxx-all}. Already in the case $\Lambda_{(\frac12,0)}\otimes \Lambda_{(1,0)} = [1]\otimes [2] = [3]+[21]+[1]$, we see that this set of representations is not closed under tensor products. While we may interpret $\Lambda_{(r,s)}$ as irreducible representations of the commutant $\uJTL_L^*(n)$, we can therefore not directly interpret this commutant as a symmetry algebra of the $O(n)$ model.

\subsection{Branching rules \texorpdfstring{$\mathcal{B}_L(n) \downarrow \uJTL_L(n)$}{}}\label{sec:obr}

We have reduced the determination of the $O(n)$ representations $\Lambda_{(r,s)}$ to the algebraic problem of decomposing representations of the Brauer algebra into representations of the unoriented Jones--Temperley--Lieb algebra. Let us now solve this problem.

\subsubsection{Brauer--Specht modules of $\mathcal{B}_L(n)$}

The irreducible representation $B^{(L)}_{\lambda}$ of the Brauer algebra is called a Brauer--Specht module. To build a basis, we start with all perfect matchings $b_i$ from $L$ lower sites to $|\lambda|$ upper sites, such that each upper site is matched with a lower site, modulo permutations of the upper sites. In the case $L=4, |\lambda|=2$, the corresponding diagrams are 
\newcommand{\bbdots}[1]{
\foreach \i in {1,...,#1}
{\path (\i, 0) coordinate (B\i);
 \fill (B\i) circle (4pt);}
}
\newcommand{\btdots}[2]{
\bbdots{#1};
\pgfmathtruncatemacro{\lb}{1+int((#1-#2)/2)};
\pgfmathtruncatemacro{\ub}{int((#1+#2)/2)};
\foreach \i in {\lb,...,\ub}
{\path (\i, 1.5) coordinate (T\i);
\fill (T\i) circle (4pt);}
}
\newcommand{\btldots}[2]{
\bbdots{#1};
\foreach \i in {1,...,#2}
{\path (\i, 1) coordinate (T\i);
\fill (T\i) circle (4pt);}
}
\newcommand{\bbline}[2]{
\draw (B#1) to [out = 90, in = 90] (B#2);
}
\newcommand{\ttline}[2]{
\draw (T#1) to [out = -90, in = -90] (T#2);
}
\newcommand{\btline}[2]{
\draw (B#1) to [out = 90, in = -90] (T#2);
}
\begin{align}
\renewcommand{\arraystretch}{1.3}
\renewcommand{\arraycolsep}{.4cm}
\begin{array}{cccccc}
 \begin{tikzpicture}[scale = .5, line width = 1pt, baseline=(current  bounding  box.center)]
 \btdots{4}{2};
\btline{1}{2};
\btline{2}{3};
\bbline{3}{4};
\end{tikzpicture}
&
\begin{tikzpicture}[scale = .5, line width = 1pt, baseline=(current  bounding  box.center)]
 \btdots{4}{2};
\btline{1}{2};
\btline{4}{3};
\bbline{3}{2};
\end{tikzpicture}
& 
\begin{tikzpicture}[scale = .5, line width = 1pt, baseline=(current  bounding  box.center)]
 \btdots{4}{2};
\btline{3}{2};
\btline{4}{3};
\bbline{1}{2};
\end{tikzpicture}
& 
\begin{tikzpicture}[scale = .5, line width = 1pt, baseline=(current  bounding  box.center)]
 \btdots{4}{2};
\btline{2}{2};
\btline{4}{3};
\bbline{3}{1};
\end{tikzpicture}
&
\begin{tikzpicture}[scale = .5, line width = 1pt, baseline=(current  bounding  box.center)]
 \btdots{4}{2};
\btline{1}{2};
\btline{3}{3};
\bbline{4}{2};
\end{tikzpicture}
&
\begin{tikzpicture}[scale = .5, line width = 1pt, baseline=(current  bounding  box.center)]
 \btdots{4}{2};
\btline{2}{2};
\btline{3}{3};
\draw (B1) to [out = 50, in = 130] (B4);
\end{tikzpicture}
\\
b_1 & b_2 & b_3 & b_4 & b_5 & b_6
\end{array}
\label{bobs}
\end{align}
The vector space $\mathbb{C}\{b_i\}$ is a representation of the Brauer algebra $\mathcal{B}_L(n)$ acting from below, and of the symmetric group $S_{|\lambda|}$ acting from above, and we have 
\begin{align}
 B^{(L)}_{\lambda} = \mathbb{C}\{b_i\}\otimes_{S_{|\lambda|}} \lambda \ , 
 \label{bll}
\end{align}
where the partition $\lambda$ now stands for a Specht module of $S_{|\lambda|}$. In our example, that Specht module is one-dimensional since $\lambda \in \{[2], [11]\}$, and $\{b_i\}$ may be considered as a basis of $B^{(4)}_{\lambda}$ itself.

As a warm-up, let us decompose $B^{(4)}_{[2]}$ and $B^{(4)}_{[11]}$ into irreducible representations of the Temperley--Lieb algebra, which are described before Eq. \eqref{dimTL-loop}. The diagrams $b_1,b_2,b_3$ do not involve any crossings, so they can be straightforwardly identified with elements of a representation of $\TL_4(n)$. The irreducible representation that is generated by diagrams with $2r$ upper sites is called $W^{(4)}_{r}$, so in our case the relevant representation is $W^{(4)}_1$. 
Since our three diagrams are permuted by the action of $\TL_L(n)$, they all belong to one copy of $W^{(4)}_{1}$, which is indeed three-dimensional according to Eq. \eqref{dimTL-loop}. Then, notice that the action of all three Temperley--Lieb generators on $b_4,b_5,b_6$ only give $b_1,b_2,b_3$: this implies that there is a three-dimensional subspace of  $\mathbb{C}\{b_i\}$ that is annihilated by $\TL_4(n)$. The one-dimensional representation $W^{(L)}_{\frac{L}{2}}$ is generated by the trivial diagram with $L$ lower and $L$ upper site; the action of any Temperley--Lieb generator yields a diagram where two upper sites are matched, which is identified with zero. 
Being annihilated by $\TL_L(n)$ is therefore the defining feature of $W^{(L)}_{\frac{L}{2}}$, and we deduce 
\begin{align}
 B^{(4)}_{[2]} \underset{\TL_4(n)}{=} B^{(4)}_{[11]}\underset{\TL_4(n)}{=} W^{(4)}_1 \oplus 3 W^{(4)}_2\quad \iff \quad c^{[2]}_1 = c^{[11]}_1 = 1 \ , \ c^{[2]}_2 = c^{[11]}_2 = 3\ ,
\end{align}
which agrees with the general formula \eqref{clr}.

\subsubsection{Diagrammatic and combinatorial subproblems}

Let us reduce our problem to a subproblem of enumerating certain diagrams, and a combinatorial subproblem of enumerating Young tableaux. 

To begin with, since the coefficient $c^\lambda_{(r,s)}$ in the branching rule \eqref{bcw} is $L$-independent, we may choose the lowest value $L=2r$ such that this coefficient appears. This reduces the problem to determining the contributions of representations of the type $W^{(L)}_{(\frac{L}{2},s)}$ in $B^{(L)}_\lambda$. From their diagrammatic construction, these representations are annihilated by the $L$ generators $e_1,e_2,\dots, e_L$ of $\uJTL_L(n)$. Since the image of the action of these generators is made of diagrams where at least two neighbouring bottom sites are connected, our representations may be built from diagrams where no two neighbouring bottom sites are connected. Let $\{b_i\}$ be the set of perfect matchings from $L$ bottom sites to $|\lambda|$ top sites such that no two neighbouring bottom sites are connected, and no two top sites are connected, modulo permutations of the top sites: in the case $L=6, |\lambda|=4$ we have $9$ such diagrams:
\begin{subequations}
\label{d3-d6}
\begin{align}
 &\begin{tikzpicture}[scale = .4, line width = 1pt, baseline=(current  bounding  box.center)]
 \btdots{6}{4};
\draw (B1) to [out = 50, in = 130] (B4);
\btline{2}{2};
\btline{3}{3};
\btline{5}{4};
\btline{6}{5};
\end{tikzpicture}
\quad
\begin{tikzpicture}[scale = .4, line width = 1pt, baseline=(current  bounding  box.center)]
 \btdots{6}{4};
\draw (B2) to [out = 50, in = 130] (B5);
\btline{1}{2};
\btline{3}{3};
\btline{4}{4};
\btline{6}{5};
\end{tikzpicture}
\quad
\begin{tikzpicture}[scale = .4, line width = 1pt, baseline=(current  bounding  box.center)]
 \btdots{6}{4};
\draw (B3) to [out = 50, in = 130] (B6);
\btline{1}{2};
\btline{2}{3};
\btline{4}{4};
\btline{5}{5};
\end{tikzpicture}
\label{d3}
\\
&\begin{tikzpicture}[scale = .4, line width = 1pt, baseline=(current  bounding  box.center)]
 \btdots{6}{4};
\bbline{1}{3};
\btline{2}{2};
\btline{4}{3};
\btline{5}{4};
\btline{6}{5};
\end{tikzpicture}
\quad
\begin{tikzpicture}[scale = .4, line width = 1pt, baseline=(current  bounding  box.center)]
 \btdots{6}{4};
\bbline{2}{4};
\btline{1}{2};
\btline{3}{3};
\btline{5}{4};
\btline{6}{5};
\end{tikzpicture}
\quad
\begin{tikzpicture}[scale = .4, line width = 1pt, baseline=(current  bounding  box.center)]
 \btdots{6}{4};
\bbline{3}{5};
\btline{1}{2};
\btline{2}{3};
\btline{4}{4};
\btline{6}{5};
\end{tikzpicture}
\quad
\begin{tikzpicture}[scale = .4, line width = 1pt, baseline=(current  bounding  box.center)]
 \btdots{6}{4};
\bbline{4}{6};
\btline{1}{2};
\btline{2}{3};
\btline{3}{4};
\btline{5}{5};
\end{tikzpicture}
\quad
\begin{tikzpicture}[scale = .4, line width = 1pt, baseline=(current  bounding  box.center)]
 \btdots{6}{4};
\draw (B1) to [out = 30, in = 150] (B5);
\btline{2}{2};
\btline{3}{3};
\btline{4}{4};
\btline{6}{5};
\end{tikzpicture}
\quad
\begin{tikzpicture}[scale = .4, line width = 1pt, baseline=(current  bounding  box.center)]
 \btdots{6}{4};
\draw (B2) to [out = 30, in = 150] (B6);
\btline{1}{2};
\btline{3}{3};
\btline{4}{4};
\btline{5}{5};
\end{tikzpicture}
\label{d6}
\end{align}
\end{subequations}
These diagrams are split into two orbits of lengths $6,3$ of the action of the generator $u$ of $\uJTL_6(n)$, which cyclically permutes the bottom sites. In general,  let us call $\Omega^{(L)}_{|\lambda|}$ the tuple of these orbit lengths, in our example $\Omega^{(6)}_4 = (6,3)$. 
Using Eq. \eqref{slrs}, we may write the number of diagrams as 
\begin{align}
 \sum_{\omega\in \Omega^{(L)}_{\ell}} \omega \underset{|\lambda|=\ell}{=} \frac{1}{f_\lambda}\sum_{\substack{s\in \frac{2}{L}\mathbb{Z}\\ -1<s\leq 1}} c^\lambda_{(\frac{L}{2},s)} = c^\ell_{\frac{L}{2}} -\delta_{L\geq 3} c^\ell_{\frac{L}{2}-1} \ ,
\end{align}
where the coefficients $c^\ell_r$ are explicitly given in Eq. \eqref{clr}.
Our diagrammatic subproblem consists in determining $\Omega^{(L)}_{\ell}$. 

Let us focus on one orbit of length $\omega$: then $u^\omega$ induces a cyclic permutation $\mu$ of the $|\lambda|$ upper sites, such that $\mu^{\frac{L}{\omega}}=1$. This implies $\omega|\lambda|\in L\mathbb{N}$. Our combinatorial subproblem is to find the eigenvalues of $\mu$ in the Specht module $\lambda$ of the permutation group $S_{|\lambda|}$. Strictly speaking, this is an algebraic problem, which amounts to decomposing $\lambda$ into representations of a cyclic subgroup of $S_{|\lambda|}$. However, this problem has a known combinatorial solution \cite{ste89}: 
\begin{align}
 \lambda \underset{\mathbb{Z}_{m}}{=} \bigoplus_{T\in T_\lambda} Z_{\frac{2}{m}\operatorname{ind}(T)}^{(m)} \ ,
 \label{sdz}
\end{align}
where $Z^{(m)}_s$ is the (one-dimensional) irreducible representation of $\mathbb{Z}_m$ where the generator of $\mathbb{Z}_m$ has the eigenvalue $e^{\pi i s}$, $T_\lambda$ is the set of standard Young tableaux of shape $\lambda$ (so that $|T_\lambda|=f_\lambda$), and the major index $\operatorname{ind}(T)\in \mathbb{Z}_{|\lambda|}$ of a tableau is the sum of the descents, i.e. the sum of the numbers $k$ in $T$ such that $k+1$ appears in a row strictly below $k$.

\subsubsection{Branching rules}

According to Eq. \eqref{us}, the subrepresentations $W^{(L)}_{(\frac{L}{2},s)}$ of $B^{(L)}_\lambda$ are characterized by the eigenvalues of $u$. 
We claim that these eigenvalues are the $\omega$-th roots of the eigenvalues of the permutations that are induced by elements of the type $u^\omega$. (See \cite{js18}(Appendix A.4) for an explanation of this point in the language of transfer matrices.) 
The resulting branching coefficients are
\begin{align}
 \boxed{c^\lambda_{(r,s)} = \sum_{T\in T_\lambda}
 \sum_{\omega\in \Omega^{(2r)}_{|\lambda|}} 
 \delta_{e^{\pi i\omega\left(s - \frac{\operatorname{ind}(T)}{r}\right)}, 1}}\ ,
 \label{clrs}
\end{align}
and therefore the branching rules
\begin{align}
 \boxed{B^{(L)}_\lambda \underset{\uJTL_L(n)}{=} \bigoplus_{r=\frac{|\lambda|}{2}}^{\frac{L}{2}}\bigoplus_{\omega\in \Omega^{(2r)}_{|\lambda|}} 
 \bigoplus_{k\in \mathbb{Z}_\omega} \bigoplus_{T\in T_\lambda} W^{(L)}_{(r,\frac{2k}{\omega} + \frac{\operatorname{ind}(T)}{r})}} \ .
 \label{bcwe}
\end{align}
Let us examine a few special cases. First we focus on maximal representations, i.e. representations $B^{(L)}_\lambda$ with $|\lambda|=L$. In this case, there is only one relevant diagram, which is trivial:
\begin{align}
 \begin{tikzpicture}[scale = .5, line width = 1pt, baseline=(current  bounding  box.center)]
  \btdots{7}{7};
  \foreach \i in {1,...,7}
  {\btline{\i}{\i};}
 \end{tikzpicture}
 \label{diag:1}
\end{align}
This implies $\Omega^{(L)}_L = (1)$, so the sum over $k$ becomes trivial, and we end up with the purely combinatorial formula 
\begin{align}
 B^{(L)}_\lambda \underset{\uJTL_L(n)}{\overset{|\lambda|=L}{=}} \bigoplus_{T\in T_\lambda} W^{(L)}_{(\frac{L}{2},\frac{2}{L}\operatorname{ind}(T))}\ .
 \label{blll}
\end{align}
This is equivalent to the branching rule $S_L\downarrow \mathbb{Z}_L$ \eqref{sdz}.

Another interesting special case is when $\lambda= [|\lambda|]$ is the trivial representation of $S_{|\lambda|}$. In this case, $T_\lambda$ contains a unique tableau whose index is $0$, and we end up with the purely diagrammatic formula  
\begin{align}
 B^{(L)}_\lambda \underset{\uJTL_L(n)}{\overset{\lambda=[|\lambda|]}{=}} \bigoplus_{r=\frac{|\lambda|}{2}}^{\frac{L}{2}}\bigoplus_{\omega\in \Omega^{(2r)}_{|\lambda|}} 
 \bigoplus_{k\in \mathbb{Z}_\omega} W^{(L)}_{(r,\frac{2k}{\omega})} \ .
\end{align}
A purely diagrammatic formula is also obtained if $\mathrm{gcd}(2r, |\lambda|)=1$, which can only happen if $2r$ is odd. In this case, $\Omega^{(2r)}_{|\lambda|}$ is a number of copies of the orbit length $2r$ -- smaller orbits cannot occur. Then $s$ no longer depends on $\operatorname{ind}(T)$, and we have 
\begin{align}
 B^{(L)}_\lambda \underset{\uJTL_L(n)}{\overset{\mathrm{gcd}(2r, |\lambda|)=1}{=}} f_\lambda \bigoplus_{r=\frac{|\lambda|}{2}}^{\frac{L}{2}} \left|\Omega^{(2r)}_{|\lambda|}\right| \bigoplus_{k\in \mathbb{Z}_{2r}}  W^{(L)}_{(r,\frac{k}{r} )} \ .
\end{align}
In other words, $c^\lambda_{(r,s)} = f_\lambda  \left|\Omega^{(2r)}_{|\lambda|}\right|$ is $s$-independent in this case. 

\subsubsection{Examples}

Let us solve the diagrammatic subproblem with $6$ bottom sites and $0$ top site. We find $c^0_3-c^0_2=4$ diagrams: 
\begin{align}
 \begin{tikzpicture}[scale = .5, line width = 1pt, baseline=(current  bounding  box.center)]
 \bbdots{6};
\bbline{1}{4};
\bbline{2}{5};
\bbline{3}{6};
\end{tikzpicture}
\qquad 
\begin{tikzpicture}[scale = .5, line width = 1pt, baseline=(current  bounding  box.center)]
 \bbdots{6};
\bbline{1}{4};
\bbline{2}{6};
\bbline{3}{5};
\end{tikzpicture}
\qquad 
\begin{tikzpicture}[scale = .5, line width = 1pt, baseline=(current  bounding  box.center)]
 \bbdots{6};
\bbline{1}{3};
\bbline{2}{5};
\bbline{4}{6};
\end{tikzpicture}
\qquad 
\begin{tikzpicture}[scale = .5, line width = 1pt, baseline=(current  bounding  box.center)]
 \bbdots{6};
\bbline{1}{5};
\bbline{2}{4};
\bbline{3}{6};
\end{tikzpicture}
\end{align}
The action of $u$ divides these diagrams into two orbits of lengths $3$ and $1$, so that $\Omega^{(6)}_0 = (3,1)$. The corresponding $u$-eigenvalues are $1,1,e^{\pm \frac{2\pi i}{3}}$, leading to  
\begin{align}
 c^{[]}_{(3,0)} = 2 \quad , \quad c^{[]}_{(3,\pm\frac13)} = c^{[]}_{(3,1)}=0 \quad , \quad c^{[]}_{(3,\pm\frac23)} = 1\ .
\end{align}
In the case of $8$ bottom sites and $0$ top site, we find $c^0_4-c^0_3=31$ diagrams that belong to $7$ different $u$-orbits. Let us give one representative of each orbit, while indicating the orbit lengths:
\begin{subequations}
\label{o80-all}
\begin{align}
 8: & \qquad 
 \begin{tikzpicture}[scale = .5, line width = 1pt]
 \bbdots{8};
\bbline{1}{3};
\bbline{2}{5};
\bbline{4}{7};
\bbline{6}{8};
 \end{tikzpicture}
 \qquad 
 \begin{tikzpicture}[scale = .5, line width = 1pt]
 \bbdots{8};
\bbline{1}{3};
\bbline{2}{6};
\bbline{4}{7};
\bbline{5}{8};
 \end{tikzpicture}
\\
 4: &  \qquad 
 \begin{tikzpicture}[scale = .5, line width = 1pt]
 \bbdots{8};
\bbline{1}{3};
\bbline{2}{4};
\bbline{5}{7};
\bbline{6}{8};
 \end{tikzpicture}
 \qquad 
 \begin{tikzpicture}[scale = .5, line width = 1pt]
 \bbdots{8};
\bbline{1}{3};
\bbline{2}{6};
\bbline{5}{7};
\bbline{4}{8};
 \end{tikzpicture}
 \qquad 
 \begin{tikzpicture}[scale = .5, line width = 1pt]
 \bbdots{8};
\bbline{1}{4};
\bbline{2}{6};
\bbline{3}{7};
\bbline{5}{8};
 \end{tikzpicture}
 \\
 2: & \qquad 
 \begin{tikzpicture}[scale = .5, line width = 1pt]
 \bbdots{8};
\bbline{1}{4};
\bbline{2}{7};
\bbline{3}{6};
\bbline{5}{8};
 \end{tikzpicture}
 \\
 1: & \qquad 
 \begin{tikzpicture}[scale = .5, line width = 1pt]
 \bbdots{8};
\bbline{1}{5};
\bbline{2}{6};
\bbline{3}{7};
\bbline{4}{8};
 \end{tikzpicture}
\end{align}
\end{subequations}
Therefore, we have $\Omega_0^{(8)}= (8,8,4,4,4,2,1)$, which leads to 
\begin{align}
 c^{[]}_{(4,0)} = 7\quad ,\quad c^{[]}_{(4,\pm\frac14)} = c^{[]}_{(4,\pm\frac34)} =2 \quad , \quad c^{[]}_{(4,\pm\frac12)} = 5 \quad , \quad c^{[]}_{(4,1)} = 6\ . 
\end{align}
Let us now consider the case with $6$ bottom sites and $4$ top sites. We already used the determination of $\Omega^{(6)}_{4}=(6,3)$ for illustrating the definition of $\Omega^{(L)}_\ell$, see the diagrams \eqref{d3-d6}. Let us focus on the combinatorial subproblem. The five Young diagrams of size $4$, together with the sets $\{\operatorname{ind}(T)|T\in T_\lambda\}$ of major indices of the corresponding Young tableaux, are:
\begin{align}
 [4]:\{0\} \quad , \quad [31]:\{1,2,3\} \quad ,\quad [22]:\{0,2\} \quad , \quad [211]: \{0,1,3\} \quad , \quad [1111]:\{2\}\ .
\end{align}
The coefficient $c^\lambda_{(r,s)}$ depends on $\operatorname{ind}(T)$ only if $\omega < 2r$, i.e. in our example if $\omega =3$. In this case, it depends only on $\operatorname{ind}(T)\bmod 2$. This implies $c^{[22]}_{(3,s)} = 2c^{[4]}_{(3,s)} = 2c^{[1111]}_{(3,s)}$, since $f_{[22]}=|T_{[22]}|=2$ whereas $f_{[4]}=f_{[1111]}=1$. In particular,
\begin{align}
 c^{[4]}_{(3,0)} = c^{[4]}_{(3,\pm \frac23)} = 2 \quad , \quad c^{[4]}_{(3,1)}=c^{[4]}_{(3,\pm \frac13)} = 1\ .
\end{align}
The cases where the dependence on $\operatorname{ind}(T)$ plays a role are $c^{[31]}_{(3,s)} = c^{[211]}_{(3,s)}$. In these cases, the odd major indices $1,3$ shift the values of $s$ by one unit. Therefore, to compute $c^{[31]}_{(3,s)}$, we cannot simply multiply $c^{[4]}_{(3,s)}$ by $f_{[31]}=3$. Instead, we obtain 
\begin{align}
 c^{[31]}_{(3,0)} = c^{[31]}_{(3,\pm \frac23)} = 4 \quad , \quad c^{[31]}_{(3,1)}=c^{[31]}_{(3,\pm \frac13)} = 5\ .
\end{align}
A longer list of $\mathcal{B}_L(n) \downarrow \uJTL_L(n)$ branching rules, expressing $B^{(L)}_\lambda$ in terms of $W^{(L)}_{(r,s)}$ for any $|\lambda| \le 4$ and $r \le 5$, can be found in Appendix~\ref{app:branching}.

\section{Algebraic approach to the \texorpdfstring{$Q$}{Q}-state Potts model}

The space of states of the $Q$-state Potts model on a lattice with $L$ sites can be constructed as 
\begin{align}
 \mathcal{S}^{\text{Potts}}_L = \Big([1]\oplus []\Big)^{\otimes L}\ .
 \label{spl}
\end{align}
On each site, we have a variable that belongs to the natural $Q$-dimensional representation of the symmetric group $S_Q$. This representation is a sum of two irreducible representations, called the trivial and standard representations. In the notations of Section \ref{sec:rcsg}, these representations correspond to the Young diagrams $[]$ and $[1]$ respectively. 

\subsection{Action of diagram algebras on the space of states}\label{sada}

We will now introduce the diagram algebras that act on the space of states of the $Q$-state Potts model. We will focus on the algebraic properties of these algebras and their representations: for more details, including the eponymous diagrams, see Appendix \ref{app:diag}.

\subsubsection{The relevant diagram algebras}

The algebra of the linear maps on $\mathcal{S}^{\text{Potts}}_L$ that commute with $S_Q$ is the partition algebra,
\begin{align}
 S_Q^* = \mathcal{P}_L(Q)\ . 
\end{align}
The partition algebra is generated by the permutations $p_i$ with $i=1,\dots, L-1$, together with generators $s_i$ with $i=1,\dots, L$ and $s_{i+\frac12}$  with $i=1,\dots, L-1$. These generators act on one or two factors of $([1]+[])^{\otimes L}$, and for $Q$ integer they can be defined as \cite{hr05}
\begin{align}
 s_i: & \quad \begin{array}{lll}
             ([1]+[])_i & \to & ([1]+[])_i 
             \\
             w_k & \mapsto & \sum_{\ell=1}^Q w_\ell 
            \end{array}
            \ , 
            \\
s_{i+\frac12}: &\quad  \begin{array}{lll}
                      ([1]+[])_i\otimes ([1]+[])_{i+1} & \to & ([1]+[])_i \otimes ([1]+[])_{i+1}
                      \\
                      w_k\otimes w_\ell & \mapsto & \delta_{k\ell} w_k\otimes w_k 
                     \end{array}
\ ,
\end{align}
where $(w_k)_k$ is a basis of $[1]+[]$ such that $S_Q$ acts by permuting the indices $k$. 
As a diagram algebra, the partition algebra exists for any $Q\in \mathbb{C}$, and the generators $p_i,s_i,s_{i+\frac12}$ are described by the diagrams \eqref{s-gen}. 

The partition algebra includes the permutation group $S_L$, and ignores the geometry of the problem. If the $L$ spins are arrayed in a chain, we can no longer permute them. The symmetries of the system must then be described by a smaller algebra, which also takes into account the boundary conditions. For an open chain with free boundary conditions, the relevant algebra is obtained by removing the permutation generators $p_i$. This leads to the Temperley--Lieb algebra $\TL_{2L}(\sqrt{Q})$, via the map $s_i\mapsto e_{2i-1}$ for $i=1,\frac32, \dots, L-\frac12, L$. 

For a closed chain with periodic boundary conditions, we should furthermore add two generators: the generator $s_{L+\frac12} \mapsto e_{2L}$, and the generator $u^2$ which translates all the sites by two units in $\TL_{2L}(\sqrt{Q})$. (This corresponds to translations by one unit in the original chain with $L$ sites.) 
In order to describe clusters rather than loops, we need to make a couple extra modifications, see Appendix \ref{app:pa}. 
Let us call the resulting algebra the Potts--Temperley--Lieb algebra $\PTL_{2L}(\sqrt{Q})$. 

\subsubsection{Representations of the partition algebra}

By Schur--Weyl duality, our space of states decomposes into irreducible representations of $\mathcal{P}_L(Q)\times S_Q$ as 
\begin{align}
 \mathcal{S}^{\text{Potts}}_L \underset{\mathcal{P}_L(Q)\times S_Q}{=} \bigoplus_{|\lambda|\leq L} P_\lambda^{(L)} \otimes \lambda \ , 
 \label{ssql}
\end{align}
where the sum is over partitions $\lambda$, which we identify with irreducible representations of $S_Q$ at generic $Q$. The irreducible finite-dimensional representations $P_\lambda^{(L)}$ of the partition algebra are also labelled by partitions: we will say more on their structures when it comes to computing branching rules in Section \ref{sec:sbr}. For the moment, let us point out that their dimensions appear as coefficients in the decomposition of $\mathcal{S}^{\text{Potts}}_L = ([1]\oplus [])^{\otimes L}$ into irreducible representations of $S_Q$, 
\begin{align}
 \Big([1]\oplus []\Big)^{\otimes L} \underset{S_Q}{=} \bigoplus_{|\lambda|\leq L} \left(\dim P_\lambda^{(L)}\right) \lambda \ .
\end{align}
For example, $([1]\oplus [])^{\otimes 2} = [2] + [11] + 3[1] + 2[]$ tells us that $\dim P^{(2)}_{[1]}=3$. 

\subsubsection{Representations of the Temperley--Lieb algebra}

From Section \ref{oada}, we remember that the standard modules $W^{(2L)}_r$ of the Temperley--Lieb algebra $\TL_{2L}(\sqrt{Q})$ are labelled by integers $r=0,1,\dots, L$. By analogy with the case of the Brauer algebra \eqref{bbtl}, we expect branching rules of the type 
\begin{align}
 P_\lambda^{(L)} \underset{\TL_{2L}(\sqrt{Q})}{=} f_\lambda \bigoplus_{r=|\lambda|}^L d_r^{|\lambda|} W_r^{(2L)}\ ,
\end{align}
with some coefficients $d^{|\lambda|}_r \in \mathbb{N}$ that depend neither on $Q$ nor on $L$, and a prefactor $f_\lambda$ given by the hook length formula \eqref{hooklf}. Inserting this into the decomposition \eqref{ssql} of the space of states, we obtain 
\begin{align}
 \boxed{\mathcal{S}^{\text{Potts}}_L \underset{\TL_{2L}(\sqrt{Q})\times S_Q}{=} \bigoplus_{r=0}^L W_r^{(2L)} \otimes \Xi_r} \qquad \text{with} \qquad \boxed{\Xi_r =\bigoplus_{|\lambda|\leq r} f_\lambda d_r^{|\lambda|} \lambda} \ .
 \label{spts}
\end{align}
We can recursively deduce the representations $\Xi_r$, by writing $\mathcal{S}^{\text{Potts}}_L \underset{S_Q}{=} \bigoplus_{r=0}^L \left(\dim W_r^{(2L)}\right) \Xi_r$ and using the dimensions \eqref{dimTL-loop} of Temperley--Lieb representations. The first few examples are 
\begin{align}
 \Xi_0 &= []\ , 
 \\
 \Xi_1 &= [1]\ , 
 \\
 \Xi_2 &= [2] + [11] \ ,
 \\
 \Xi_3 &= [3]+2[21]+[111] + [2]+[11]+[1]\ . 
\end{align}
We can also deduce the recursion relation 
\begin{align}
 [1]\otimes \Xi_r = \Xi_{r-1}\oplus \Xi_r \oplus \Xi_{r+1}\ . 
\end{align}
Using the associativity of the tensor product, this implies 
\begin{align}
 \Xi_{r_1}\otimes \Xi_{r_2} = \bigoplus_{r=|r_1-r_2|}^{r_1+r_2} \Xi_r \ .
 \label{xrxr}
\end{align}
We may interpret $\Xi_r$ as an irreducible representation of the commutant $\TL_{2L}^*(\sqrt{Q})$. 

\subsubsection{Representations of the Potts--Temperley--Lieb algebra}

For $r\geq 2$, the irreducible representations of the Potts--Temperley--Lieb algebra $\PTL_{2L}(\sqrt{Q})$ simply coincide with irreducible representations $W_{(r,s)}^{(2L)}$ of the unoriented algebra $\uJTL_{2L}(\sqrt{Q})$, which we described in Eq. \eqref{wrswr}. However, in the unoriented case, the pseudo-momentum $s$ was defined modulo $2$ because the generator $u$ of $\uJTL_{2L}(\sqrt{Q})$ had eigenvalues $e^{\pi is}$. In the case of $\PTL_{2L}(\sqrt{Q})$, only $u^2$ exists, and $s$ is now defined modulo integers. In other words, the following inequivalent representations of $\uJTL_{2L}(\sqrt{Q})$ become equivalent as representations of $\PTL_{2L}(\sqrt{Q})$:
\begin{align}
 W_{(r,s)}^{(2L)} \underset{\PTL_{2L}(\sqrt{Q})}{\simeq} W_{(r,s+1)}^{(2L)} \ .
\end{align}
Therefore, we assume $s\in \frac{1}{r}\mathbb{Z} \cap (-\frac12, \frac12]$ for representations of $\PTL_{2L}(\sqrt{Q})$. As reviewed in Appendix \ref{app:pa}, $\PTL_{2L}(\sqrt{Q})$ has no representation with $r=1$, and two representations with $r=0$, which we call $W^{(2L)}_{(0,0)}$ and $W^{(2L)}_{(0,\frac12)}$. 
We expect that representations of the partition algebra decompose as 
\begin{align}
 \boxed{P^{(L)}_\lambda \underset{\PTL_{2L}(\sqrt{Q})}{=} \delta_{\lambda,[]} W^{(2L)}_{(0,0)} \oplus \delta_{\lambda, [1]} W^{(2L)}_{(0,\frac12)} \oplus \bigoplus_{r=\max(2, |\lambda|)}^L \bigoplus_{\substack{s\in\frac{1}{r}\mathbb{Z} \\ -\frac12 < s \leq \frac12}} d^\lambda_{(r,s)} W^{(2L)}_{(r,s)}}\ ,
 \label{pdw}
\end{align}
where the coefficients $d^\lambda_{(r,s)}\in\mathbb{N}$ depend neither on $Q$ nor on $L$. We will explain how to compute them in Section \ref{sec:sbr}. For the moment, let us proceed with decomposing the space of states. Inserting our decomposition of $P^{(L)}_\lambda$ into Eq. \eqref{ssql}, we find 
\begin{multline}
 \boxed{\mathcal{S}_L^{\text{Potts}} \underset{\PTL_{2L}(\sqrt{Q})\times S_Q}{=} 
 W^{(2L)}_{(0,0)}\otimes [] \oplus W^{(2L)}_{(0,\frac12)} \otimes [1]\oplus
 \bigoplus_{r=2}^L \bigoplus_{\substack{s\in\frac{1}{r}\mathbb{Z} \\ -\frac12 < s \leq \frac12}} 
 W^{(2L)}_{(r,s)}\otimes \Xi_{(r,s)}} \\
 \text{with} \qquad \boxed{\Xi_{(r,s)} = \bigoplus_{|\lambda|\leq r} d^\lambda_{(r,s)} \lambda} \ .
 \label{stos}
\end{multline}
The representation $\Xi_{(r,s)}$ of $S_Q$ that we just defined should coincide with the representation $\Xi_{(r,s)}$ that appeared in the spectrum \eqref{spotts} of the Potts CFT. Moreover, the large $L$ limit of the representation $W^{(2L)}_{(r,s)}$ should coincide with a representation of the interchiral algebra, 
\begin{align}
 \lim_{\text{critical}} W^{(2L)}_{(r,s)} = \doublewidetilde{\mathcal{W}}_{(r,s)}\ .
\end{align}
We can obtain some nontrivial information on $\Xi_{(r,s)}$ by comparing our decomposition \eqref{stos} of the space of states with the decomposition \eqref{spts} into representions of the Temperley--Lieb algebra. Using the relations \eqref{wrswr} and \eqref{w00w0h} between modules of $\PTL_{2L}(\sqrt{Q})$ and $\TL_{2L}(\sqrt{Q})$, we obtain 
\begin{align}
 \bigoplus_{\substack{s\in\frac{1}{r}\mathbb{Z} \\ -\frac12 < s \leq \frac12}} \Xi_{(r,s)} = \Xi_r - \delta_{r\geq 1} \Xi_{r-1} +(-1)^r [1]\ . 
\end{align}
The algebraic definition of the representations $\Xi_{(r,s)}$ of the group $S_Q$ that appear in the space of states also allows us to rederive their invariances under shifts \eqref{xspt} and reflection \eqref{lms}, by the same arguments as in the $O(n)$ model. 

Unlike the representations $\Xi_r$ with their tensor products \eqref{xrxr}, the representations $\Xi_{(r,s)}$ \eqref{x-all} do not close under tensor products. While we may interpret $\Xi_{(r,s)}$ as irreducible representations of the commutant $\PTL^*_{2L}(\sqrt{Q})$, we can therefore not interpret this commutant as a symmetry algebra of the $Q$-state Potts model. 

\subsection{Branching rules \texorpdfstring{$\mathcal{P}_L(Q)\downarrow \PTL_{2L}(\sqrt{Q})$}{}}\label{sec:sbr}

We have reduced the determination of the $S_Q$ representations $\Xi_{(r,s)}$ to the algebraic problem of decomposing representations of the partition algebra into representations of the Potts--Temperley--Lieb algebra. Let us now solve this problem.

\subsubsection{Diagrammatic and combinatorial subproblems}

Let us describe the structure of irreducible representation $P_\lambda^{(L)}$ of the partition algebra. To build a basis, we start with the set of all partitions of $L$ lower sites and $|\lambda|$ upper sites, such that each upper site belongs to a subset that contains no other upper site, and at least one lower site. To draw such diagrams, we draw lines between elements of a given subset, without necessarily drawing all possible lines:
\begin{align}
 \begin{tikzpicture}[scale = .4, line width = 1pt, baseline=(current  bounding  box.center)]
  \btdots{2}{1};
  \btline{1}{1};
  \btline{2}{1};
  \bbline{1}{2};
 \end{tikzpicture}
 \quad = \quad 
 \begin{tikzpicture}[scale = .4, line width = 1pt, baseline=(current  bounding  box.center)]
  \btdots{2}{1};
  \btline{2}{1};
  \bbline{1}{2};
 \end{tikzpicture}
 \quad = \quad 
 \begin{tikzpicture}[scale = .4, line width = 1pt, baseline=(current  bounding  box.center)]
  \btdots{2}{1};
  \btline{1}{1};
  \bbline{1}{2};
 \end{tikzpicture}
 \quad = \quad 
 \begin{tikzpicture}[scale = .4, line width = 1pt, baseline=(current  bounding  box.center)]
  \btdots{2}{1};
  \btline{1}{1};
  \btline{2}{1};
 \end{tikzpicture}
\end{align}
In the case $L=4,|\lambda|=2$, modulo permutations of the lower sites, the possible diagrams are 
\begin{align}
 \begin{tikzpicture}[scale = .4, line width = 1pt, baseline=(current  bounding  box.center)]
  \btdots{4}{2};
  \btline{1}{2};
  \btline{2}{3};
 \end{tikzpicture}
 \qquad 
 \begin{tikzpicture}[scale = .4, line width = 1pt, baseline=(current  bounding  box.center)]
  \btdots{4}{2};
  \btline{1}{2};
  \btline{2}{3};
  \bbline{3}{4};
 \end{tikzpicture}
 \qquad 
 \begin{tikzpicture}[scale = .4, line width = 1pt, baseline=(current  bounding  box.center)]
  \btdots{4}{2};
  \btline{1}{2};
  \btline{2}{3};
  \btline{3}{3};
 \end{tikzpicture}
 \qquad 
 \begin{tikzpicture}[scale = .4, line width = 1pt, baseline=(current  bounding  box.center)]
  \btdots{4}{2};
  \btline{1}{2};
  \btline{2}{3};
  \btline{3}{3};
  \btline{4}{3};
 \end{tikzpicture}
 \qquad 
 \begin{tikzpicture}[scale = .4, line width = 1pt, baseline=(current  bounding  box.center)]
  \btdots{4}{2};
  \btline{1}{2};
  \btline{2}{2};
  \btline{3}{3};
  \btline{4}{3};
 \end{tikzpicture}
\end{align}
Of these five types of diagrams, only the second one appeared in the basis of the Brauer--Specht module $B_\lambda^{(4)}$ of the Brauer algebra. Modulo permutations of the upper sites, there are six diagrams of this type, see Eq. \eqref{bobs}.
The irreducible representation $P_\lambda^{(L)}$ is obtained by tensoring the space of diagrams $\mathbb{C}\{b_i\}$ with the Specht module $\lambda$ of the symmetric group $S_{|\lambda|}$, 
\begin{align}
 P_\lambda^{(L)} = \mathbb{C}\{b_i\}\otimes _{S_{|\lambda|}}\lambda \ . 
\end{align}

Now, to compute the coefficient $d_{(r,s)}^\lambda$ of the branching rule \eqref{pdw}, we choose the lowest value $L=r$ such that this coefficient appears. This reduces the problem to determining the contributions of representations of the type $W^{(2L)}_{(L,s)}$ in $P^{(L)}_\lambda$. From their diagrammatic constructions, such representations are annihilated by the $2L$ generators $e_1,e_2,\dots, e_{2L}$ of $\PTL_{2L}(\sqrt{Q})$. By the embedding $e_i\mapsto s_{\frac{i+1}{2}}$ of  $\PTL_{2L}(\sqrt{Q})$ into $\mathcal{P}_L(Q)$, this means that they are annihilated by $s_1,s_{\frac32},s_2,\dots, s_{L+\frac12}$. The image of $s_1,\dots , s_L$ is made of diagrams with isolated bottom sites, while the image of $s_\frac32,\dots, s_{L+\frac12}$ is made of diagrams where two neighbouring bottom sites are connected. To build a basis of $W^{(2L)}_{(L,s)}$, we may work modulo such diagrams, and restrict to diagrams such that bottom sites are neither isolated nor connected to their neighbours. 

The diagrammatic subproblem consists in determining the permutations of the upper sites that are induced by permutations of the lower sites in this set of diagrams. To a diagram $b$ we associate a cycle length $\omega$ and permutation $\mu\in S_{|\lambda|}$ such that $u^\omega b \mu = b$, where $u\in \mathcal{P}_L(Q)$ and $\mu\in S_{|\lambda|}$ act from the bottom and top of the diagram respectively. The permutation $\mu$ must therefore obey $\mu^{\frac{L}{\omega}}=1$. More precisely, we are interested in the conjugacy class of $\mu$ in $S_{|\lambda|}$, i.e. in its cycle type -- the list of lengths of its cycles, which we still denote by $\mu$. Let $\Upsilon_{|\lambda|}^{(L)}$ be the set of pairs $(\omega, \mu)$ that describe the action of $u$ on our set of diagrams, with each orbit of length $\omega$ giving rise to only one pair. In particular, there is a natural embedding $
 \Omega_{|\lambda|}^{(L)}  \hookrightarrow \Upsilon_{|\lambda|}^{(L)}$.

Once we know $\Upsilon_{|\lambda|}^{(L)}$, the combinatorial subproblem is to determine the eigenvalues of $\mu$ in the Specht module $\lambda$. This problem has a known combinatorial solution \cite{ste89}. Each Young tableau $T\in T_\lambda$ gives rise to an eigenvalue $e^{2\pi i \frac{\operatorname{ind}_\mu(T)}{|\mu|}}$, where $|\mu|$ is the order of the permutation $\mu$, and the cyclic exponent $ \operatorname{ind}_\mu(T)\in \mathbb{Z}_{|\mu|}$ is the $\mu$-index of $T$. To compute $\mu$-indices, let us write $\mu=(\mu_1,\dots ,\mu_k)$ the cycle lengths of $\mu$, with $|\mu|= \text{lcm}(\mu_i)$. Let us define a tuple of $|\lambda|$ integers $(b_\mu(1),\dots,b_\mu(|\lambda|)) = \left(\frac{|\mu|}{\mu_1},2\frac{|\mu|}{\mu_1}, \dots, |\mu|, \frac{|\mu|}{\mu_2},2\frac{|\mu|}{\mu_2},\dots, |\mu|, \dots\right)$. Then we have 
\begin{align}
 \operatorname{ind}_\mu(T) = \sum_{j\in \{\text{descents}(T)\}} b_\mu(j)\bmod |\mu|\ .
\end{align}

\subsubsection{Branching rules}

We now claim that the eigenvalues of $u$ in the subrepresentations  $W^{(2L)}_{(L,s)}$ of $P_\lambda^{(L)}$ are the $\omega$-th roots of the eigenvalues of $u^\omega$ in $\lambda$. The resulting branching coefficients are 
\begin{align}
 \boxed{d^\lambda_{(r,s)} = \sum_{T\in T_\lambda}\sum_{(\omega,\mu)\in \Upsilon_\lambda^{(r)}}
 \delta_{e^{2\pi i \left(\omega s-\frac{\operatorname{ind}_\mu(T)}{|\mu|}\right)}, 1}} \ .
 \label{dlrs}
\end{align}
Equivalently, the branching rules are 
\begin{align}
 \boxed{P_\lambda^{(L)} \underset{\PTL_{2L}(\sqrt{Q})}{=}  \bigoplus_{r=|\lambda|}^L \bigoplus_{(\omega,\mu)\in \Upsilon_{|\lambda|}^{(L)}} \bigoplus_{k\in\mathbb{Z}_\omega} \bigoplus_{T\in T_\lambda} W^{(2L)}_{(r,\frac{k}{\omega}+\frac{\operatorname{ind}_\mu(T)}{\omega|\mu|})}}\ .
 \label{plld}
\end{align}
Let us examine a few special cases. First we focus on maximal representations, i.e. representations $P_\lambda^{(L)}$ with $|\lambda|=L$. In this case there is only one relevant diagram, which is trivial, see Eq. \eqref{diag:1}. This implies $\Upsilon_L^{(L)}=((1, 1))$, so the sum over $k$ becomes trivial, and we end up with the purely combinatorial formula
\begin{align}
 P_\lambda^{(L)} \underset{\PTL_{2L}(\sqrt{Q})}{\overset{|\lambda|=L}{=}} \bigoplus_{T\in T_\lambda} W^{(2L)}_{(L,\frac{1}{L}\operatorname{ind}(T))}\ .
\end{align}
This is equivalent to the analogous formula \eqref{blll} for the case of the $O(n)$ model, which explains the coincidence of the coefficients 
\begin{align}
 d^\lambda_{(|\lambda|, s)} = c^\lambda_{(\frac{|\lambda|}{2}, 2s)} \ .
\end{align}
Another interesting special case is when $\lambda= [|\lambda|]$ is the trivial representation of $S_{|\lambda|}$. In this case, there is only one standard Young tableau, such that $\forall\mu,\operatorname{ind}_\mu(T)=0$ and we end up with the purely diagrammatic formula 
\begin{align}
 P_\lambda^{(L)} \underset{\PTL_{2L}(\sqrt{Q})}{\overset{\lambda= [|\lambda|]}{=}} \bigoplus_{r=|\lambda|}^L \bigoplus_{(\omega,\mu)\in \Upsilon_{|\lambda|}^{(L)}} \bigoplus_{k\in\mathbb{Z}_\omega}
 W^{(2L)}_{(r,\frac{k}{\omega})}\ .
\end{align}

\subsubsection{Examples}

We first focus on the trivial representation $\lambda = []$. For $L=1,2,3,5$, there are no allowed diagrams. For $L=4$, we have $\Upsilon^{(4)}_0 = \Omega^{(4)}_0$ with one diagram 
$\begin{tikzpicture}[scale = .4, line width = 1pt]
  \bbdots{4};
 \bbline{1}{3};
 \bbline{2}{4};
 \end{tikzpicture}$, which is invariant under translation by one unit. For $L=6$, we find $\Upsilon^{(6)}_0 = \Omega^{(6)}_0 \cup (1) = (3,1,1)$, where the additional translation-invariant diagram is 
 $\begin{tikzpicture}[scale = .4, line width = 1pt]
  \bbdots{6};
 \bbline{1}{3};
 \bbline{3}{5};
 \bbline{2}{4};
 \bbline{4}{6};
 \end{tikzpicture}$. This leads to 
 \begin{align}
  d^{[]}_{(6, 0)} = 3 \quad , \quad d^{[]}_{(6,\frac16)} = d^{[]}_{(6,\frac12)}=0 \quad ,\quad d^{[]}_{(6,\frac13)} = 1\ .
 \end{align}
For $L=7$ we find $\Upsilon^{(7)}_0=(7,7)$, leading to $\forall s, d^{[]}_{(7,s)}=2$. Let us draw one representative of each one of the two cycles of length $7$:
\begin{align}
7: & \qquad 
 \begin{tikzpicture}[scale = .4, line width = 1pt]
  \bbdots{7};
  \bbline{1}{4};
 \bbline{4}{6};
 \bbline{2}{7};
 \bbline{3}{5};
 \end{tikzpicture}
 \qquad 
 \begin{tikzpicture}[scale = .4, line width = 1pt]
  \bbdots{7};
  \bbline{1}{4};
 \bbline{4}{6};
 \bbline{2}{5};
 \bbline{3}{7};
 \end{tikzpicture}
\end{align}
For $L=8$ we find $\Upsilon^{(8)}_0 = \Omega^{(8)}_0 \cup (8,8,8,4,4,2,1)$. 
The diagrams that contribute to $\Omega^{(8)}_0$ are drawn in Eq. \eqref{o80-all}, let us 
draw the remaining diagrams:
\begin{subequations}
\label{u80-all}
\begin{align}
 8: & \qquad 
 \begin{tikzpicture}[scale = .5, line width = 1pt]
 \bbdots{8};
\bbline{1}{4};
\bbline{2}{5};
\bbline{5}{7};
\bbline{3}{6};
\bbline{6}{8};
 \end{tikzpicture}
 \qquad 
 \begin{tikzpicture}[scale = .5, line width = 1pt]
 \bbdots{8};
\bbline{1}{4};
\bbline{3}{5};
\bbline{5}{7};
\bbline{2}{6};
\bbline{6}{8};
 \end{tikzpicture}
 \qquad 
 \begin{tikzpicture}[scale = .5, line width = 1pt]
 \bbdots{8};
\bbline{1}{3};
\bbline{2}{5};
\bbline{5}{7};
\bbline{4}{6};
\bbline{6}{8};
 \end{tikzpicture}
\\
 4: &  \qquad 
 \begin{tikzpicture}[scale = .5, line width = 1pt]
 \bbdots{8};
\bbline{1}{5};
\bbline{2}{4};
\bbline{4}{7};
\bbline{3}{6};
\bbline{6}{8};
 \end{tikzpicture}
 \qquad 
 \begin{tikzpicture}[scale = .5, line width = 1pt]
 \bbdots{8};
\bbline{1}{3};
\bbline{3}{5};
\bbline{5}{7};
\bbline{2}{4};
\bbline{6}{8};
 \end{tikzpicture}
 \\
 2: & \qquad 
 \begin{tikzpicture}[scale = .5, line width = 1pt]
 \bbdots{8};
\bbline{1}{3};
\bbline{3}{5};
\bbline{5}{7};
\bbline{2}{6};
\bbline{4}{8};
 \end{tikzpicture}
 \\
 1: & \qquad 
 \begin{tikzpicture}[scale = .5, line width = 1pt]
 \bbdots{8};
\bbline{1}{3};
\bbline{3}{5};
\bbline{5}{7};
\bbline{2}{4};
\bbline{4}{6};
\bbline{6}{8};
 \end{tikzpicture}
\end{align}
\end{subequations}
This leads to 
\begin{align}
 d^{[]}_{(8,0)} = 14 \quad ,\quad d^{[]}_{(8,\frac18)} = 5 \quad , \quad d^{[]}_{(8,\frac14)} &= 10 \quad ,\quad d^{[]}_{(8,\frac12)} = 12\ .
\end{align}
Let us now consider the case of $\Upsilon^{(6)}_1$, which is still a purely diagrammatic problem. We find the following five orbits:
\begin{subequations}
 \label{o51-all}
 \begin{align}
  6: & \qquad 
 \begin{tikzpicture}[scale = .5, line width = 1pt]
 \btldots{6}{1};
\btline{1}{1};
\bbline{1}{3};
\bbline{2}{5};
\bbline{4}{6};
 \end{tikzpicture}
 \qquad 
 \begin{tikzpicture}[scale = .5, line width = 1pt]
 \btldots{6}{1};
\btline{1}{1};
\bbline{2}{4};
\bbline{3}{5};
\bbline{4}{6};
 \end{tikzpicture}
\\
 3: &  \qquad 
 \begin{tikzpicture}[scale = .5, line width = 1pt]
 \btldots{6}{1};
\btline{1}{1};
\bbline{1}{4};
\bbline{3}{5};
\bbline{2}{6};
 \end{tikzpicture}
 \qquad 
 \begin{tikzpicture}[scale = .5, line width = 1pt]
 \btldots{6}{1};
\btline{1}{1};
\bbline{1}{4};
\bbline{2}{5};
\bbline{3}{6};
 \end{tikzpicture}
\\
 2: &  \qquad 
 \begin{tikzpicture}[scale = .5, line width = 1pt]
 \btldots{6}{1};
\btline{1}{1};
\bbline{1}{3};
\bbline{3}{5};
\bbline{2}{4};
\bbline{4}{6};
 \end{tikzpicture}
 \end{align}
\end{subequations}
This leads to 
\begin{align}
 d^{[1]}_{(6,0)} = 5 \quad ,\quad d^{[1]}_{(6,\frac16)} = 2 \quad , \quad d^{[1]}_{(6,\frac13)} = 4 \quad ,\quad d^{[1]}_{(6,\frac12)} = 3\ .
\end{align}
Finally, let us deal with the example of $\Upsilon^{(4)}_3$. We have only one cycle of length $\omega = 2$. Let us draw one representative diagram, together with the action of $u$ and $u^2$:
\begin{align}
2: & \qquad 
 \begin{tikzpicture}[scale = .5, line width = 1pt]
 \btldots{4}{3};
\btline{1}{1};
\bbline{1}{3};
\btline{2}{2};
\btline{4}{3};
 \end{tikzpicture}
 \quad \overset{u}{\longrightarrow} \quad 
 \begin{tikzpicture}[scale = .5, line width = 1pt]
 \btldots{4}{3};
\btline{2}{1};
\draw (B2) to [out = 30, in = 150] (B4);
\btline{3}{2};
\draw (B1) to [out = 40, in = -90] (T3);
 \end{tikzpicture}
 \quad \overset{u}{\longrightarrow} \quad 
 \begin{tikzpicture}[scale = .5, line width = 1pt]
 \btldots{4}{3};
\btline{1}{1};
\draw (B1) to [out = 30, in = 150] (B3);
\btline{2}{3};
\draw (B4) to [out = 140, in = -90] (T2);
 \end{tikzpicture}
\end{align}
The action of $u^2$  induces a permutation of the three top sites that is neither trivial nor even cyclic. This permutation is actually of cycle type $\mu = (2, 1)$,  so that $\Upsilon^{(4)}_3=((2, (2, 1)))$. The $\mu$-indices of the Young tableaux are 
\begin{align}
\renewcommand{\arraystretch}{1.3}
 \begin{array}{|c|ccc|}
 \hline
  \lambda & [3] & [21] & [111]
  \\
  \hline
  \{\operatorname{ind}_{(2,1)}(T)\}_{T\in T_\lambda} & \{0\} & \{0, 1\} & \{1\}
  \\
  \hline
 \end{array}
\end{align}
This leads to 
\begin{align}
 & d^{[3]}_{(4,0)}=d^{[3]}_{(4,\frac12)}=1 \quad , \quad d^{[3]}_{(4,\frac14)}=0\ ,
 \\
 & d^{[21]}_{(4,0)}=d^{[21]}_{(4,\frac14)}=d^{[21]}_{(4,\frac12)} = 1\ ,
 \\
 & d^{[111]}_{(4,0)}=d^{[111]}_{(4,\frac12)}=0 \quad , \quad d^{[111]}_{(4,\frac14)}=1\ .
\end{align}

\section{Concluding remarks}

\subsubsection{Extended symmetries and commutant algebras}

In the spaces of states of the Potts and $O(n)$ models, any representation $\mathcal{W}_{(r,s)}$ of the conformal algebra comes with a representation $\Xi_{(r,s)}$ or $\Lambda_{(r,s)}$ of the global symmetry group $S_Q$ or $O(n)$. That global representation is in general not irreducible: rather, it is a sum of irreducible representations, whose number increases with $r$. 

Such global representations can be interpreted as irreducible representations of an algebra that includes $S_Q$ or $O(n)$. In the language of the introduction, this larger algebra is the commutant algebra $\mathfrak{A}^\star_L$.
In \cite{rs07}, this commutant algebra was constructed by adding generators to the relevant group. However, this larger algebra is unlikely to be a symmetry algebra. This is because the underlying periodic geometry prevents us from defining a natural tensor product, in contrast with the case of open boundary conditions. And in fact, our global representations do not close under the ordinary $S_Q$ or $O(n)$ tensor product. 
By Schur--Weyl duality, this is related to the problem of defining the fusion of representations of the affine Temperley--Lieb algebra \cite{gjs18,im21}. 

It is therefore important to distinguish an algebra that merely acts on the space of states, from a true symmetry algebra that also constrains the dynamics of the theory, and whose representations must therefore close under tensor products. Actually, in the Potts model, this distinction is also relevant for the interchiral algebra. In the space of states, this algebra acts as $s\to s+1$. In general correlation functions, the existence of degenerate fields only imply relations $s\to s+2$. Therefore, the interchiral algebra is only partly a dynamical symmetry algebra.

\subsubsection{The limits of global symmetry}

When undertaking our study of the action of $O(n)$ and $S_Q$ on the spaces of states, our main motivation was to solve the $O(n)$ and Potts CFTs. The idea is that a primary field would be labelled not only by its conformal dimension, but also by an irreducible representation of the global symmetry group $O(n)$ or $S_Q$. For example, our result for $\Lambda_{(\frac32,0)}$ \eqref{l320} implies that the $O(n)$ CFT has two primary fields of dimension $\Delta=\bar\Delta=\Delta_{(\frac32,0)}$, which we may denote as $V_{(\frac32,0)}^{[3]}$ and $V_{(\frac32,0)}^{[1^3]}$. 

Correlation functions of such fields are then constrained by global symmetry. In particular, for any given correlation function, we can compare the number of invariant tensors of the global symmetry group, with the number of linearly independent solutions of crossing symmetry \cite{gnjrs21}. For example, in the case of four-point connectivities of the Potts CFT, we have $4$ invariant tensors in $[1]^{\otimes 4}$, corresponding to the $4$ terms in $[1]\otimes [1]$ \eqref{ooto}. And we also find $4$ solutions of crossing symmetry \cite{nr20}. This agreement apparently extends to $N$-point connectivities: while we do not know the number of solutions of crossing symmetry for $N>4$, we do know the number of linearly independent connectivities \cite{dv11}, and we find that it coincides with the number of invariant tensors in $[1]^{\otimes N}$ \cite{rz22}. (This is not specific to the two-dimensional model.)

However, for general four-point functions in the $O(n)$ and Potts CFTs, we find much fewer solutions of crossing symmetry than invariant tensors \cite{gnjrs21, niv22}. This would be natural if the CFTs had extended global symmetries beyond the groups $O(n)$ and $S_Q$. Natural candidates for extended symmetries are the commutant algebras that act on the spaces of states, but we have argued that these are actually not dynamical symmetries. It might be that in higher dimensions, we have just as many solutions of crossing symmetry as invariant tensors, and that some solutions coincide (or become linearly dependent) in two dimensions -- just like conformal dimensions of some primary fields coincide. 

For global symmetry, there are more bad news from numerical bootstrap studies. An unexpected and non-trivial numerical result is that we can compute correlation functions that mix fields from the $O(n)$ and Potts CFTs, such as $\left<V_{(1,0)}V_{(1,0)}V_{(0,\frac12)}V_{(0,\frac12)}\right>$ \cite{niv22}. There is no hope of interpreting such correlation functions from the point of view of either $O(n)$ or $S_Q$: both symmetries are broken, both groups now appear too large. In order to understand such correlation functions, diagram algebras look more promising. It would be interesting to investigate whether such hybrid correlation functions also exist in higher dimensions.

\subsubsection{Universality and non-critical models}

We have described the spaces of states of the Potts and $O(n)$ models in terms of representations $\Xi_{(r,s)}$ or $\Lambda_{(r,s)}$ of their global symmetry groups. These representations do not depend on the lattice size. Therefore, their relevance is not limited to the CFT, or even to the critical lattice model. And indeed, as we show in Appendix \ref{app:brtm}, they appear in numerical studies of lattice models at finite $L$. 

In fact, these representations follow from the action of the algebra $\mathfrak{A}_L^*$, which is the commutant of $\mathfrak{A}_L$ in the space $\mathcal{S}_L$. 
This means that any version of the  $O(n)$ or $Q$-state Potts models whose transfer matrix can be expressed in terms of the 
Temperley--Lieb algebra $\uJTL_L(n)$ or $\PTL(\sqrt{Q})$ will have a similar space of states. 
Examples include the  $O(n)$ and Potts models with non-critical couplings (i.e. perturbed by the energy operator), or even with random, edge-dependent couplings. 

On the other hand, interactions that preserve the global symmetry $S_Q$ or $O(n)$, but are not expressible in terms of the appropriate diagram algebra, would break the representations $\Xi_{(r,s)}$ or $\Lambda_{(r,s)}$ into irreducible representations of $S_Q$ or $O(n)$. And of course, breaking the $S_Q$ or $O(n)$ symmetry itself (for example by applying a magnetic field) would completely destroy the structure of the space of states. 

It would therefore be interesting to study which interactions can be expressed in terms of the Temperley--Lieb algebra. On the resulting phase space, it should be possible to compute free energies and determine the phase diagram --- such an analysis has recently been performed in the case of the Heisenberg models \cite{brr22}, and our case is similar although a bit more complicated. 

\subsubsection{Dilute $O(n)$ model}

In this article, we have been focussing on the dense $O(n)$ model, whose lattice space of states \eqref{SLOn} is based on the defining representation $[1]$. Using $[1]\oplus []$ instead, we would obtain the dilute $O(n)$ model, whose lattice space of states is 
\begin{align}
 \mathcal{S}_L^{O(n)\text{, dilute}} = \Big([1]\oplus []\Big)^{\otimes L}\ .
 \label{sond}
\end{align}
In the critical limit, we expect that the dilute model is described by the same CFT, with however $1<\Re\beta^2 <2$ instead of $0<\Re\beta^2<1$ for the dense model \cite{gnjrs21}. This means in particular that the space of states is the same. To support this claim, let us now sketch how the representations $\Lambda_{(r,s)}$ of $O(n)$ could emerge in the algebraic approach to the dilute $O(n)$ model. 

In terms of diagrams, the presence of the trivial representation $[]$ in the space of states amounts to allowing partitions with blocks of size one or two, rather than just blocks of size two. As reviewed in Appendix \ref{app:npa}, this amounts to replacing the Brauer algebra with the Rook Brauer algebra in the decomposition \eqref{sonl} of $\mathcal{S}_L^{O(n)\text{, dilute}}$ \cite{dh12}. 
Then, in order to describe our two-dimensional model with its non-crossing lines, we should restrict to a planar algebra: the Motzkin algebra $\MM_L(\n+1)$, which is the dilute analog of the Temperley--Lieb algebra \cite{bh11}. 

Using the branching rules from the Temperley--Lieb algebra to the Motzkin algebra \cite{bs13}, it can be checked that the decomposition of $\mathcal{S}_L^{O(n)\text{, dilute}}$ into representations of $\MM_L(\n+1)\times O(n)$ involves the same $O(n)$ representations $\Lambda_r$ as the decomposition \eqref{sto} of $\mathcal{S}_L^{O(n)}$ into representations of the Temperley--Lieb algebra. Similarly, we expect that the decomposition of $\mathcal{S}_L^{O(n)\text{, dilute}}$ into the dilute analog of the unoriented Jones--Temperley--Lieb algebra would involve the same $O(n)$ representations $\Lambda_{(r,s)}$ \eqref{stou} as in the dense case. 

There is however a significant algebraic difference between the dense and dilute cases: while in the dense case we had a restriction $|\lambda|\equiv L\bmod 2$ on the representations of the Brauer algebra, we do not have such a restriction on the representations of the Rook Brauer algebra, because that algebra can involve blocks of size one. In the dilute case, the restriction $|\lambda|\equiv 2r\bmod 2$ in the structures of $\Lambda_r$ and $\Lambda_{(r,s)}$ should follow from the branching rules from the Rook Brauer to the Motzkin algebra.

\section*{Acknowledgments} 

We are grateful to Azat Gainutdinov, Victor Gorbenko, Yifei He, Allysa Lumley, Slava Rychkov, Bernardo Zan, and Jean-Bernard Zuber, for helpful discussions and correspondence. We are particularly indebted to Mike Zabrocki for his explanations on the representations of $S_Q$ at generic $Q$, for implementing their tensor products in SageMath, for collaborating on the code \cite{rz22}, and for pointing out that the spaces of states of the $O(n)$ and Potts models are related by a plethystic substitution. We thank Linnea Grans-Samuelsson and Rongvoram Nivesvivat for stimulating collaboration on closely related subjects. We are grateful to Ivan Kostov, Jules Lamers, Paul Roux, Mike Zabrocki and Bernardo Zan for helpful comments on the draft article. We thank the anonymous SciPost referees for helpful comments and suggestions.  

This work is partly a result of
the ERC-SyG project, Recursive and Exact New Quantum Theory (ReNewQuantum) which received funding from the European Research Council (ERC) under the European Union's Horizon 2020 research and innovation programme under grant agreement No 810573. This work was also supported by the ERC NuQFT Project, which received similar funding under grant agreement No 669205. This work was also supported by the French Agence Nationale de la Recherche (ANR) under grant ANR-21-CE40-0003 (project CONFICA).

\appendix

\section{Diagram algebras}\label{app:diag}

A diagram algebra  has a basis made of diagrams, with a multiplication coming from diagram concatenation.
We consider diagrams that have $L$ labelled sites at the bottom of a frame,
another $L$ labelled sites at the top of the frame, and connections inside
the frame. A diagram is planar if the connections do not cross. The sites are aligned in the
space direction, which is perpendicular to the imaginary time direction. 

The frame can be either a rectangle, or an annulus where the left and right
sides are identified.
The annulus geometry describes models with periodic boundary
conditions in the space direction. The distinction between the annulus and the rectangle only matters for planar diagrams. 

We will first review non-periodic algebras, whose frame is the rectangle. We will then deal with periodic algebras, whose frame is the annulus. Periodic algebras are more complicated and less well-known. 

\subsection{Non-periodic algebras}\label{app:npa}

\subsubsection{The partition algebra}

The largest algebra that we will consider is the partition algebra $\PP_L(Q)$ \cite{hr05}. In this case, the diagrams represent set partitions of the $2L$ sites.
Subsets of the partition correspond to connected components of the diagrams, called blocks, which are made of points connected by edges.
(Two edges that cross are not considered as connected.) 
These
edges can be drawn in different ways, so that a partition is associated with an equivalence class of diagrams. An equivalent option, coming from the theory of hypergraphs, is to consider that a block of size $k$ is represented by a $k$-hyperedge---a generalized edge that connects $k$ distinct sites. We shall however only use ordinary edges in our figures. Here is an example with $L=8$:
\begin{align}
\begin{tikzpicture}[scale=.55,line width=1.15pt] 
\foreach \i in {1,...,8} 
{ \path (\i,2) coordinate (T\i); \path (\i,0) coordinate (B\i); } 
[line width=4pt]  (T1) -- (T8) -- (B8) -- (B1) -- (T1);
\draw (T3) -- (B4);
\draw (T2) -- (B1);
\draw (T6) -- (B8);
\draw (T4)  .. controls +(.1,-1) and +(-.1,.7) .. (B6);
\draw (T1) .. controls +(.1,-.8) and +(-.1,-.8) .. (T3) ;
\draw (T4) .. controls +(.1,-.8) and +(-.1,-1) .. (T7) ;
\draw (T7) .. controls +(.1,-.6) and +(-.1,-.6) .. (T8) ;
\draw (B2) .. controls +(.1,.6) and +(-.1,.6) .. (B3) ;
{\draw (B5) .. controls +(.1,.8) and +(-.1,.8) .. (B7) ;}
\foreach \i in {1,...,8}
\foreach \i in {1,...,8} 
{ \fill (T\i) circle (4pt); 
 \node[above] at (T\i) {$\i'$};
\fill (B\i) circle (4pt); 
 \node[below] at (B\i) {$\i$};
} 
\draw[latex-latex] (-1, 3) node[left]{imaginary time} -- (-1, -2) -- (10, -2) node[below right]{space};
\end{tikzpicture}
\end{align}
Labelling the points at the bottom and top of  the diagrams as $1,\ldots,8$ and $1',\ldots,8'$, the corresponding set partition is $(12')(23)(41'3')(57)(64'7'8')(86')(5')$.

Multiplication is defined by concatenation---placing two rectangles on top of each other and identifying the top points of one with the bottom points of the other. Connected components  (clusters) that are disconnected from the top and bottom of the combined diagram are eliminated, but each of them contributes a numerical factor $Q$ to the product. 
For example:
\begin{equation}\label{eq:diagrammultexample}
\begin{array}{r}
\begin{array}{c}
\scalebox{0.75}{\begin{tikzpicture}[scale=.55,line width=1.35pt] 
\foreach \i in {1,...,12} 
{ \path (\i,2) coordinate (T\i); \path (\i,0) coordinate (B\i); } 
[line width=4pt] 
(T1) -- (T12) -- (B12) -- (B1) -- (T1);
\draw (T2) .. controls +(.1,-.6) and +(-.1,-.6) .. (T3) ;
\draw (T6) .. controls +(.1,-.6) and +(-.1,-.6) .. (T7) ;
\draw (T3) .. controls +(.1,-.8) and +(-.1,-.8) .. (T5) ;
\draw (T9) .. controls +(.1,-.8) and +(-.1,-.8) .. (T11) ;
\draw (T10) .. controls +(.1,-.8) and +(-.1,-.8) .. (T12) ;
\draw (B3) .. controls +(.1,.8) and +(-.1,.8) .. (B5) ;
\draw (B8) .. controls +(.1,.8) and +(-.1,.8) .. (B10) ;
\draw (B7) .. controls +(.1,1.3) and +(-.1,1.3) .. (B12) ;
\draw (B1) -- (T4);
\draw (B2) -- (T1);
\draw (B6) .. controls +(.1,1) and +(-.1,-.8) .. (T9) ;
\draw (B9) -- (T8);
\draw (B11) .. controls +(.1,.6) and +(-.1,-.6) .. (T12) ;
\foreach \i in {1,...,12}  { \fill (T\i) circle (4.5pt); \fill (B\i) circle (4.5pt); } 
\end{tikzpicture}}\end{array} \\
\begin{array}{c}
\scalebox{0.75}{\begin{tikzpicture}[scale=.55,line width=1.35pt] 
\foreach \i in {1,...,12} 
{ \path (\i,2) coordinate (T\i); \path (\i,0) coordinate (B\i); } 
\filldraw[fill=gray!25,draw=gray!25,line width=4pt]  (T1) -- (T12) -- (B12) -- (B1) -- (T1);
\draw (T1) .. controls +(.1,-.6) and +(-.1,-.6) .. (T2) ;
\draw (T7) .. controls +(.1,-.6) and +(-.1,-.6) .. (T8) ;
\draw (T10) .. controls +(.1,-.8) and +(-.1,-.8) .. (T12) ;
\draw (T3) .. controls +(.1,-1) and +(-.1,-1) .. (T6) ;
\draw (B3) .. controls +(.1,.6) and +(-.1,.6) .. (B4) ;
\draw (B6) .. controls +(.1,.6) and +(-.1,.6) .. (B7) ;
\draw (B10) .. controls +(.1,.6) and +(-.1,.6) .. (B11) ;
\draw (B1) .. controls +(.1,.8) and +(-.1,.8) .. (B3) ;
\draw (B9) .. controls +(.1,1) and +(-.1,1) .. (B12) ;
\draw (B2) -- (T2);
\draw (B6) -- (T5);
\draw (B8) .. controls +(.1,1) and +(-.1,-.8) .. (T11) ;
\draw (B10) .. controls +(-.1,.6) and +(.1,-.6) .. (T9) ;
\foreach \i in {1,...,12}  { \fill (T\i) circle (4.5pt); \fill (B\i) circle (4.5pt); } 
\end{tikzpicture}}\end{array} \end{array} 
= Q^2
\begin{array}{c}  
\scalebox{0.75}{\begin{tikzpicture}[scale=.55,line width=1.35pt] 
\foreach \i in {1,...,12} 
{ \path (\i,2) coordinate (T\i); \path (\i,0) coordinate (B\i); } 
[line width=4pt]  (T1) -- (T12) -- (B12) -- (B1) -- (T1);
\draw (T2) .. controls +(.1,-.6) and +(-.1,-.6) .. (T3) ;
\draw (T6) .. controls +(.1,-.6) and +(-.1,-.6) .. (T7) ;
\draw (T3) .. controls +(.1,-.8) and +(-.1,-.8) .. (T5) ;
\draw (T9) .. controls +(.1,-.8) and +(-.1,-.8) .. (T11) ;
\draw (T10) .. controls +(.1,-.8) and +(-.1,-.8) .. (T12) ;
\draw (B3) .. controls +(.1,.6) and +(-.1,.6) .. (B4) ;
\draw (B6) .. controls +(.1,.6) and +(-.1,.6) .. (B7) ;
\draw (B10) .. controls +(.1,.6) and +(-.1,.6) .. (B11) ;
\draw (B1) .. controls +(.1,.8) and +(-.1,.8) .. (B3) ;
\draw (B9) .. controls +(.1,1) and +(-.1,1) .. (B12) ;
\draw (T1) -- (B2) -- (T4);
\draw (B7) .. controls +(.1,.6) and +(-.1,-.6) .. (T9) ;
\draw (B8) .. controls +(.1,.6) and +(-.1,-.6) .. (T10) ;
\draw (B10) .. controls +(-.1,.6) and +(.1,-.6) .. (T8) ;
\foreach \i in {1,...,12} { \fill (T\i) circle (4.5pt); \fill (B\i) circle (4.5pt); } 
\end{tikzpicture} }
 \end{array} 
 \end{equation} 
A propagating block is a block that contains vertices both from the top and from the bottom rows. Concatenation never increases the number of propagating blocks.
 
It is convenient to introduce the elements $p_i, s_i, s_{i+\frac12} \in \PP_L(Q)$, which generate the partition algebra:
 \begin{equation}
\label{s-gen}
\begin{array}{ccc}
p_i=\!
\begin{array}{c}\scalebox{.75}{\begin{tikzpicture}[scale=.55,line width=1.35pt] 
\foreach \i in {1,...,8} 
{ \path (\i,2) coordinate (T\i); \path (\i,0) coordinate (B\i); } 
[line width=4pt] 
(T1) -- (T8) -- (B8) -- (B1) -- (T1);
\draw (T1) -- (B1);
\draw (T3) -- (B3);
\draw (T4) -- (B5);
\draw (T5) -- (B4);
\draw (T6) -- (B6);
\draw (T8) -- (B8);
\foreach \i in {1,3,4,5,6,8} { \fill (T\i) circle (4.5pt); \fill (B\i) circle (4.5pt); } 
\draw (T2) node  {$\cdots$}; \draw (B2) node  {$\cdots$}; \draw (T7) node  {$\cdots$}; \draw (B7) node  {$\cdots$}; 
\draw  (T4)  node[black,above=0.05cm]{$\scriptstyle{i}$};
\draw  (T5)  node[black,above=0.0cm]{$\scriptstyle{i+1}$};
\end{tikzpicture}}\end{array},
&
\ s_i=\!
\begin{array}{c}\scalebox{0.75}{\begin{tikzpicture}[scale=.55,line width=1.35pt] 
\foreach \i in {1,...,8} 
{ \path (\i,2) coordinate (T\i); \path (\i,0) coordinate (B\i); } 
[line width=4pt]  (T1) -- (T8) -- (B8) -- (B1) -- (T1);
\draw (T1) -- (B1);
\draw (T3) -- (B3);
\draw (T5) -- (B5);
\draw (T6) -- (B6);
\draw (T8) -- (B8);
\foreach \i in {1,3,4,5,6,8} { \fill (T\i) circle (4.5pt); \fill (B\i) circle (4.5pt); } 
\draw (T2) node  {$\cdots$}; \draw (B2) node  {$\cdots$}; \draw (T7) node  {$\cdots$}; \draw (B7) node  {$\cdots$}; 
\draw  (T4)  node[black,above=0.05cm]{$\scriptstyle{i}$};
\end{tikzpicture}}\end{array},
&
\ s_{i+\frac12}=\!
\begin{array}{c}\scalebox{0.75}{\begin{tikzpicture}[scale=.55,line width=1.35pt] 
\foreach \i in {1,...,8} 
{ \path (\i,2) coordinate (T\i); \path (\i,0) coordinate (B\i); } 
[line width=4pt]  (T1) -- (T8) -- (B8) -- (B1) -- (T1);
\draw (T1) -- (B1);
\draw (T3) -- (B3);
\draw (T4) .. controls +(.1,-.6) and +(-.1,-.6) .. (T5);
\draw (B4) .. controls +(.1,+.60) and +(-.1,+.6) .. (B5);
\draw (T4) -- (B4);
\draw (T5) -- (B5);
\draw (T6) -- (B6);
\draw (T8) -- (B8);
\foreach \i in {1,3,4,5,6,8} { \fill (T\i) circle (4.5pt); \fill (B\i) circle (4.5pt); } 
\draw (T2) node  {$\cdots$}; \draw (B2) node  {$\cdots$}; \draw (T7) node  {$\cdots$}; \draw (B7) node  {$\cdots$}; 
\draw  (T4)  node[black,above=0.05cm]{$\scriptstyle{i}$};
\draw  (T5)  node[black,above=0.0cm]{$\scriptstyle{i+1}$};
\end{tikzpicture}}\end{array} \\
\hskip0.3in 1 \leq i \leq L-1 &\hskip.3in 1 \leq i \leq L &\hskip.6in 1 \leq i \leq L-1
\end{array}
\end{equation}
In order to study subalgebras of $\PP_L(Q)$, it is convenient to define
$e_i=s_{i+\frac12}s_is_{i+1}s_{i+\frac12}$,  
$l_i=p_is_i$, and $r_i=s_ip_i$,  so that diagrammatically
\begin{equation}
\label{extra-gen}
\begin{array}{ccc}
e_i= \!
\begin{array}{c}\scalebox{0.75}{\begin{tikzpicture}[scale=.55,line width=1.35pt] 
\foreach \i in {1,...,8} 
{ \path (\i,2) coordinate (T\i); \path (\i,0) coordinate (B\i); } 
[line width=4pt]  (T1) -- (T8) -- (B8) -- (B1) -- (T1);
\draw (T1) -- (B1);
\draw (T3) -- (B3);
\draw (T4) .. controls +(.1,-.6) and +(-.1,-.6) .. (T5);
\draw (B4) .. controls +(.1,+.6) and +(-.1,+.6) .. (B5);
\draw (T6) -- (B6);
\draw (T8) -- (B8);
\foreach \i in {1,3,4,5,6,8} { \fill (T\i) circle (4.5pt); \fill (B\i) circle (4.5pt); } 
\draw (T2) node  {$\cdots$}; \draw (B2) node  {$\cdots$}; \draw (T7) node  {$\cdots$}; \draw (B7) node  {$\cdots$}; 
\draw  (T4)  node[black,above=0.05cm]{$\scriptstyle{i}$};
\draw  (T5)  node[black,above=0.0cm]{$\scriptstyle{i+1}$};
\end{tikzpicture}}\end{array},
&
l_i= \!
\begin{array}{c}\scalebox{.75}{\begin{tikzpicture}[scale=.55,line width=1.35pt] 
\foreach \i in {1,...,8} 
{ \path (\i,2) coordinate (T\i); \path (\i,0) coordinate (B\i); } 
[line width=4pt]  (T1) -- (T8) -- (B8) -- (B1) -- (T1);
\draw (T1) -- (B1);
\draw (T3) -- (B3);
\draw (T4) -- (B5);
\draw (T6) -- (B6);
\draw (T8) -- (B8);
\foreach \i in {1,3,4,5,6,8} { \fill (T\i) circle (4.5pt); \fill (B\i) circle (4.5pt); } 
\draw (T2) node  {$\cdots$}; \draw (B2) node  {$\cdots$}; \draw (T7) node  {$\cdots$}; \draw (B7) node  {$\cdots$}; 
\draw  (T4)  node[black,above=0.05cm]{$\scriptstyle{i}$};
\draw  (T5)  node[black,above=0.0cm]{$\scriptstyle{i+1}$};
\end{tikzpicture}}\end{array},
&
r_i=\!
\begin{array}{c}\scalebox{.75}{\begin{tikzpicture}[scale=.55,line width=1.35pt] 
\foreach \i in {1,...,8} 
{ \path (\i,2) coordinate (T\i); \path (\i,0) coordinate (B\i); } 
[line width=4pt]  (T1) -- (T8) -- (B8) -- (B1) -- (T1);
\draw (T1) -- (B1);
\draw (T3) -- (B3);
\draw (T5) -- (B4);
\draw (T6) -- (B6);
\draw (T8) -- (B8);
\foreach \i in {1,3,4,5,6,8} { \fill (T\i) circle (4.5pt); \fill (B\i) circle (4.5pt); } 
\draw (T2) node  {$\cdots$}; \draw (B2) node  {$\cdots$}; \draw (T7) node  {$\cdots$}; \draw (B7) node  {$\cdots$}; 
\draw  (T4)  node[black,above=0.05cm]{$\scriptstyle{i}$};
\draw  (T5)  node[black,above=0.0cm]{$\scriptstyle{i+1}$};
\end{tikzpicture}}\end{array} \\
\hskip.3in 1 \leq i \leq L-1 &\hskip.3in 1 \leq i \leq L-1 &\hskip.3in 1 \leq i \leq L-1
\end{array}.
\end{equation}
In particular, the  $e_i$s  satisfy  the  Temperley--Lieb relations
\begin{subequations}
\label{TL-def-rels}
\begin{eqnarray}
e_i^2&=&Q e_i \ , \\
e_ie_{i\pm 1}e_i&=&e_i \ , \\
\left[e_i,e_j\right]&=&0 \ , \quad \mbox{for } |i-j|\geq 2 \ . 
\end{eqnarray}
\end{subequations}

\subsubsection{Irreducible modules of the partition algebra}
 
Irreducible modules $P_{\lambda}^{(L)}$ of $\PP_L(Q)$ are labelled by integer partitions $\lambda$ with $|\lambda|\leq L$. The partition $\lambda$ corresponds to a representation of the symmetric group $S_{|\lambda|}$. A  basis of $P_{\lambda}^{(L)}$ can be built by considering diagrams with $L$ lower sites, $|\lambda|$ upper sites, and 
$|\lambda|$ propagating blocks. In other words, each upper site belongs to a subset that includes at least one lower site, and no other upper site. For example, in the case $L=12$ and $|\lambda|=5$, here is one possible diagram:
\begin{align}
 \begin{tikzpicture}[scale=.55,line width=1.35pt]
 \btdots{12}{5};
 \bbline{1}{2};
 \btline{4}{4};
 \bbline{3}{4};
 \btline{5}{5};
 \btline{7}{6};
 \btline{8}{7};
 \btline{9}{7};
 \btline{10}{8};
 \bbline{11}{10};
 \bbline{11}{12};
 \end{tikzpicture}
\end{align}
In the representation $P_{\lambda}^{(L)}$, the partition algebra acts on such diagrams from below, while the symmetric group $S_{|\lambda|}$ acts from above according to the representation $\lambda$. 
In other words, if $\{b_i\}$ is the set of such diagrams, we have $P_{\lambda}^{(L)} = \mathbb{C}\{b_i\} \otimes_{S_{|\lambda|}} \lambda$. The action of the partition algebra can lead to diagrams with fewer than $|\lambda|$ propagating blocks, which are then modded out.

The dimension of $P_{\lambda}^{(L)}$ is the number of diagrams, times $f_\lambda = \dim \lambda$. To count the diagrams, remember that the number of partitions of $L$ elements into $i$ non-empty subsets is given by the Stirling number of the second kind $\stirling{L}{i}$. To build a diagram, we start with one such partition with $|\lambda|\leq i\leq L$, and choose $|\lambda|$ subsets that we connect to the upper sites. Since the number of such choices is $\binom{i}{|\lambda|} = \frac{i!}{|\lambda|!(i-|\lambda|)!}$, we obtain
\begin{equation}
\dim P_{\lambda}^{(L)}=f_{\lambda}\times \sum_{i=|\lambda|}^L \binom{i}{|\lambda|} \stirling{L}{i}\ . \label{dimpartrep}
\end{equation}
An alternate derivation proceeds by noticing that a set partition with $|\lambda|$ marked
blocks consists of unmarked blocks, which are a set of non-empty sets, hence with exponential generating function $\exp({\rm e}^x - 1)$, and marked blocks, which are $|\lambda|$ indistinguishable non-empty sets,
hence with exponential generating function $\frac{1}{|\lambda|!} ({\rm e}^x - 1)^{|\lambda|}$. Therefore 
$\frac{1}{f_\lambda} \dim P_{\lambda}^{(L)}$ has the exponential generating function $\exp({\rm e}^x - 1) \times \frac{1}{|\lambda|!}  ({\rm e}^x - 1)^{|\lambda|}$.

\subsubsection{The Brauer subalgebra and its representations}

The Brauer subalgebra $\B_L(\n\equiv Q)\subset \PP_L(Q)$ is generated by pairwise partitions. This means that all blocks must have size two, and  in particular there  are no isolated sites. This algebra is generated by the  $e_i$s and $p_i$s.
In diagram concatenation, the clusters that are eliminated are closed loops, and we denote their weight by $\n$ instead of $Q$.
Propagating blocks become propagating lines, called through-lines.

Irreducible modules $B_\lambda^{(L)}$ of $\B_L(\n)$ are labelled by partitions $\lambda$ with $|\lambda|\leq L$ and $L\equiv |\lambda|\bmod 2$. They are built from diagrams and from irreducible representations of $S_{|\lambda|}$, just like the irreducible modules of the partition algebra. The relevant diagrams have $L$ bottom sites, $|\lambda|$ top sites, and $|\lambda|$ through-lines. For example, in the case $L=10$ and $|\lambda|=4$:
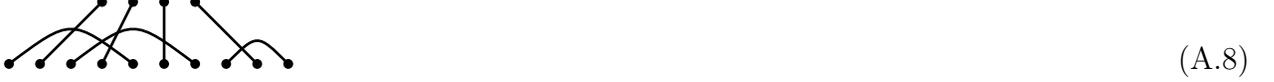
\begin{equation}
\label{B10pic}
\scalebox{0.75}{\begin{tikzpicture}[scale=.55,line width=1.35pt] 
\foreach \i in {1,2,3,4,5,6,7,8,9,10}
{ \fill (\i,0) circle (4.5pt); }
\draw (1,0) .. controls +(2,1.5) .. (5,0);
\draw (3,0) .. controls +(2,1.5) .. (7,0);
\draw (8,0) .. controls +(1,1.0) .. (10,0);
\draw (2,0) -- (4,2);
\draw (4,0) -- (5,2);
\draw (6,0) -- (6,2);
\draw (9,0) -- (7,2);
\foreach \i in {4,5,6,7}
{ \fill (\i,2) circle (4.5pt); }
\end{tikzpicture}}
\end{equation}
To enumerate such diagrams, first choose the $|\lambda|$ bottom sites that are connected to top sites, and then connect the remaining bottom sites pairwise. This leads to 
\begin{equation}
\dim B^{(L)}_\lambda=\binom{L}{|\lambda|}(L-|\lambda|-1)!! f_\lambda \ ,
\label{dimBrauer}
\end{equation}
where, by convention, $(-1)!!=1$.

\subsubsection{The Temperley--Lieb subalgebra and its representations}

The planar subalgebra of the Brauer algebra is the Temperley--Lieb algebra $\TL_L(\n)$.
Within $\TL_L(\n)$, every site is connected with one other site by a single edge, all edges are drawn inside the rectangular frame, and the edges can be drawn so that there are no crossings. When concatenating diagrams, each loop is replaced by a numerical factor $\n$. This algebra is generated by the identity and the $e_i$s.

Irreducible modules $W_r^{(L)}$ of $\TL_L(\n)$ are parametrized simply by the number of through-lines, which we will denote here by $2r$, where $r \in \frac12\mathbb{N}$ and $L\equiv 2r\bmod 2$. This no longer involves a representation of the symmetric group, because permuting the through-lines would lead to crossings. 
The dimension of $W_r^{(L)}$ is the number of diagrams---often called link patterns---that can be drawn by marking $2r$ out of $L$ points and pairing the remaining points without crossings and without enclosing any marked point:
 \begin{equation}
\dim W_r^{(L)}=\binom{L}{\frac{L}{2}+r}-\binom{L}{\frac{L}{2}+r+1} \ .
\label{dimTL-loop}
 \end{equation}
 An example of a basis element in $W^{(10)}_1$ would 
look like:
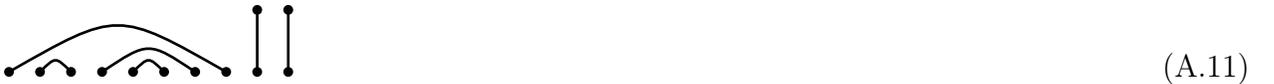
\begin{equation}
\scalebox{0.75}{\begin{tikzpicture}[scale=.55,line width=1.35pt] 
\foreach \i in {1,2,3,4,5,6,7,8,9,10}
{ \fill (\i,0) circle (4.5pt); }
\draw (1,0) .. controls +(3.5,+2.0) .. (8,0);
\draw (2,0) .. controls +(0.5,+0.5) .. (3,0);
\draw (4,0) .. controls +(1.5,+1.0) .. (7,0);
\draw (5,0) .. controls +(0.5,+0.5) .. (6,0);
\draw (9,0) -- (9,2);
\draw (10,0) -- (10,2);
\foreach \i in {9,10}
{ \fill (\i,2) circle (4.5pt); }
\end{tikzpicture}}
\label{tlrd}
\end{equation}

\subsubsection{Another Temperley--Lieb subalgebra: the join-detach algebra}

Somewhat confusingly, there is another Temperley--Lieb subalgebra $\TL_{2L}(\sqrt{Q})\subset \PP_L(Q)$ of the partition algebra, with twice as many sites and a different parameter $\sqrt{Q}$. 
This follows from the relations
\begin{align}
s_i^2=Q^{1-2(i\bmod 1)}s_i \quad , \quad 
s_is_{i\pm\frac12}s_i =Q s_i \ ,
\end{align}
for $i=1,\frac32,2,\ldots,L-\frac12,L$, which imply that 
\begin{equation}
 \hat{e}_{2i-1} = Q^{-\frac12+2(i\bmod 1)} s_i 
 \label{hes}
\end{equation}
satisfy the Temperley--Lieb relations \eqref{TL-def-rels} with the parameter $\sqrt{Q}$.
This Temperley--Lieb subalgebra is called the join-detach algebra, with $s_{i\in \mathbb{N}}$ detaching the site $i$, while $s_{i\in\mathbb{N}+\frac12}$ joins the two sites $i\pm\frac12$.
The random cluster formulation of the Potts-model partition function \eqref{FKmod} can be constructed in the transfer matrix formalism from this join-detach algebra, with the second term in the expansion \eqref{fkbol} corresponding to the join operator for a space-like edge, and the first term corresponding to the detach operator for a time-like edge \cite{js18}.

\subsubsection{Rook Brauer algebra and Motzkin algebra}

Let the two-parameter Rook Brauer algebra $\RB_L(x,y)$
be generated by the diagrams whose blocks have size one or two \cite{mm12}. 
When multiplying diagrams, we eliminate the clusters that are disconnected from the top and bottom of the combined diagram, while giving a weight $x$ to each loop and $y$ to each open path or isolated site:
\begin{equation}
 \begin{array}{c}
\begin{tikzpicture}[scale=.4,line width=1pt] 
\foreach \i in {1,...,11} 
{ \path (\i,1) coordinate (T\i); \path (\i,-1) coordinate (B\i); } 
\draw[black] (T7) -- (B10);
\draw[black] (T11) -- (B8);
\draw[black] (T2) -- (B4);
\draw[black] (T1) .. controls +(.1,-.75) and +(-.1,-.75) .. (T3) ;
\draw[black] (T4) .. controls +(.1,-1.1) and +(-.1,-1.1) .. (T8) ;
\draw[black] (T5) .. controls +(.1,-.5) and +(-.1,-.5) .. (T6) ;
\draw[black] (B1) .. controls +(.1,.5) and +(-.1,.5) .. (B3) ;
\draw[black] (B5) .. controls +(.1,1.1) and +(-.1,1.1) .. (B7) ;
\draw[black] (B2) .. controls +(.1,1.1) and +(-.1,1.1) .. (B6) ;
\foreach \i in {1,...,11} 
{ \fill (T\i) circle (4pt); \fill (B\i) circle (4pt); } 
\end{tikzpicture} \\
\begin{tikzpicture}[scale=.4,line width=1pt] 
\fill[black!10] (1, -1) -- (1, 1) -- (11, 1) -- (11, -1) -- cycle;
\foreach \i in {1,...,10} 
{ \path (\i,1) coordinate (T\i); \path (\i,-1) coordinate (B\i); } 
\draw[black] (T10) -- (B5);
\draw[black] (T11) -- (B8);
\draw[black] (T1) -- (B4);
\draw[black] (T3) -- (B1);
\draw[black] (T4) -- (B3);
\draw[black] (T8) -- (B10);
\draw[black] (T2) .. controls +(.1,-.75) and +(-.1,-.75) .. (T5) ;
\draw[black] (T6) .. controls +(.1,-.5) and +(-.1,-.5) .. (T7) ;
\draw[black] (B7) .. controls +(.1,1.5) and +(-.1,1.5) .. (B11) ;
\foreach \i in {1,...,11} 
{ \fill (T\i) circle (4pt); \fill (B\i) circle (4pt); } 
\end{tikzpicture}\end{array}
= xy
\begin{array}{c}
\begin{tikzpicture}[scale=.4,line width=1pt] 
\foreach \i in {1,...,11} 
{ \path (\i,1) coordinate (T\i); \path (\i,-1) coordinate (B\i); } 
\draw[black] (T7) -- (B5);
\draw[black] (T11) -- (B10);
\draw[black] (T2) -- (B3);
\draw[black] (T1) .. controls +(.1,-.75) and +(-.1,-.75) .. (T3) ;
\draw[black] (T4) .. controls +(.1,-1.1) and +(-.1,-1.1) .. (T8) ;
\draw[black] (T5) .. controls +(.1,-.5) and +(-.1,-.5) .. (T6) ;
\draw[black] (B1) .. controls +(.1,.75) and +(-.1,.75) .. (B4) ;
\draw[black] (B7) .. controls +(.1,1.5) and +(-.1,1.5) .. (B11) ;
\foreach \i in {1,...,11} 
{ \fill (T\i) circle (4pt); \fill (B\i) circle (4pt); } 
\end{tikzpicture}\end{array}.
\end{equation}
Then $\RB_L(Q,Q)$ is a subalgebra of the partition algebra $\PP_L(Q)$, 
generated by $e_i, p_i$ and $s_i$.
Moreover, $\RB_L(\n+1)\equiv \RB_L(\n+1, 1)$ is in Schur-Weyl duality
with $O(\n)$ in the tensor product $\left([]\oplus [1]\right)^{\otimes L}$ \cite{dh12}.

The planar version of the Rook Brauer algebra $\RB_L(\n+1)$ is the Motzkin algebra $\MM_L(\n+1)$ \cite{bh11}. It is generated by $e_i$, $l_i$ and $r_i$.
We believe that $\MM_L(\n+1)$
is in fact isomorphic to the dilute Temperley--Lieb algebra  $\dTL_L(\n)$ \cite{gri95, bs13}, in which
we only allow loop and isolated sites, but not open paths, and where loops have fugacity $\n$ instead of $\n+1$. 

\subsection{Periodic algebras}\label{app:pa}

Let us now consider periodic algebras, i.e.\ algebras of diagrams that can be drawn without crossings in an annular frame. 

\subsubsection{Affine Temperley--Lieb algebra and unoriented Jones--Temperley--Lieb algebra}

The affine Temperley--Lieb algebra is a subalgebra of the Brauer algebra that is larger than the Temperley--Lieb algebra, $\TL_L(\n) \subset \ATL_L(\n) \subset \B_L(\n)$. It is defined from $\TL_L(\n)$ by adding the two generators 
\begin{align}
e_L&=\ 
\begin{tikzpicture}[scale=.55,line width=1.35pt, baseline=(current  bounding  box.center)] 
\foreach \i in {1,...,5}
{\path (\i,2) coordinate (T\i); 
\path (\i,0) coordinate (B\i); } 
\foreach \i in {1,2,4,5}
{\fill (T\i) circle (4.5pt);
\fill (B\i) circle (4.5pt);}
\draw (T1) .. controls +(-.1,-.5)  .. (-.,1.3) ;
\draw (T5) .. controls +(.1,-.5) .. (6.1, 1.3);
\draw (T2) --  (B2) ;
\draw (B1) .. controls +(-.1,.5)  .. (-.,.6);
\draw (T4) -- (B4) ;
\draw (B5) .. controls +(.1,.5)  .. (6.1,.6);
\draw (T3) node  {$\cdots$}; 
\draw (B3) node  {$\cdots$}; 
\end{tikzpicture} 
\label{el}
\\
u&=\ 
\begin{tikzpicture}[scale=.55,line width=1.35pt, baseline=(current  bounding  box.center)] 
\foreach \i in {1,...,8} 
{ \path (\i,2) coordinate (T\i); 
\path (\i,0) coordinate (B\i); } 
\draw (T2) -- (B1);
\draw (T3) -- (B2);
\draw (T5) -- (B4);
\draw (T7) -- (B6);
\foreach \i in {1,2,3,5,7} { \fill (T\i) circle (4.5pt);}
\foreach \i in {1,2,4,6,7} {\fill (B\i) circle (4.5pt); } 
\draw (B7) -- (7.5,.9);
\draw (T1) -- (.4,.8);
\draw (T4) node  {$\cdots$}; \draw (B3) node  {$\cdots$}; \draw (T6) node  {$\cdots$}; \draw (B5) node  {$\cdots$}; 
\end{tikzpicture}
\label{u}
\end{align}
The generator $e_L$ satisfies the Temperley--Lieb relations \eqref{TL-def-rels}, where we now identify $e_{L+1}=e_1$ and consider all indices modulo $L$. Moreover, the generator $u$ satisfies
\begin{subequations}
\label{ATL-u-rels}
\begin{align}
ue_ju^{-1}&=e_{j+1} \ , \\
u^2e_{L-1}&=e_1 e_2 \cdots e_{L-1} \ ,
\end{align}
\end{subequations}
and $u^L$ is a central element. 

Although making the system periodic  seems rather innocuous, it has the profound consequence that $\ATL_L$ becomes infinite-dimensional. This is a consequence of the fact that  it is possible to cyclically permute the upper sites carrying through-lines without otherwise changing the diagrams. Translating through-lines by one site to the right is called the pseudo-translation $t$,
\begin{align}
 \begin{tikzpicture}[scale=.55,line width=1.35pt,, baseline=(current  bounding  box.center)]
  \btdots{7}{7};
  \ttline{1}{2};
  \ttline{5}{6};
  \bbline{2}{3};
  \bbline{6}{7};
  \btline{1}{3};
  \btline{4}{4};
  \btline{5}{7};
 \end{tikzpicture}
 \quad \overset{t}{\longrightarrow} \quad 
 \begin{tikzpicture}[scale=.55,line width=1.35pt,, baseline=(current  bounding  box.center)]
  \btdots{7}{7};
  \ttline{1}{2};
  \ttline{5}{6};
  \bbline{2}{3};
  \bbline{6}{7};
  \btline{1}{4};
  \btline{4}{7};
  \draw (T3) to [out = -90, in = 30] (1,.75);
  \draw (B5) to [out = 90, in = -150] (7, .75);
 \end{tikzpicture}
 \label{t}
\end{align}
The pseudo-translation $t$ differs from the translation $u$, and its expression in terms of algebra generators  depends on the diagram it acts on. In particular, the action of $t^N$ on a diagram in general yields diagrams that differ for all $N\in\mathbb{Z}$, in other words we can have the through-lines wind around the annulus an arbitrary number of times. 
Therefore, even for $L$ finite, we have $\dim \ATL_L(\n) = \infty$. In fact, another reason why the algebra is infinite-dimensional is the existence of non-contractible loops. For example, the following diagram concatenation leads to one non-contractible loop:
\begin{align}
\renewcommand{\arraystretch}{2}
 \begin{array}{c}
  \begin{tikzpicture}[scale = .6, line width=1.15pt, baseline=(current  bounding  box.center)]
  \btdots{4}{4};
  \ttline{1}{2};
  \ttline{3}{4};
  \bbline{1}{2};
  \bbline{3}{4};
 \end{tikzpicture}
 \\
 \begin{tikzpicture}[scale = .6, line width=1.15pt, baseline=(current  bounding  box.center)]
  \fill[black!10] (.5, -.1) -- (.5, 1.6) -- (4.5, 1.6) -- (4.5, -.1) -- cycle;
  \btdots{4}{4};
  \ttline{2}{3};
  \bbline{3}{2};
  \draw (T4) to [out = -90, in = 180] (4.5, 1);
  \draw (T1) to [out = -90, in = 0] (.5, 1);
  \draw (B4) to [out = 90, in = 180] (4.5, .5);
  \draw (B1) to [out = 90, in = 0] (.5, .5);
 \end{tikzpicture}
 \end{array}
 \quad = \quad 
 \begin{tikzpicture}[scale = .6, line width=1.15pt, baseline=(current  bounding  box.center)]
  \btdots{4}{4};
  \ttline{1}{2};
  \ttline{3}{4};
  \bbline{3}{2};
  \draw (.5, .8) to (4.5, .8);
  \draw (B4) to [out = 90, in = 180] (4.5, .5);
  \draw (B1) to [out = 90, in = 0] (.5, .5);
 \end{tikzpicture}
\end{align}
Iterating, it is possible to generate an arbitrary number of non-contractible loops. 

We define the unoriented Jones--Temperley--Lieb algebra  $\uJTL_L(\n)$ as a finite-dimensional quotient of the affine Temperley--Lieb algebra $\ATL_L(\n)$, by identifying infinite families of diagrams. To begin with, we identify diagrams that are related by having all through-lines wind once around the annulus, i.e.\ we impose that the pseudo-translation obeys 
\begin{align}
 t^{2r} \underset{\uJTL_L(\n)}{=} 1\ .
 \label{ttro}
\end{align}
We then deal with the non-contractible loops, by eliminating them with the same weight $\n$ as the contractible loops. In other words, adding a contractible loop to a diagram, which is possible if there are no through-lines, amounts to multiplying this diagram by $\n$.

\subsubsection{Representations of the unoriented Jones--Temperley--Lieb algebra}

Representations of $\uJTL_L(\n)$ are generated by diagrams with $L$ lower sites, $2r\leq L$ upper sites, and $2r$ through-lines with $r\in \frac12\mathbb{N}$. On this set of diagrams, we can act by cyclic permutations of the upper sites or equivalently of the through-lines. This action is described by the group $\mathbb{Z}_{2r}$ generated by $t$. Irreducible finite-dimensional modules $W^{(L)}_{(r,s)}$ are therefore parameterized by $(r,s)$, where 
$e^{\pi is}$ is the eigenvalue of $t$. Equivalently, the number $s$ itself, which we call the pseudo-momentum, is defined modulo $2$ and obeys $rs\in \mathbb{Z}$, so that 
\begin{align}
 \left(t - e^{\pi is}\right) W^{(L)}_{(r,s)} = 0 \ .
\end{align}
The dimensions of these modules are obtained by diagram counting, so they are $s$-independent:
\begin{equation}
 \dim \, W_{(r,s)}^{(L)}=\binom{L}{\frac{L}{2}+r} \,, \qquad \mbox{for } r>0 \,.
 \label{dimATL}
\end{equation}
A sample basis element in $W^{(10)}_{(1,s)}$ is:
\begin{equation}
\scalebox{0.75}{\begin{tikzpicture}[xscale=.55, yscale=-.55, line width=1.35pt] 
\foreach \i in {1,2,3,4,5,6,7,8,9,10}
{ \fill (\i,2) circle (4.5pt); }
\draw (1,2) arc (0:-90:1 and 0.5);
\draw (2,2) .. controls +(0.5,-0.5) .. (3,2);
\draw (4,2) arc (0:-90:4 and 1.5);
\draw (6,2) .. controls +(0.5,-0.5) .. (7,2);
\draw (5,2) -- (5,0);
\draw (8,2) -- (8,0);
\draw (9,2) arc (180:270:2 and 1.5);
\draw (10,2) arc (180:270:1 and 0.5);
\end{tikzpicture}}
\end{equation}
where the arcs going through the periodic boundary condition illustrate the notion of non-crossing with an annular
frame. 

Let us now focus on representations with no through-lines, i.e.\ $r=0$. This implies that $L$ is even. The corresponding diagrams can then have non-contractible loops. Representations of the affine Temperley--Lieb algebra are parametrized by the weight of these loops, which can be any complex number. For generic values of this weight, the $\binom{L}{\frac{L}{2}}$ diagrams with no through-lines and no non-contractible loops form a basis of an irreducible finite-dimensional representation. 
In the case of the unoriented Jones--Temperley--Lieb algebra, the weight of non-contractible loops is $\n$. The diagrams with no through-lines and no non-contractible loops now generate a reducible representation, which has an irreducible quotient $W^{(L)}_{(0,0)}$, whose dimension is  
\begin{equation}
 \dim W^{(L)}_{(0,0)} = \binom{L}{\frac{L}{2}}-\binom{L}{\frac{L}{2}+1} \,.
 \label{dimATL0}
\end{equation}
In this quotient, we have identities of the type
\begin{equation}
\begin{array}{cc}
\begin{array}{c}\scalebox{0.75}{\begin{tikzpicture}[scale=.55,line width=1.35pt] 
\draw (B2) .. controls +(.1,.6) and +(-.1,.6) .. (B3) ;
\draw (B1) .. controls +(.1,.9) 
and +(-.1,.9) .. (B4) ;
%
\foreach \i in {1,...,4} 
{ \fill (B\i) circle (4.5pt); } 
\end{tikzpicture}}\end{array}
=&
\begin{array}{c}
\scalebox{0.75}{\begin{tikzpicture}[scale=.55,line width=1.35pt] 
\draw (B1) .. controls +(-.1,.5)  .. (-.,.6);
\draw (B2) .. controls +(.1,.6) and +(-.1,.6) .. (B3) ;
\draw (B4) .. controls +(.1,.5)  .. (5.1,.6);
\foreach \i in {1,...,4} 
{ \fill (B\i) circle (4.5pt); } 
\end{tikzpicture}} \end{array}\end{array}
\label{eee}
\end{equation}
and in general the quotient module does not distinguish whether an edge connecting two given sites
goes through the periodic boundary condition or not. Each equivalence class of diagrams in
$W^{(L)}_{(0,0)}$ can thus be drawn with a rectangular frame.

\subsubsection{The Jones--Temperley--Lieb algebra and the Potts--Temperley--Lieb algebra}
 
The Jones--Temperley--Lieb algebra $\JTL_{2L}(\sqrt{Q})$  exists only for $L$ even. It stems from the reformulation of the random cluster model as a loop model, with the loops turning around the clusters as in Figure \eqref{plat}. Algebraically, it is a version of the join-detach algebra $\TL_{2L}(\sqrt{Q})$ that lives on an annulus. To this algebra, we add a generator $e_{2L}$ \eqref{el}, together with a generator $u^2$ rather than $u$ \eqref{u}, since two sites of the loop model correspond to one site of the cluster model. In this algebra, we do not have a pseudo-translation $t$ but rather its square $t^2$, which still obeys the periodicity condition $\eqref{ttro}$. This implies in particular that the number of through-lines $2r$ is even, i.e.\ $r\in\mathbb{N}$, and that the pseudo-momentum $s$ is now defined modulo $1$. 

The Jones--Temperley--Lieb algebra respects an alternating orientation of the lattice sites, such that through-lines preserve their orientation under imaginary time evolution. As a consequence, it is related by Schur--Weyl duality to the group $U(n)$, rather than $O(n)$ for the unoriented Jones--Temperley--Lieb algebra \cite{rs07,vjs16}. 

Although motivated by the properties of clusters, the Jones--Temperley--Lieb algebra does not go all the way to describing the Potts model, because it still describes loops. To properly describe cluster connectivities, we have to slightly modify it into what we will call the Potts--Temperley--Lieb algebra $\PTL_{2L}(\sqrt{Q})$. Unfortunately, we do not know how to define this algebra by generators and relations, or by a set of diagrams. The problem is that we need the weight of non-contractible loops to depend on the representation, and detecting the representation involves algebra elements such as the pseudo-translation $t$ \eqref{t}, which has no simple expression in terms of the generators $e_i$. To define the Potts--Temperley--Lieb algebra, we will fall back to the Wedderburn--Artin theorem, which allows us to characterize a semisimple finite-dimensional associative algebra by its finite-dimensional irreducible representations. 

We define $\PTL_{2L}(\sqrt{Q})$ to have the same representations $W_{(r,s)}^{(2L)}$ (with $r\in\mathbb{N}$ and $s\in\frac{1}{r}\mathbb{Z}\bmod 1$) as $\JTL_{2L}(\sqrt{Q})$ for $r\geq 2$. This is because the
embedding $\TL_{2L}(\sqrt{Q})\subset \PP_L(Q)$ \eqref{hes} relates $2r$ through-lines (described by $\TL_{2L}(\sqrt{Q})$) to $r$ propagating clusters (described by $\PP_L(Q)$). On the annulus, this relation no longer holds for $r\leq 1$, which is why $\PTL_{2L}(\sqrt{Q})\neq \JTL_{2L}(\sqrt{Q})$. The existence of one propagating cluster indeed does not imply the existence of two through-lines, because the cluster could have cross topology, i.e. wind around the space direction as in Figure \eqref{cwct}. On the other hand, a propagating cluster prevents the existence of non-contractible loops.
Therefore, we define $\PTL_{2L}(\sqrt{Q})$ to have no representation with $r=1$, but a representation with $r=0$ where non-contractible loops have weight $0$, which we will call $W^{(2L)}_{(0,\frac12)}$.

This representation is however not generated by all diagrams with no through-lines, but only by half of them. This is because the number $N_\text{boundary}$ of lines that cross the left boundary and come back on the right (by the space periodicity of the annulus) is conserved modulo $2$. The only ways to change $N_\text{boundary}$ by one unit would indeed have been to act with $u$ (which does not belong to $\PTL_{2L}(\sqrt{Q})$), or to replace a non-contractible loop with a non-zero factor. We therefore define the irreducible representation $W^{(2L)}_{(0,\frac12)}$ by keeping only the diagrams with $N_\text{boundary}\equiv 1\bmod 2$. This enforces the propagation of one cluster, while $N_\text{boundary}\equiv 0\bmod 2$ would enforce the propagation of one dual cluster \cite{js18,jacobsen15}. Our representation has the dimension 
\begin{align}
 \dim W^{(2L)}_{(0,\frac12)}= \frac12\binom{2L}{L}\ . 
 \label{dimATLh}
 \end{align}

\subsubsection{Branching rules from periodic to non-periodic algebras}

Let us discuss how representations of periodic algebras decompose into representations of the corresponding non-periodic algebras. The basic idea is that we should cut all lines that cross the left boundary and come back on the right, so that each cut line becomes a pair of through-lines. For example, the following diagram shows that the representation $W^{(10)}_{(1,s)}$ of $\uJTL_{10}(n)$ contains the representation $W^{(10)}_{3}$ of $\TL_{10}(n)$:
\begin{equation}
\scalebox{0.75}{\begin{tikzpicture}[xscale=.55, yscale=-.55, line width=1.35pt, baseline=(current  bounding  box.center)] 
\foreach \i in {1,...,10}
{ \fill (\i,2) circle (4.5pt); }
\draw (1,2) arc (0:-90:1 and 0.5);
\draw (2,2) .. controls +(0.5,-0.5) .. (3,2);
\draw (4,2) arc (0:-90:4 and 1.5);
\draw (6,2) .. controls +(0.5,-0.5) .. (7,2);
\draw (5,2) -- (5,0);
\draw (8,2) -- (8,0);
\draw (9,2) arc (180:270:2 and 1.5);
\draw (10,2) arc (180:270:1 and 0.5);
\end{tikzpicture}}
\qquad \longrightarrow \qquad 
\scalebox{0.75}{\begin{tikzpicture}[xscale=.55, yscale=-.55, line width=1.35pt, baseline=(current  bounding  box.center)] 
\foreach \i in {1,...,10}
{ \fill (\i,2) circle (4.5pt); }
\draw (1,2) -- (1, 0);
\draw (2,2) .. controls +(0.5,-0.5) .. (3,2);
\draw (4,2) -- (4, 0);
\draw (6,2) .. controls +(0.5,-0.5) .. (7,2);
\draw (5,2) -- (5,0);
\draw (8,2) -- (8,0);
\draw (9,2) -- (9, 0);
\draw (10,2) -- (10, 0);
\end{tikzpicture}}
\end{equation}
In the case of a representation $ W^{(L)}_{(r>0,s)}$ of the affine Temperley--Lieb algebra, this leads to
\begin{align}
 W^{(L)}_{(r,s)} \underset{\TL_L(n)}{=} \bigoplus_{r'=r}^{\frac{L}{2}} W^{(L)}_{r'} \ ,
 \label{wrswr}
\end{align}
where the sum runs by increments of $1$. This result also holds if $W^{(L)}_{(r,s)}$ is an irreducible representation of the Jones--Temperley--Lieb algebra, unoriented Jones--Temperley--Lieb algebra or of the Potts--Temperley--Lieb algebra. 

The representations $W^{(L)}_{(0,0)}$ and $W^{(2L)}_{(0,\frac12)}$ behave differently, because they are respectively a coset and a subrepresentation of the reducible representation generated by diagrams with no through-lines. In the case of $W^{(L)}_{(0,0)}$, identities of the type \eqref{eee} allow us to eliminate all lines that cross the left boundary. In the case of $W^{(2L)}_{(0,\frac12)}$, the number of such lines is odd. This leads to 
\begin{subequations}
\label{w00w0h}
\begin{align}
 W^{(L)}_{(0,0)} & \underset{\TL_L(n)}{=} W^{(L)}_0 \ , \\
 \quad W^{(2L)}_{(0,\frac12)} & \underset{\TL_L(n)}{=} \bigoplus_{r'\overset{2}{=}1}^{L} W^{(2L)}_{r'}\ ,
\end{align}
\end{subequations}
where the sum over $r'$ runs by increments of $2$.

\subsection{Summary}\label{app:sum}



\subsubsection{Algebras}

In the following list of diagram algebras, 
$n$ is the loop weight, $Q$ the cluster weight. We indicate whether algebras are planar or periodic. (Periodicity only makes sense for planar algebras.) The generators are defined in Equations \eqref{s-gen}, \eqref{extra-gen}, \eqref{u}.
\begin{center}
\renewcommand{\arraystretch}{1.3}
 \begin{tabular}{|l|l|c|l|}
  \hline
  Notation & Name & Property & Generators
  \\
  \hline 
  $\PP_L(Q)$ & Partition &  & $p_i,s_i,s_{i+\frac12} $
  \\
  $\B_L(\n)$ & Brauer &  & $p_i, e_i$
  \\
  $\RB_L(\n)$ & Rook Brauer &  & $p_i,s_i$
  \\
  \hline
  $\TL_L(\n)$ & Temperley--Lieb & Planar & $e_i$
  \\
  $\MM_L(\n)$ & Motzkin & Planar & $e_i, l_i, r_i$
  \\
  $\dTL_L(\n)$ & dilute Temperley--Lieb & Planar & $e_i, l_i, r_i$
  \\
  \hline 
  $\ATL_L(\n)$ & affine Temperley--Lieb & Periodic & $e_i, u$
  \\
  $\uJTL_L(\n)$ & unoriented Jones--Temperley--Lieb & Periodic & $e_i, u$
  \\
  $\JTL_{2L}(\n)$ & Jones--Temperley--Lieb & Periodic & $e_i, u^2$
  \\
  $\PTL_{2L}(\n)$ & Potts--Temperley--Lieb & Periodic & $e_i, u^2$
  \\
  \hline 
 \end{tabular}
\end{center}
If the Motzkin and dilute Temperley--Lieb algebras appear the same, it is because they are identical. On the other hand, the Jones--Temperley--Lieb and Potts--Temperley--Lieb are subtly different.

\subsubsection{Irreducible modules}

For some algebras, we list the irreducible finite-dimensional representations.  We recall that $f_\lambda$ is the dimension of a representation of the symmetric group $S_{|\lambda|}$, given by the hook length formula.
\begin{align}
\renewcommand{\arraystretch}{1.5}
\begin{array}{|l|l|l|l|l|}
\hline 
 \text{Algebra} & \text{Representation} & \text{Parameters} & \text{Dimension} & \text{Equation} 
 \\
 \hline 
 \PP_L(Q) & P_{\lambda}^{(L)} & |\lambda|\leq L & f_{\lambda} \sum_{i=|\lambda|}^L \binom{i}{|\lambda|} \stirling{L}{i} & \eqref{dimpartrep}
 \\
 \hline 
 \B_L(\n) & B^{(L)}_\lambda & \begin{array}{l}|\lambda|\leq L \\ |\lambda|\equiv L\bmod 2 \end{array} & f_\lambda \binom{L}{|\lambda|}(L-|\lambda|-1)!!  & \eqref{dimBrauer}
 \\
 \hline 
 \TL_L(\n) & W_r^{(L)} & \begin{array}{l} 0\leq r\leq \frac{L}{2} \\ r\equiv \frac{L}{2}\bmod 1 \end{array} &  \binom{L}{\frac{L}{2}+r}-\binom{L}{\frac{L}{2}+r+1} & \eqref{dimTL-loop} 
 \\
 \hline 
 \multirow{2}{*}{$\uJTL_L(\n)$} 
 & W_{(r,s)}^{(L)} & \begin{array}{l} \frac12\leq r\leq \frac{L}{2} \\ r\equiv \frac{L}{2}\bmod 1 \\ s\equiv \frac{1}{r}\mathbb{Z} \bmod 2 \end{array}  & \binom{L}{\frac{L}{2}+r} & \eqref{dimATL} 
 \\ \cline{2-5}  
 & W_{(0,0)}^{(L)} & & \binom{L}{\frac{L}{2}}-\binom{L}{\frac{L}{2}+1} & \eqref{dimATL0}
 \\
 \hline 
 \multirow{2}{*}{$\JTL_L(\n)$} 
 & W_{(r,s)}^{(2L)} & \begin{array}{l} 1\leq r\leq 2L \\ s\equiv \frac{1}{r}\mathbb{Z} \bmod 1 \end{array}  & \binom{2L}{L+r} & \eqref{dimATL} 
 \\ \cline{2-5}  
 & W_{(0,0)}^{(2L)} & & \binom{2L}{L}-\binom{2L}{L+1} & \eqref{dimATL0}
 \\
 \hline 
 \multirow{3}{*}{$\PTL_L(\n)$} 
 & W_{(r,s)}^{(2L)} & \begin{array}{l} 2\leq r\leq 2L \\ s\equiv \frac{1}{r}\mathbb{Z} \bmod 1 \end{array}  & \binom{2L}{L+r} & \eqref{dimATL} 
 \\ \cline{2-5}  
 & W_{(0,0)}^{(2L)} & & \binom{2L}{L}-\binom{2L}{L+1} & \eqref{dimATL0}
 \\ \cline{2-5}  
 & W_{(0,\frac12)}^{(2L)} & & \frac12\binom{2L}{L} & \eqref{dimATLh}
 \\
 \hline 
\end{array}
\end{align}

\section{Branching rules from transfer matrices}\label{app:brtm}

In this appendix we will explain how the branching rules of Sections \ref{sec:obr} and \ref{sec:sbr} can be numerically determined by diagonalizing transfer matrices. This method is useful for checking our results. Moreover, it is of rather general validity. 

\subsection{Branching rules \texorpdfstring{$\mathcal{B}_L(n) \downarrow \uJTL_L(n)$}{}} \label{app:branching}

We consider a spin chain of
$L$ sites with $2r$ through-lines. Brauer modules $B^{(L)}_\lambda$ carry an irreducible representation of the symmetric group
$S_{|\lambda|}$ on the $|\lambda|=2r$ through-lines. 
We want to decompose them into representations $W^{(L)}_{(r,s)}$ of the unoriented Jones--Temperley--Lieb algebra. Diagrammatically, this amounts to understanding what happens when we forbid line crossings.

As a function of $L$, $\dim B^{(L)}_\lambda$ \eqref{dimBrauer} increases super-exponentially while $\dim W^{(L)}_{(r,s)}$ \eqref{dimATL} increases only exponentially. Therefore, as $L$ increases, more and more $\uJTL_L(n)$ modules appear in a given $\mathcal{B}_L(n)$ at fixed $\lambda$. However, we know that the coefficient $c^\lambda_{(r,s)}$ of a given $\uJTL_L(n)$ module is $L$-independent: we do not get more copies of the same modules, but extra modules with increasing values of $r$ within the bounds $|\lambda|\leq 2r\leq L$. We may therefore take the limit $L\to \infty$ of the branching rule \eqref{bcw}, and we obtain
\begin{align}
B_{\lambda} \underset{\uJTL(n)}{=} \bigoplus_{r=\frac{|\lambda|}{2}}^{\infty} \bigoplus_{\substack{s\in\frac{1}{r}\mathbb{Z} \\ -1< s\leq 1}} c^\lambda_{(r,s)} W_{(r,s)}
 \ .
\end{align}
where we omit the dependence on $L=\infty$.
For example, $B_{[3,1]} \underset{\uJTL(n)}{=} W_{(2,\pm \frac12)} \oplus W_{(2,1)} \oplus \ldots$, where we introduce the notation 
\begin{align}
 W_{(r,\pm s)} = W_{(r,s)} \oplus W_{(r,-s)}\ .
\end{align}

\subsubsection{Algorithm}

We first construct a basis for $B^{(L)}_\lambda$. A suitable linear operator ${\cal O}^{(L)}$ belonging to
$\uJTL_L$ is then numerically diagonalised within the corresponding space, giving rise to a spectrum
$\Lambda_{\cal O}(L,\lambda)$. Diagonalising next the same operator ${\cal O}^{(L)}$ within each of the representations
$W^{(L)}_{(r,s)}$ with $2r \le L$ and any allowed $s$ gives rise to the spectra $\Lambda_{\cal O}(L,r,s)$. It is then found that $\Lambda_{\cal O}(L,\lambda)$
is the union of all the $\Lambda_{\cal O}(L,r,s)$, and the multiplicity of each eigenvalue in $\Lambda_{\cal O}(L,\lambda)$
reveals the corresponding integer coefficient $c_{(r,s)}^\lambda$ on the right-hand side of the branching rule.

For this strategy to be fully successful we should choose ${\cal O}^{(L)}$ such that the spectrum
$\Lambda_{\cal O}(L,r,s)$ is simple for any $(r,s)$ and such that the different spectra $\Lambda_{\cal O}(L,r,s)$ have
no coinciding eigenvalues. It turns out a good choice to take ${\cal O}^{(L)}$ as the row-to-row 
transfer matrix of the $\uJTL(n)$ loop model on the square lattice.
For a generic value $n>0$ of the loop weight, it turns out that the spectra $\Lambda_{\cal O}(L,r,s)$ are real, simple and
non-overlapping, except for the symmetry $\Lambda_{\cal O}(L,r,s) = \Lambda_{\cal O}(L,r,-s)$. This is however not a problem,
since the branching coefficients themselves also do not depend on the sign of $s$.

We now give a few more details. Each element of the basis for $B^{(L)}_\lambda$ consists of a link pattern in which crossings are allowed, plus some extra information that will be constructed to impose that the irreducible representation $\lambda$ lives on the $2r=|\lambda|$ through-lines. We first construct the
set of link patterns consisting of $2r$ through-lines and a perfect matching (which may include crossings)
of the remaining $L-2r$ points. This vector space of link patterns is denoted ${\cal S}_0$ and has dimension
$\binom{L }{L-|\lambda|} (L-|\lambda|-1)!!$. We may endow each of the link patterns in ${\cal S}_0$ with an extra label,
which can take $f_\lambda$ possible values, in order to have a space of total dimension \eqref{dimBrauer}.
We call this the vector space of labelled link patterns and denote it ${\cal S}_1$.

To diagonalise ${\cal O}^{(L)}$ within
${\cal S}_1$, we use the Arnoldi method for non-symmetric matrices, which is an iterative eigenvalue method
that only relies on providing a method (or function, in algorithmic parlance) that computes the vector $v_2 = {\cal O}^{(L)} v_1$
from any given vector $v_1 \in {\cal S}_1$. To provide this method we shall need to introduce
a larger space, the space of link patterns with distinguishable through-lines, denoted ${\cal S}_2$,
which will be used in intermediary stages of the computation of $v_2$. The idea is that each state
in ${\cal S}_2$ is a link pattern in which any one of its through-lines carries a distinct ``mark'' from the set
$\{1,2,\ldots,2r\}$. Here is an illustration of two such link patterns from $B^{(10)}_\lambda$ with
$|\lambda|=4$, differing only by the marks on the through-lines:
\begin{equation}
\scalebox{0.75}{\begin{tikzpicture}[scale=.55,line width=1.35pt] 
\foreach \i in {1,2,3,4,5,6,7,8,9,10}
{ \fill (\i,2) circle (4.5pt); }
\draw (1,2) .. controls +(2,-1.5) .. (5,2);
\draw (3,2) .. controls +(2,-1.5) .. (7,2);
\draw (8,2) .. controls +(1,-1.0) .. (10,2);
\draw (2,2) -- (2,0);
\draw (4,2) -- (4,0);
\draw (6,2) -- (6,0);
\draw (9,2) -- (9,0);
\draw (2,0) node[below] {$1$};
\draw (4,0) node[below] {$2$};
\draw (6,0) node[below] {$3$};
\draw (9,0) node[below] {$4$};
\end{tikzpicture}} \qquad \qquad
\scalebox{0.75}{\begin{tikzpicture}[scale=.55,line width=1.35pt] 
\foreach \i in {1,2,3,4,5,6,7,8,9,10}
{ \fill (\i,2) circle (4.5pt); }
\draw (1,2) .. controls +(2,-1.5) .. (5,2);
\draw (3,2) .. controls +(2,-1.5) .. (7,2);
\draw (8,2) .. controls +(1,-1.0) .. (10,2);
\draw (2,2) -- (2,0);
\draw (4,2) -- (4,0);
\draw (6,2) -- (6,0);
\draw (9,2) -- (9,0);
\draw (2,0) node[below] {$3$};
\draw (4,0) node[below] {$2$};
\draw (6,0) node[below] {$4$};
\draw (9,0) node[below] {$1$};
\end{tikzpicture}}
\end{equation}
In this way, each link pattern in ${\cal S}_2$ is endowed with an element of the symmetric group
$S_{2r}$, enabling us to follow how the (now distinguishable) through-lines are permuted by the action of ${\cal O}^{(L)}$.
In the space of permutations of $\{1,2,\ldots,2r\}$ we finally construct a $(2r)!$-dimensional matrix
representation $\Pi_\lambda$ of the Young projector on $\lambda$. Choosing an overall
normalising factor of $\frac{f_\lambda}{|\lambda|!}$ we find that $\Pi_\lambda$ is an idempotent of rank
$f_\lambda$ (it has $f_\lambda$ unit eigenvalues). Let $\{w^{\Pi_\lambda}_i\}_{i=1}^{f_\lambda}$ be the corresponding
set of orthonormalised eigenvectors.

The multiplication method of the Arnoldi algorithm is now decomposed into the following steps.
Given the vector $v_1 \in {\cal S}_1$, we map each of the corresponding basis states with label $i$ 
into a linear combination of $(2r)!$ basis states in ${\cal S}_2$, using the coefficients coming from
the eigenvector $w^{\Pi_\lambda}_i$. In this way $v_1$ has been ``unfolded'' into the larger space ${\cal S}_2$. We then multiply
this vector by ${\cal O}^{(L)}$, using the hash-table data structures and the sparse-matrix factorisation
techniques which have been extensively described in \cite{js18} (Appendix A). In this process the
through-lines carrying their respective marks will undergo permutations.
The result after multiplication by ${\cal O}^{(L)}$ is the vector $v_2$,
but living in the enlarged space ${\cal S}_2$. To complete the computation, we ``refold'' $v_2$ into the
smaller space ${\cal S}_1$ by projecting on the eigenvectors $\{w^{\Pi_\lambda}_i\}$.

The result of this diagonalisation of ${\cal O}^{(L)}$ within ${\cal S}_1$ is the spectrum
$\Lambda_{\cal O}(L,\lambda)$ corresponding to the representation $B^{(L)}_\lambda$.
The spectra $\Lambda_{\cal O}(L,r,s)$ corresponding to each representation $W^{(L)}_{(r,s)}$
are obtained by more elementary means, which have been well described in \cite{js18} (Appendix A).
Comparing the two sets of spectra we finally deduce the branching rules.

\subsubsection{Results}

We have computed the branching rules of $B^{(L)}_\lambda$ up to size $L=10$
and for all representations with $|\lambda| \le 4$, with the results:
\begin{subequations}
\begin{align}
 B_{[]} &\underset{\uJTL(n)}{=} W_{(0,0)} \oplus W_{(2,0)} \oplus 2 W_{(3,0)} \oplus W_{(3,\pm \frac23)} \\
 & \oplus \ 7 W_{(4,0)} \oplus 2 W_{(4,\pm \frac14)} \oplus 5 W_{(4,\pm \frac12)} \oplus 2 W_{(4,\pm \frac34)} \oplus 6 W_{(4,1)} \nonumber \\
 & \oplus \ 36 W_{(5,0)} \oplus 24 W_{(5,\pm \frac15)} \oplus 34 W_{(5,\pm \frac25)} \oplus 24 W_{(5,\pm \frac35)} \oplus 34 W_{(5,\pm \frac45)} \oplus 25 W_{(5,1)} \oplus \ldots \nonumber \\
 B_{[1]} &\underset{\uJTL(n)}{=} W_{(\frac12,0)} \oplus W_{(\frac52,0)} \oplus W_{(\frac52,\pm \frac25)} \oplus W_{(\frac52,\pm \frac45)} \\
 & \oplus \ 5 W_{(\frac72,0)} \oplus 5 W_{(\frac72,\pm \frac27)} \oplus 5 W_{(\frac72,\pm \frac47)} \oplus 5 W_{(\frac72,\oplus \frac67)} \nonumber \\
 & \oplus \ 36 W_{(\frac92,0)} \oplus 36 W_{(\frac92,\pm \frac29)} \oplus 36 W_{(\frac92,\pm \frac49)} \oplus 36 W_{(\frac92,\pm \frac23)} \oplus 36 W_{(\frac92,\pm \frac89)} \oplus \ldots \nonumber \\
 B_{[2]} &\underset{\uJTL(n)}{=} W_{(1,0)} \oplus W_{(2,0)} \oplus W_{(2,1)} \oplus 4 W_{(3,0)} \oplus 2 W_{(3,\pm \frac13)} \oplus 4 W_{(3,\pm \frac23)} \oplus 2 W_{(3,1)} \\
 & \oplus \ 23 W_{(4,0)} \oplus 18 W_{(4,\pm \frac14)} \oplus 23 W_{(4,\pm \frac12)} \oplus 18 W_{(4,\pm \frac34)} \oplus 23 W_{(4,1)} \nonumber \\
 & \oplus \ 191 W_{(5,0)} \oplus 174 W_{(5,\pm \frac15)} \oplus 191 W_{(5,\pm \frac25)} \oplus 174 W_{(5,\pm \frac35)} \oplus 191 W_{(5,\pm \frac45)} \oplus 174 W_{5,1} \oplus \ldots \nonumber \\
 B_{[1^2]} &\underset{\uJTL(n)}{=} W_{(1,1)} \oplus W_{(2,\pm \frac12)} \oplus 2 W_{(3,0)} \oplus 4 W_{(3,\pm \frac13)} \oplus 2 W_{(3,\pm \frac23)} \oplus 4 W_{(3,1)} \\
 & \oplus \ 18 W_{(4,0)} \oplus 23 W_{(4,\pm \frac14)} \oplus 18 W_{(4,\pm \frac12)} \oplus 23 W_{(4,\pm \frac34)} \oplus 18 W_{(4,1)} \nonumber \\
 & \oplus \ 174 W_{(5,0)} \oplus 191 W_{(5,\pm \frac15)} \oplus 174 W_{(5,\pm \frac25)} \oplus 191 W_{(5,\pm \frac35)} \oplus 174 W_{(5,\pm \frac45)} \oplus 191 W_{(5,1)} \oplus \ldots \nonumber \\
 B_{[3]} &\underset{\uJTL(n)}{=} W_{(\frac32,0)} \oplus W_{(\frac52,0)} \oplus W_{(\frac52,\pm \frac25)} \oplus W_{(\frac52,\pm \frac45)} \\
 & \oplus \ 7 W_{(\frac72,0)} \oplus 7 W_{(\frac72,\pm \frac27)} \oplus 7 W_{(\frac72,\pm \frac47)} \oplus 7 W_{(\frac72,\pm \frac67)} \nonumber \\
 & \oplus \ 63 W_{(\frac92,0)} \oplus 61 W_{(\frac92,\pm \frac29)} \oplus 61 W_{(\frac92,\pm \frac49)} \oplus 63 W_{(\frac92,\pm \frac23)} \oplus 61 W_{(\frac92,\pm \frac89)} \oplus \ldots \nonumber \\
 B_{[21]} &\underset{\uJTL(n)}{=} W_{(\frac32,\pm \frac23)} \oplus 2 W_{(\frac52,0)} \oplus 2 W_{(\frac52,\pm \frac25)} \oplus 2 W_{(\frac52,\pm \frac45)} \\
 & \oplus \ 14 W_{(\frac72,0)} \oplus 14 W_{(\frac72,\pm \frac27)} \oplus 14 W_{(\frac72,\pm \frac47)} \oplus 14 W_{(\frac72,\pm \frac67)} \nonumber \\
 & \oplus \ 122 W_{(\frac92,0)} \oplus 124 W_{(\frac92,\pm \frac29)} \oplus 124 W_{(\frac92,\pm \frac49)} \oplus 122 W_{(\frac92,\pm \frac23)} \oplus 124 W_{(\frac92,\pm \frac89)} \oplus \ldots \nonumber \\
 B_{[1^3]} &\underset{\uJTL(n)}{=} W_{(\frac32,0)} \oplus W_{(\frac52,0)} \oplus W_{(\frac52,\pm \frac25)} \oplus W_{(\frac52,\pm \frac45)} \\
 & \oplus \ 7 W_{(\frac72,0)} \oplus 7 W_{(\frac72,\pm \frac27)} \oplus 7 W_{(\frac72,\pm \frac47)} \oplus 7 W_{(\frac72,\pm \frac67)} \nonumber \\
 & \oplus \ 63 W_{(\frac92,0)} \oplus 61 W_{(\frac92,\pm \frac29)} \oplus 61 W_{(\frac92,\pm \frac49)} \oplus 63 W_{(\frac92,\pm \frac23)} \oplus 61 W_{(\frac92,\pm \frac89)} \oplus \ldots \nonumber \\
 B_{[4]} &\underset{\uJTL(n)}{=} W_{(2,0)} \oplus 2 W_{(3,0)} \oplus W_{(3,\pm \frac13)} \oplus 2 W_{(3,\pm \frac23)} \oplus W_{(3,1)} \\
 & \oplus \ 16 W_{(4,0)} \oplus 12 W_{(4,\pm \frac14)} \oplus 15 W_{(4,\pm \frac12)} \oplus 12 W_{(4,\pm \frac34)} \oplus 16 W_{(4,1)} \nonumber \\
 & \oplus \ 158 W_{(5,0)} \oplus 147 W_{(5,\pm \frac15)} \oplus 158 W_{(5,\pm \frac25)} \oplus 147 W_{(5,\pm \frac35)} \oplus 158 W_{(5,\pm \frac45)} \oplus 147 W_{(5,1)} \oplus \ldots \nonumber \\
 B_{[31]} &\underset{\uJTL(n)}{=} W_{(2,\pm \frac12)} \oplus W_{(2,1)} \oplus 4 W_{(3,0)} \oplus 5 W_{(3,\pm \frac13)} \oplus 4 W_{(3,\pm \frac23)} \oplus 5 W_{(3,1)} \\
 & \oplus \ 39 W_{(4,0)} \oplus 43 W_{(4,\pm \frac14)} \oplus 40 W_{(4,\pm \frac12)} \oplus 43 W_{(4,\pm \frac34)} \oplus 39 W_{(4,1)} \nonumber \\
 & \oplus \ 452 W_{(5,0)} \oplus 463 W_{(5,\pm \frac15)} \oplus 452 W_{(5,\pm \frac25)} \oplus 463 W_{(5,\pm \frac35)} \oplus 452 W_{(5,\pm \frac45)} \oplus 463 W_{(5,1)} \oplus \ldots \nonumber \\
 B_{[2^2]} &\underset{\uJTL(n)}{=} W_{(2,0)} \oplus W_{(2,1)} \oplus 4 W_{(3,0)} \oplus 2 W_{(3,\pm \frac13)} \oplus 4 W_{(3,\pm \frac23)} \oplus 2 W_{(3,1)} \\
 & \oplus \ 31 W_{(4,0)} \oplus 24 W_{(4,\pm \frac14)} \oplus 31 W_{(4,\pm \frac12)} \oplus 24 W_{(4,\pm \frac34)} \oplus 31 W_{(4,1)} \nonumber \\
 & \oplus \ 316 W_{(5,0)} \oplus 294 W_{(5,\pm \frac15)} \oplus 316 W_{(5,\pm \frac25)} \oplus 294 W_{(5,\pm \frac35)} \oplus 316 W_{(5,\pm \frac45)} \oplus 294 W_{(5,1)} \oplus \ldots \nonumber \\
 B_{[21^2]} &\underset{\uJTL(n)}{=} W_{(2,0)} \oplus W_{(2,\pm \frac12)} \oplus 4 W_{(3,0)} \oplus 5 W_{(3,\pm \frac13)} \oplus 4 W_{(3,\pm \frac23)} \oplus 5 W_{(3,1)} \\
 & \oplus \ 40 W_{(4,0)} \oplus 43 W_{(4,\pm \frac14)} \oplus 39 W_{(4,\pm \frac12)} \oplus 43 W_{(4,\pm \frac34)} \oplus 40 W_{(4,1)} \nonumber \\
 & \oplus \ 452 W_{(5,0)} \oplus 463 W_{(5,\pm \frac15)} \oplus 452 W_{(5,\pm \frac25)} \oplus 463 W_{(5,\pm \frac35)} \oplus 452 W_{(5,\pm \frac45)} \oplus 463 W_{(5,1)} \oplus \ldots \nonumber \\
 B_{[1^4]} &\underset{\uJTL(n)}{=} W_{(2,1)} \oplus 2 W_{(3,0)} \oplus W_{(3,\pm \frac13)} \oplus 2 W_{(3,\pm \frac23)} \oplus W_{(3,1)} \\
 & \oplus \ 15 W_{(4,0)} \oplus 12 W_{(4,\pm \frac14)} \oplus 16 W_{(4,\pm \frac12)} \oplus 12 W_{(4,\pm \frac34)} \oplus 15 W_{(4,1)} \nonumber \\
 & \oplus \ 158 W_{(5,0)} \oplus 147 W_{(5,\pm \frac15)} \oplus 158 W_{(5,\pm \frac25)} \oplus 147 W_{(5,\pm \frac35)} \oplus 158 W_{(5,\pm \frac45)} \oplus 147 W_{(5,1)} \oplus \ldots \nonumber 
\end{align}
\end{subequations}

\subsubsection{Observations}

We recognise here the coefficients $c^\lambda_{(r,s)}$ computed in section~\ref{sec:obr}.
In addition we make a number of observations:
\begin{itemize}
 \item The results for $B_{[3]}$ and $B_{[1^3]}$ are identical. This is because the even and odd permutations have the same sign in the Young projector $\Pi_{[3]}$ and opposite signs in $\Pi_{[1^3]}$, and therefore cannot mix under the action of the cyclic group.
 \item The results for $B_{[4]}$ and $B_{[1^4]}$ coincide for $r$ odd. The results for $B_{[2^2]}$ and $r$ odd have coefficients that are larger by a factor of ${\rm dim} \, [2^2] = 2$.
 \item The results for $B_{[31]}$ and $B_{[21^2]}$ coincide for $r$ odd.
\end{itemize}

\subsection{Branching rules \texorpdfstring{$\mathcal{P}_L(Q)\downarrow \PTL_{2L}(\sqrt{Q})$}{}}

Let us briefly describe our algorithm in this case, while focussing on the differences with respect to the case of $\mathcal{B}_L(n) \downarrow \uJTL_L(n)$ in Appendix~\ref{app:branching}.

We consider a system of $L$ sites (or Potts spins). Partition-algebra modules $P^{(L)}_\lambda$ have
a basis of set partitions of the $L$ sites, with $r\in \mathbb{N}$ propagating blocks. The propagating blocks carry an irreducible representation
of the symmetric group $S_{|\lambda|}$ with $|\lambda| = r$.
We want to decompose $P^{(L)}_\lambda$ into representations $W^{(2L)}_{(r,s)}$ of $\PTL_{2L}(\sqrt{Q})$, with $|\lambda|\leq r\leq L$. 
We consider the limit $L\to \infty$ and omit the dependence on $L$. 
We have $r\geq 2$ except in the cases $P_{[]} \underset{\PTL(\sqrt{Q})}{=} W_{(0, 0)} \oplus \cdots$ and $P_{[1]} \underset{\PTL(\sqrt{Q})}{=} W_{(0,\frac12)} \oplus \cdots$.

\subsubsection{Algorithm}

We first construct a basis of set partitions of $\{1,2,\ldots,L\}$ with $|\lambda|$
propagating (but indistinguishable) blocks, defining the vector space ${\cal S}_0$. Each of the set partitions in ${\cal S}_0$
is then endowed with an extra label which can take $f_\lambda$ possible values, defining the vector space
of labelled set partitions ${\cal S}_1$. This ${\cal S}_1$ has the same dimension \eqref{dimpartrep} as
the module $P^{(L)}_\lambda$. We then wish to diagonalise a suitable linear operator ${\cal O}^{(L)}$
within ${\cal S}_1$, in order to obtain the spectrum $\Lambda_{\cal O}(L,\lambda)$ corresponding to the
representation $P^{(L)}_\lambda$. This is done, as in Appendix~\ref{app:branching}, by providing the
multiplication method of the Arnoldi algorithm. In this method, ${\cal S}_1$ is first unfolded---by means of
the orthonormal eigenvectors $\{w^{\Pi_\lambda}_i\}_{i=1}^{f_\lambda}$ of the Young projector $\Pi_\lambda$---into
an enlarged space ${\cal S}_2$ in which each of the propagating blocks carries a distinct mark (each
of the sites in a given block carries the same mark). After the multiplication by ${\cal O}^{(L)}$, acting
within ${\cal S}_2$, the result is folded back into ${\cal S}_1$ by projecting on the eigenvectors
$w^{\Pi_\lambda}_i$.

We find that an appropriate choice of the linear operator ${\cal O}^{(L)}$ is that of the row-to-row transfer
matrix of the $Q$-state cluster model on an axially oriented square lattice with periodic boundary conditions.
Notice that this is a product of $L$ operators adding the horizontal edges and another $L$ operators adding
the vertical ones, each edge operator being a sparse matrix expressed in terms of join and detach operators \cite{js18}.
For generic values of $Q>0$, the operator ${\cal O}^{(L)}$ has the advantage that the spectra do not contain any accidental degeneracies. The branching rules can therefore be unambiguously read off from the multiplicities in
$\Lambda_{\cal O}(L,\lambda)$ and by comparison with the spectra $\Lambda_{\cal O}(2L,r,s)$ of the Potts model.

\subsubsection{Results}

For the fully symmetric representations $P_{[p]}$ we find then the branching rules
\begin{subequations}
\label{symdown}
\begin{align}
 P_{[]} &\underset{\PTL(\sqrt{Q})}{=} W_{(0,0)} \oplus W_{(4,0)} \oplus 3 W_{(6,0)} \oplus W_{(6,\pm \frac13)} \oplus \cdots \\
 P_{[1]} &\underset{\PTL(\sqrt{Q})}{=} W_{(0,\frac12)} \oplus W_{(4,0)} \oplus W_{(4,\frac12)} \oplus W_{(5,0)} \oplus W_{(5,\pm \frac15)} \oplus W_{(5,\pm \frac25)} \\
  & \oplus \ 5 W_{(6,0)} \oplus 2 W_{(6,\pm \frac16)} \oplus 4 W_{(6,\pm \frac13)} \oplus 3 W_{(6,\frac12)} \oplus \cdots \nonumber \\
 P_{[2]} &\underset{\PTL(\sqrt{Q})}{=} W_{(2,0)} \oplus 2 W_{(4,0)} \oplus W_{(4,\frac12)} \oplus 2 W_{(5,0)} \oplus 2 W_{(5,\pm \frac15)} \oplus 2W_{(5,\pm \frac25)} \oplus \\
 & \oplus \ 10 W_{(6,0)} \oplus 5 W_{(6,\pm \frac16)} \oplus 9 W_{(6,\pm \frac13)} \oplus 5 W_{(6,\frac12)} \oplus \cdots \nonumber \\
 P_{[3]} &\underset{\PTL(\sqrt{Q})}{=} W_{(3,0)} \oplus W_{(4,0)} \oplus W_{(4,\frac 12)} \oplus 2 W_{(5,0)} \oplus 2 W_{(5,\pm \frac15)} \oplus 2 W_{(5,\pm \frac25)} \\
 & \oplus \ 11 W_{(6,0)} \oplus 6 W_{(6,\pm \frac16)} \oplus 9 W_{(6,\pm \frac13)} \oplus 7 W_{(6,\frac 12)} \oplus \cdots \nonumber \\
 P_{[4]} &\underset{\PTL(\sqrt{Q})}{=} W_{(4,0)} \oplus W_{(5,0)} \oplus W_{(5,\pm \frac15)} \oplus W_{(5,\pm \frac25)} \\
 & \oplus \ 7 W_{(6,0)} \oplus 3 W_{(6,\pm \frac16)} \oplus 6 W_{(6,\pm \frac13)} \oplus 4 W_{(6,\frac12)} \oplus \cdots \nonumber \\
 P_{[5]} &\underset{\PTL(\sqrt{Q})}{=} W_{(5,0)} \oplus 2 W_{(6,0)} \oplus W_{(6,\pm \frac16)} \oplus 2 W_{(6,\pm \frac13)} \oplus W_{(6,\frac12)} \oplus \cdots
\end{align}
\end{subequations}
For the fully antisymmetric representations $P_{[1^p]}$ we find
\begin{subequations}
\label{asymdown}
\begin{align}
 P_{[1^2]} &\underset{\PTL(\sqrt{Q})}{=} W_{(2,\frac12)} \oplus W_{(4,\pm \frac14)} \oplus W_{(4,\frac12)} \oplus 
 2 W_{(5,0)} \oplus 2 W_{(5,\pm \frac15)} \oplus 2 W_{(5,\pm \frac25)} \\
 & \oplus \ 6 W_{(6,0)} \oplus 8 W_{(6,\pm \frac16)} \oplus 6 W_{(6,\pm \frac13)} \oplus 9 W_{(6,\frac12)} \oplus \cdots \nonumber \\
 P_{[1^3]} &\underset{\PTL(\sqrt{Q})}{=} W_{(3,0)} \oplus W_{(4,\pm \frac14)} \oplus 2 W_{(5,0)} \oplus 2 W_{(5,\pm \frac15)} \oplus 2 W_{(5,\pm \frac25)} \\
 & \oplus \ 8 W_{(6,0)} \oplus 9 W_{(6,\pm \frac16)} \oplus 6 W_{(6,\pm \frac13)} \oplus 10 W_{(6,\frac12)} \oplus \cdots \nonumber \\
 P_{[1^4]} &\underset{\PTL(\sqrt{Q})}{=} W_{(4,\frac12)} \oplus W_{(5,0)} \oplus W_{(5,\pm \frac15)} \oplus W_{(5,\pm \frac25)} \\
 & \oplus \ 6 W_{(6,0)} \oplus 4 W_{(6,\pm \frac16)} \oplus 5 W_{(6,\pm \frac13)} \oplus 5 W_{(6,\frac12)} \oplus \cdots \nonumber \\
 P_{[1^5]} &\underset{\PTL(\sqrt{Q})}{=} W_{(5,0)} \oplus 2 W_{(6,0)} \oplus W_{(6,\pm \frac16)} \oplus 2 W_{(6,\pm \frac13)} \oplus W_{(6,\frac12)} \oplus \cdots 
\end{align}
\end{subequations}
We have also computed some, but not all, of the cases with $f_\lambda > 1$ and $|\lambda| \le 6$:
\begin{subequations}
\label{mixsymdown}
\begin{align}
 P_{[21]} &\underset{\PTL(\sqrt{Q})}{=} W_{(3,\pm \frac13)} \oplus W_{(4,0)} \oplus W_{(4,\pm \frac14)} \oplus W_{(4,\frac12)} \oplus  4 W_{(5,0)} \oplus 4 W_{(5,\pm \frac15)} \oplus 4 V_{(5,\pm \frac25)} \\
 & \oplus \ 15 W_{(6,0)} \oplus 16 W_{(6,\pm \frac16)} \oplus 17 W_{(6,\pm \frac13)} \oplus 15 W_{(6,\frac12)} \oplus \cdots \nonumber \\
 P_{[31]} &\underset{\PTL(\sqrt{Q})}{=} W_{(4,\pm \frac14)} \oplus W_{(4,\frac12)} \oplus 3 W_{(5,0)} + 3 W_{(5,\pm \frac15)} + 3 V_{(5,\pm \frac25)} \\
 & \oplus \ 14 W_{(6,0)} \oplus 15 W_{(6,\pm \frac16)} \oplus 14 W_{(6,\pm \frac13)} \oplus 15 W_{(6,\frac12)} \oplus \cdots \nonumber \\
 P_{[2^2]} &\underset{\PTL(\sqrt{Q})}{=} W_{(4,0)} \oplus W_{(4,\frac12)} \oplus 2 W_{(5,0)} + 2 W_{(5,\pm \frac15)} + 2 W_{(5,\pm \frac25)} \\
 & \oplus \ 11 W_{(6,0)} \oplus 8 W_{(6,\pm \frac16)} \oplus 12 W_{(6,\pm \frac13)} \oplus 7 W_{(6,\frac12)} \oplus \cdots \nonumber \\
 P_{[21^2]} &\underset{\PTL(\sqrt{Q})}{=} W_{(4,0)} \oplus W_{(4,\pm \frac14)} \oplus 3 W_{(5,0)} + 3 W_{(5,\pm \frac15)} + 3 V_{(5,\pm \frac25)} \\
 & \oplus \ 13 W_{(6,0)} \oplus 16 W_{(6,\pm \frac16)} \oplus 13 W_{(6,\pm \frac13)} \oplus 16 W_{(6,\frac12)} \oplus \cdots \nonumber \\
 P_{[41]} &\underset{\PTL(\sqrt{Q})}{=} W_{(5,\pm \frac15)} \oplus W_{(5,\pm \frac25)} \oplus 6 W_{(6,0)} \oplus 6 W_{(6,\pm \frac16}) \oplus 6 W_{(6,\pm \frac13)} \oplus 6 W_{(6,\frac12)} \oplus \cdots \\
 P_{[3,2]} &\underset{\PTL(\sqrt{Q})}{=} W_{(5,0)} \oplus W_{(5,\pm \frac15)} \oplus W_{(5,\pm \frac25)} \oplus 8 W_{(6,0)} \oplus 7 W_{(6,\pm \frac16)} \oplus 8 W_{(6,\pm \frac13)} \oplus 7 W_{(6,\frac12)} \oplus \cdots \\
 P_{[31^2]} &\underset{\PTL(\sqrt{Q})}{=} 2 W_{(5,0)} \oplus W_{(5,\pm \frac15)} \oplus W_{(5,\pm \frac25)} \oplus 8 W_{(6,0)} \oplus 10 W_{(6,\pm \frac16)} \oplus 8 W_{(6,\pm \frac13)} \oplus 10 W_{(6,\frac12)} \oplus \cdots \\
 P_{[2^21]} &\underset{\PTL(\sqrt{Q})}{=} W_{(5,0)} \oplus W_{(5,\pm \frac15)} \oplus W_{(5,\pm \frac25)} \oplus 8 W_{(6,0)} \oplus 7 W_{(6,\pm \frac16)} \oplus 8 W_{(6,\pm \frac13)} \oplus 7 W_{(6,\frac12)} \oplus \cdots \\
 P_{[21^3]} &\underset{\PTL(\sqrt{Q})}{=} W_{(5,\pm \frac15)} \oplus W_{(5,\pm \frac25)} \oplus 6 W_{(6,0)} \oplus 6 W_{(6,\pm \frac16)} \oplus 6 W_{(6,\pm \frac13)} \oplus 6 W_{(6,\frac12)} \oplus \cdots \\
 P_{[51]} &\underset{\PTL(\sqrt{Q})}{=} W_{(6,\pm \frac16)} \oplus W_{(6,\pm \frac13)} \oplus W_{(6,\frac12)} \oplus \cdots \\
 P_{[42]} &\underset{\PTL(\sqrt{Q})}{=} 2 W_{(6,0)} \oplus W_{(6,\pm \frac16)} \oplus 2 W_{(6,\pm \frac13)} \oplus W_{(6,\frac12)} \oplus \cdots
\end{align}
\end{subequations}

\subsubsection{Observations}

We recognise here the coefficients $d^\lambda_{(r,s)}$ computed in Section~\ref{sec:sbr}.
In particular, combining \eqref{symdown}--\eqref{mixsymdown}
with a dimensional count is enough to uniquely determine the $\Lambda_{(r,s)}$ for all $r \le 6$, thus recovering
all of the results displayed in \eqref{x-all}.

Concerning the terms on the right-hand side with the minimal value of $r$, that is $r=|\lambda|$, we can go a bit further.
In that case we can infer some families of branching coefficients, in which $\lambda$ consists of one long
row followed by one or two short rows. This proceeds by studying carefully the corresponding Young projector
$\Pi_\lambda$, either by hand or aided by {\sc Mathematica} for the more complicated cases, but without
employing the above method of numerically diagonalizing the transfer matrix.

We first find that
\begin{equation}
 d^{[L-1,1]}_{(L,\frac{k}{L})} = \begin{cases} 0 & \mbox{for } k = 0 \,, \\ 1 & \mbox{otherwise} \,. \end{cases}
\end{equation}
The sum of multiplicities is of course $\sum_{k=0}^{L-1} d^{[L-1,1]}_{(L,\frac{k}{L})} = L-1$, the dimension of the representation $\lambda = [L-1,1]$ on the left-hand side.
Similarly we find for $P_{[L-2,2]}$ the multiplicities
\begin{equation}
\label{dLm22}
 d^{[L-2,2]}_{(L,\frac{k}{L})} =
 \begin{cases} \left\lfloor \frac{L-2}{2} \right\rfloor & \mbox{for } k \mbox{ even} \,, \\
 \left\lfloor \frac{L-3}{2} \right\rfloor & \mbox{for } k \mbox{ odd} \,, \end{cases}
\end{equation}
where $\lfloor x \rfloor$ denotes the integer part of $x$. Their sum is $\frac12 L(L-3)$, as it should be by
the hook length formula \eqref{hooklf}. The result \eqref{dLm22} has been checked up to $L=8$.
For $P_{[L-2,1,1]}$ we find
\begin{equation}
 d^{[L-2,1,1]}_{(L,\frac{k}{L})} =
 \begin{cases}
 1 + \left\lfloor \frac{L-3}{2} \right\rfloor & \mbox{for } k=0 \,, \\
 \left\lfloor \frac{L-3}{2} \right\rfloor & \mbox{for } k \neq 0 \mbox{ even} \,, \\
 \left\lfloor \frac{L-2}{2} \right\rfloor & \mbox{for } k \mbox{ odd} \,, \end{cases} 
\end{equation}
and the sum of multiplicities is $\frac12 (L-1)(L-2)$. This result has been checked up to $L=7$.
And finally we obtain
\begin{equation}
 d^{[L-3,2,1]}_{(L,\frac{k}{L})} =
 \begin{cases}
 \left\lfloor \frac{(L-3)^2}{3} \right\rfloor & \mbox{for } 3 \nmid k \,, \\[6pt]
 \left\lfloor \frac{(L-3)^2}{3} \right\rfloor & \mbox{for } 3 \mid k \mbox{ and } 3 \nmid L \,, \\[6pt]
 \left\lfloor \frac{(L-3)^2}{3} - 1 \right\rfloor & \mbox{for } 3 \mid k \mbox{ and } 3 \mid L \,, \end{cases} 
\end{equation}
with total multiplicity $\frac13 L(L-2)(L-4)$. This has been checked up to $L=7$.

\bibliographystyle{../../inputs/morder7}
\bibliography{../../inputs/992}

\end{document}